\documentclass{elsarticle}
\usepackage[utf8]{inputenc}
\usepackage[font=normalsize, labelfont=bf]{caption}
\usepackage{subcaption}

\usepackage{graphicx}
\usepackage{subcaption}
\usepackage{caption}
\usepackage{setspace}
\usepackage{lineno}
\usepackage{amssymb}
\usepackage{amsmath}
\usepackage{upgreek}
\usepackage{pgfplots}
\usepackage{bm}
\usepackage{soul,color}
\usepackage{epstopdf}
\usepackage{units}
\usepackage{lscape}
\usepackage{booktabs}
\usepackage{gensymb}
\usepackage{multirow}
\usepackage{floatrow}
\usepackage[flushleft]{threeparttable}
\usepackage{tikz}
\usepackage{placeins}
\usepackage{natbib}
\usepackage{array}
\usepackage{tabularx}
\usepackage{algorithm}
\usepackage{algpseudocode}
\usepackage[a4paper]{geometry}
\usepackage{hyperref}
\usepackage[colorinlistoftodos]{todonotes}

\modulolinenumbers[5]
\pgfplotsset{compat=1.5}
\usepgfplotslibrary{units}
\usetikzlibrary{spy}
\usetikzlibrary{pgfplots.groupplots}

\soulregister\cite7
\soulregister\ref7
\soulregister\pageref7

\hypersetup{
    colorlinks=true,
    linkcolor=blue,
    citecolor=blue,
    urlcolor=blue,
    pdftitle={Wall Wettability Control of Cavitation Patterns and Flow Stability},
    pdfauthor={Mahmood Mousavi et al.}
}

\journal{----}

\begin{document}

\begin{frontmatter}

\title{Wall Wettability Control of Cavitation Patterns and Stability}
\author[1]{Parisa Tayerani}
\author[2]{Mahmood Mousavi\corref{mycorrespondingauthor}}
\cortext[mycorrespondingauthor]{Corresponding author: \href{mailto:mousas@bgsu.edu}{mousas@bgsu.edu}}
\author[1]{Mahmoud Pasandidehfard}

\author[3]{Ehsan Roohi}

\address[1]{Department of Mechanical Engineering, Ferdowsi University of Mashhad, Mashhad, 91775-1111, Iran}

\address[2]{Mechanical and Manufacturing Engineering Program, School of Engineering, Bowling Green State University, Bowling Green, Ohio 43403, USA}
\address[3]{Mechanical and Industrial Engineering, University of Massachusetts, 160 Governors Dr., Amherst, MA, 01003, USA}
\begin{abstract}
This study investigates the role of wall wettability, characterized by the wall contact angle (WCA), in controlling cavitation dynamics and stability around a Clark Y hydrofoil. High-fidelity Large Eddy Simulations (LES) coupled with a dynamic contact angle model were employed within the OpenFOAM framework to explore WCAs ranging from hydrophilic ($0^\circ$) to superhydrophobic ($160^\circ$) under distinct cavitation numbers ($\sigma = 1.6$, $0.8$, and $0.4$), representing incipient, cloud, and supercavitation regimes, respectively. The results show that increasing WCA consistently promotes earlier cavitation inception, thicker cavity development, and greater flow unsteadiness. For $\sigma = 1.6$, higher WCAs led to smaller, detached vapor bubbles and localized pressure fluctuations. At $\sigma = 0.8$, superhydrophobic surfaces caused more extensive vapor structures, intensified shedding dynamics, and stronger pressure fluctuations. For $\sigma = 0.4$, high WCAs facilitated stable, wall-adhered cavities that suppressed re-entrant jet activity and reduced unsteady loading. These findings demonstrate that surface wettability serves as an effective passive control mechanism for tailoring cavitation behavior and optimizing flow stability in engineering applications.
\end{abstract}


\end{frontmatter}
\section{Introduction}

Cavitation is the formation of vapor cavities (bubbles) in a liquid when the local pressures drop below the liquid vapor pressure \cite{franc2005introduction, roohi2013numerical}. The cavitation number $(\sigma)$ is a dimensionless parameter that characterizes the propensity for cavitation to occur in fluid flow \cite{ tupper2013introduction}. Different cavitation regimes emerge as the cavitation number decreases \cite{brandao2020numerical} including incipient cavitation, sheet cavitation, cloud cavitation, and supercavitation. Incipient cavitation is an initial stage that occurs when the cavitation number reaches a critical value where the vapor bubbles first begin to form \cite{sou2014numerical}. In sheet cavitation regime, vapor cavities coalesce into continuous sheets attached to solid surfaces, such as the suction side of the hydrofoils or blades \cite{pelz2017transition}. With a further reduction in the cavitation number, the attached vapor sheets become unstable, breaking off into clusters or clouds of cavitation bubbles \cite{ge2022dynamic}. These clouds can collapse violently, generating strong shock waves that can cause significant damage to nearby surfaces. Moreover, at very low cavitation numbers, a single, large vapor cavity envelops the object, such as a hydrofoil or projectile, effectively reducing contact between the liquid and the object surface \cite{li2008structures}. This regime can drastically reduce drag, which is advantageous in specific applications, such as high-speed marine vessels. 

It is important to note that the exact cavitation number thresholds for these regimes depend on various factors, including fluid properties, flow velocity, and the geometry of the object in the flow \cite{tupper2013introduction}. Furthermore, some studies have observed transitional behaviors between these regimes, influenced by factors such as flow instability and turbulence \cite{omelyanyuk2022experimental}.

Among different cavitation regimes, supercavitation is a unique mode in which the cavitation bubble extends beyond the length of the body, forming a reentrant liquid jet in the cavity closure region \cite{ xu2018supercavitating}. Controlling cavitation is crucial in many hydrodynamic applications to mitigate its detrimental effects and enhance system performance \cite{khare2021prediction}. Cavitation control strategies are broadly classified into active and passive methods \cite{zaresharif2021cavitation, li2024active}. Active flow control techniques involve external energy inputs, such as thermal and mass transfer mechanisms, plasma actuators, smart surfaces, and flow fluctuation techniques \cite{kamali2016three}. In contrast, passive flow control relies on geometric or material modifications that alter flow behavior without requiring an external energy source. These methods include vortex generators \cite{qiu2023effect, che2019control}, surface topography modifications (e.g., bumps, grooves, and cavities) \cite{ghasemnezhad2024large, tavakoli2024numerical}, wavy leading edges \cite{pendar2020les}, porous surfaces \cite{yu2024investigation}, complex surface structure \cite{el2025cavitation}, and biomimetic riblets \cite{kadivar2024riblets}, and changes in wall wettability by manipulation of wall contact angle (WCA) \cite{sotoudeh2023natural, mousavi2024potential}.

WCA manipulation has been extensively explored in droplet dynamics to control spreading, splashing, and rebound behavior. For instance, Zhang et al. \cite {zhang2024droplet} used five different WCAs ranging from superhydrophilic to superhydrophobic to study droplet impact dynamics. They showed that the larger the WCA, the smaller the minimum height factor. Furthermore, during the oscillating stage, the oscillation cycle time increases with larger WCA, while it first rises and then decreases as the Weber number increases. Mousavi and Faroughi \cite{mousavi2025spreading} investigated how hybrid WCA affects the spreading and bouncing behavior of viscoelastic droplets. Their findings revealed that surfaces with reduced WCA result in conventional droplet bouncing with some adhesion, whereas surfaces with elevated WCA cause droplet spreading over reduced contact regions and enhanced bouncing. Additionally, hybrid WCA creates intricate, non-uniform droplet behavior that differs significantly from uniformly wettable surfaces. Zhang et al. \cite{zhang2024prediction} examined droplet impact behavior on surfaces with hybrid water contact angles spanning from hydrophilic to superhydrophobic regions to determine the duration of contact between droplets and the superhydrophobic area. Their results showed that at intermediate offset distances, the movement of liquid from hydrophilic regions toward superhydrophobic areas extends the contact duration.

In addition to its role in droplet dynamics, WCA manipulation has been applied to control various flow phenomena such as drag reduction \cite{ma2025drag}, bubble generation \cite{mousavi2024bubble}, droplet breakup \cite{mousavi2023impact}. Superhydrophobic and hybrid surfaces have been shown to reduce drag and alter boundary layer behavior, offering significant advantages for both laminar and turbulent flows \cite{rothstein2010slip, bhushan2009natural, ou2004laminar}. For instance, Rezaei\cite{rezaei2023investigating} demonstrated that integrating hybrid superhydrophobic-hydrophilic surfaces on wind turbine blades operating in rainy conditions can effectively regulate boundary layer thickness, delay flow separation, and enhance aerodynamic performance. Patterned wettability has also been utilized to minimize pressure losses and control flow reattachment points \cite{lv2014novel}. Inspired by these findings, the concept of manipulating WCA can be extended to cavitating flows to influence bubble dynamics, cavity formation, and closure characteristics. 

Recent studies have shown that WCA affects cavitation dynamics and can alter cavitation regimes \cite{mousavi2023effects}. Modifying WCA directly impacts vortex structures and shear stress distribution near solid boundaries. Numerical simulations have shown that both superhydrophobic surfaces (high WCA) and hydrophilic surfaces (low WCA) increase the spatial extent of wall-attached vortices, altering bubble collapse behavior \cite{yuan2023fluctuation, huang2023effects}.

Furthermore, Huang et al. \cite{huang2023effects} observed that superhydrophobic surfaces increase the area and duration of wall shear stress, with peak values reaching 174.41 kPa, compared to 103.12 kPa on neutral surfaces. They also noticed that hydrophilic surfaces extend the spatial influence of shear stress but reduce their temporal persistence, with peak values around 131.82 kPa. Saini et al. \cite{saini2022dynamics} highlighted that when the WCA is less than $90^\circ$, a classical re-entrant jet directed toward the wall is typically observed. In contrast, for WCA greater than $90^\circ$, an annular re-entrant jet parallel to the wall is formed, fundamentally altering the interaction mechanism. In addition, Yuan et al. \cite{yuan2021study} indicated that changing the enhancement in WCA reduces the maximum pressure and microjet velocity but enhances the cavitation bubble lifetime. They also observed that the range of wall effects is smaller for a hydrophilic wall than for a hydrophobic one. Emadi and Mirjalili \cite{emadi2025passive} showed that WCA of $80^\circ$ reduces the cavitation bubbles more effectively compared to the angles of 0 and $160^\circ$. Mousavi and Roohi \cite{mousavi2023effects} concluded that a specific pattern in contact angle distribution reduces the cavitation volume behind the hydrofoil and decreases the flow instability.

This study investigates the influence of wall wettability on cavitation dynamics and flow stability around a Clark Y hydrofoil using LES methodology. By analyzing how WCA affects cavitation inception, vapor structure evolution, pressure fluctuations, flow separation, and velocity deficit, this research addresses a critical gap in understanding the role of surface properties in multiphase flows. 

The paper is organized as follows: Section 2 details the governing equations and numerical methods; Section 3 describes the computational domain’s geometry and boundary conditions; Section 4 validates the numerical approach with experimental data; Section 5 presents the results and analysis of WCA effects; and Section 6 summarizes key findings and their implications for future research and applications.


\section{Governing Equations and Numerical Methods}
The complex nature of cavitating flows with dynamic WCAs requires advanced numerical techniques to capture the multiscale physics involved. LES is employed in this study to resolve the large, energy-containing eddy structures on the computational grid, while the smaller, more isotropic subgrid-scale (SGS) structures are modeled. The development of LES encounters a primary challenge due to the strong coupling between SGS modeling and the truncation error of the numerical discretization scheme. For a review of recent SGS models, including their application in cavitating flows, see \cite{larkermani2018modulated}. In the present study, the dynamic one equation eddy-viscosity SGS approach is employed within the OpenFOAM framework \cite{kim1995new}.

The finite volume discretization inherently applies a top-hat-shaped kernel filter, which provides filtered flow quantities. The filtered quantity for a cell $\Omega_p$ is expressed as:
\begin{equation} \label{eq:filtered_quantity}
\bar{f}_p = \frac{1}{V_p} \int_{\Omega_p} f \, dV,
\end{equation}
where the overbar denotes a filtered quantity, $f$ represents a generic flow quantity, and $V_p$ is the cell volume. In this explicit LES approach, the subgrid stress terms are not dependent on truncation errors but are explicitly modeled through the \texttt{dynamicKEqn} formulation. The filtered governing equations for mass and momentum conservation in LES are:

\begin{equation} \label{eq:mass_conservation}
\frac{\partial \rho}{\partial t} + \nabla \cdot (\rho \bar{v}) = 0,
\end{equation}
\begin{equation} \label{eq:momentum_conservation}
\frac{\partial (\rho \bar{v})}{\partial t} + \nabla \cdot (\rho \bar{v} \otimes \bar{v}) = -\nabla \bar{p} + \nabla \cdot \left( \bar{\mathbf{s}} - \mathbf{B} \right) + \mathbf{f}_\sigma,
\end{equation}
where $\mathbf{f}_\sigma$ represents the surface tension force, modeled using the Continuum Surface Force (CSF) model as $\mathbf{f}_\sigma = \sigma \kappa \nabla \alpha$, with $\sigma$ being the surface tension coefficient, $\kappa$ the curvature of the interface, and $\alpha$ the volume fraction of the liquid phase. This term is essential for capturing the dynamics of multiphase flows such as cavitation.
To account for wall wettability effects, a dynamic wall contact angle model is employed at the solid boundaries. The dynamic WCA varies with the contact line velocity, defining the angle at which the fluid interface meets the wall. This variation is critical for accurately simulating the interaction between the cavitating flow and the surface, influencing cavitation behavior based on surface properties.

Moreover, in Eq. (\ref{eq:momentum_conservation}), $\bar{\mathbf{v}}$ is the filtered velocity field, $\bar{p}$ is the filtered pressure, and $\bar{\mathbf{s}} = 2\mu \mathbf{D}$ is the viscous stress tensor, where $\mathbf{D} = \frac{1}{2} \left( \nabla \bar{\mathbf{v}} + (\nabla \bar{\mathbf{v}})^T \right)$ is the rate-of-strain tensor. The term $\mathbf{B}$ denotes the SGS stress tensor resulting from the filtering operation. The SGS stress tensor $\mathbf{B}$ is expressed as:
\begin{equation} \label{eq:sgs_stress_tensor}
\mathbf{B} = \rho \left( \overline{v \otimes v} - \bar{v} \otimes \bar{v} + \widetilde{B} \right),
\end{equation}
where only $\widetilde{B}$ requires modeling, as the other terms can be directly computed from the filtered quantities. In the present study, the SGS terms are modeled using the \texttt{dynamicKEqn} approach. This SGS approach belongs to the class of one-equation eddy-viscosity models and solves a modeled transport equation for the SGS kinetic energy, $k_{\text{sgs}}$ (Eq. (\ref{eq:sgs_kinetic_energy})), allowing the turbulent viscosity ($\nu_{\text{SGS}}$) to be dynamically computed based on the local flow field and grid resolution.

\begin{equation} \label{eq:sgs_kinetic_energy}
\frac{\partial (\rho k_{\text{sgs}})}{\partial t} + \nabla \cdot (\rho \bar{v} k_{\text{sgs}}) = \nabla \cdot \left( \rho D_{\text{k}} \nabla k_{\text{sgs}} \right) + 2 \mu_{\text{SGS}} \mathbf{D} : \mathbf{D} - \rho \epsilon,
\end{equation}
where $\mu_{\text{SGS}}$ is the eddy viscosity, $D_{\text{k}}$ is the diffusivity of $k_{\text{sgs}}$, and $\epsilon$ represents the dissipation rate. The eddy viscosity $\mu_{\text{SGS}}$ is related to $k_{\text{sgs}}$ and a dynamically computed model coefficient.

\subsection{Multiphase Flow and Cavitation Modeling}
The present study addresses cavitating flows, which are inherently multiphase. To capture the interface between the liquid and vapor phases, the Volume of Fluid (VOF) method \cite{hirt1981volume} is employed. In the VOF method, a single set of momentum equations is shared by both phases, and the interface is tracked by solving a transport equation for the volume fraction, $\alpha$. The volume fraction field $\alpha$ is defined such that $\alpha = 1$ in the liquid phase, $\alpha = 0$ in the vapor phase, and values between 0 and 1 represent the interface. The transport equation for the liquid volume fraction is given by:

\begin{equation} \label{eq:alpha_transport}
\frac{\partial \alpha}{\partial t} + \nabla \cdot (\bar{\mathbf{v}} \alpha) + \nabla \cdot (\bar{\mathbf{v}}_r \alpha (1-\alpha)) = S_{\dot{m}},
\end{equation}
where $\bar{\mathbf{v}}$ is the filtered velocity field, and $\bar{\mathbf{v}}_r$ is an artificial compression velocity, used to maintain a sharp interface without explicitly resolving its fine structure. The last term on the left-hand side, involving $\bar{\mathbf{v}}_r$, is often referred to as the interface compression term. The term $S_{\dot{m}}$ on the right-hand side represents the source/sink terms due to phase change, which couples the cavitation model to the VOF equation.

The density $\rho$ and dynamic viscosity $\mu$ of the mixture are calculated as weighted averages of the liquid and vapor properties based on the volume fraction:
\begin{align}
\rho &= \alpha \rho_l + (1-\alpha) \rho_v, \label{eq:mixture_density} \\
\mu &= \alpha \mu_l + (1-\alpha) \mu_v, \label{eq:mixture_viscosity}
\end{align}
where subscripts $l$ and $v$ denote liquid and vapor properties, respectively. 

\subsection{Cavitation Model: Kunz Model}
For modeling the mass transfer between the liquid and vapor phases during cavitation, the Kunz cavitation model \cite{kunz2000preconditioned} is employed. This is an empirical barotropic model that relates the rate of phase change to the local pressure relative to the saturation pressure, $p_{sat}$. The model assumes that cavitation occurs when the local pressure falls below $p_{sat}$, and re-condensation occurs when the pressure rises above it.

The net mass transfer rate $S_{\dot{m}}$ (source/sink term in Eq. \ref{eq:alpha_transport}) is calculated as:
\begin{equation} \label{eq:kunz_mass_transfer}
S_{\dot{m}} = \dot{m}^+ - \dot{m}^-
\end{equation}
where $\dot{m}^+$ represents the mass transfer from liquid to vapor (vaporization) and $\dot{m}^-$ represents the mass transfer from vapor to liquid (condensation). These rates are formulated as:

For vaporization ($\bar{p} < p_{sat}$):
\begin{equation} \label{eq:kunz_vaporization}
\dot{m}^+ = C_v \frac{\rho_l}{\rho_v} \frac{(\bar{p} - p_{sat})}{0.5 \rho_l U_{\text{inf}}^2} \frac{(1-\alpha)}{t_{\text{inf}}} \quad \text{if } \bar{p} < p_{sat}
\end{equation}

For condensation ($\bar{p} > p_{sat}$):
\begin{equation} \label{eq:kunz_condensation}
\dot{m}^- = C_c \frac{(\bar{p} - p_{sat})}{0.5 \rho_l U_{\text{inf}}^2} \frac{\alpha}{t_{\text{inf}}} \quad \text{if } \bar{p} > p_{sat}
\end{equation}
where $\bar{p}$ is the filtered local pressure.

The coefficients $C_v$ and $C_c$ are empirical constants that control the rates of vaporization and condensation, respectively. In this study, the values used are $C_v = 1000$ and $C_c = 1000$. The characteristic flow velocity $U_{\text{inf}}$ is set to $10.0 \, \text{m/s}$, and the characteristic time $t_{\text{inf}}$ is set to $0.005 \, \text{s}$ \cite{roohi2013numerical}. 

\subsection{Numerical Schemes}

In this study, specific numerical schemes are employed to discretize the governing equations. The temporal discretization utilizes the first-order implicit Euler scheme, which can be expressed as
\begin{equation}
\frac{\partial \phi}{\partial t} \approx \frac{\phi^{n+1} - \phi^n}{\Delta t}
\label{eq:temporal}
\end{equation}
This scheme is known for its robustness and simplicity in transient simulations.

Spatial gradients are approximated using the Gauss linear method, which computes the gradient as
\begin{equation}
\nabla \phi \approx \sum_f \phi_f \mathbf{S}_f
\label{eq:gradient}
\end{equation}
where $\phi_f$ represents the linearly interpolated value at the face $f$, and $\mathbf{S}_f$ is the face area vector. This method ensures second-order accuracy for the gradient terms.

The divergence terms are discretized using different schemes tailored to each specific field. For the convection term involving the momentum equation, a linear scheme is used, represented by
\begin{equation}
\nabla \cdot (\rho \mathbf{U} \otimes \mathbf{U}) \approx \sum_f \rho_f \mathbf{U}_f (\mathbf{U} \cdot \mathbf{S})_f
\label{eq:divmomentum}
\end{equation}
For convection terms involving the volume fraction $\alpha$, the Van Leer total variation diminishing (TVD) scheme is employed to capture sharp interfaces. This is expressed as
\begin{equation}
\nabla \cdot (\phi \alpha) \approx \sum_f \phi_f \, \text{VanLeer}(\alpha)
\label{eq:divalpha}
\end{equation}
Additionally, the interface compression scheme is used to maintain sharp interfaces in multiphase flows and is given by
\begin{equation}
\nabla \cdot (\phi_r \alpha) \approx \sum_f \phi_{r,f} \alpha_f
\label{eq:divinterface}
\end{equation}
For turbulence-related quantities, such as $k$ and $\epsilon$, an upwind scheme is employed to ensure numerical stability. For example, the convection of $k$ is written as
\begin{equation}
\nabla \cdot (\phi k) \approx \sum_f \phi_f k_{\text{upwind}}
\label{eq:divk}
\end{equation}

The Laplacian terms are discretized using the Gauss linear corrected scheme, which accounts for mesh non-orthogonality. The discretized Laplacian operator is represented as
\begin{equation}
\nabla \cdot (\Gamma \nabla \phi) \approx \sum_f \Gamma_f (\nabla \phi)_f \cdot \mathbf{S}_f
\label{eq:laplacian}
\end{equation}
Here, $(\nabla \phi)_f$ is corrected for non-orthogonality to enhance accuracy, particularly in unstructured or non-orthogonal meshes.

Field values at cell faces are interpolated using the linear interpolation scheme, ensuring second-order spatial accuracy. The interpolation is represented by
\begin{equation}
\phi_f = \frac{\phi_P + \phi_N}{2}
\label{eq:interpolation}
\end{equation}
where $\phi_P$ and $\phi_N$ are the values at the adjacent cell centers $P$ and $N$.

The surface-normal gradients are computed using a corrected scheme to account for mesh non-orthogonality. The corrected surface-normal gradient is calculated as
\begin{equation}
(\nabla \phi)_f \cdot \hat{n}_f \approx \frac{\phi_N - \phi_P}{|\mathbf{d}|} + \text{correction}
\label{eq:snGrad}
\end{equation}
where $\mathbf{d}$ is the displacement vector between cell centers $P$ and $N$.

Flux calculations are required for specific fields to ensure conservation and accuracy. In this case, fluxes are computed for $p_{rgh}$, $p_{corr}$, and $\alpha_1$. These flux requirements contribute to maintaining consistency in the flux fields and improve the overall stability and accuracy of the numerical solution.

The numerical schemes applied in this study are selected to balance accuracy, stability, and computational efficiency.


\section{Geometry and boundary condition}
\begin{figure}[h]
    \centering
    \includegraphics[width=0.8\linewidth]{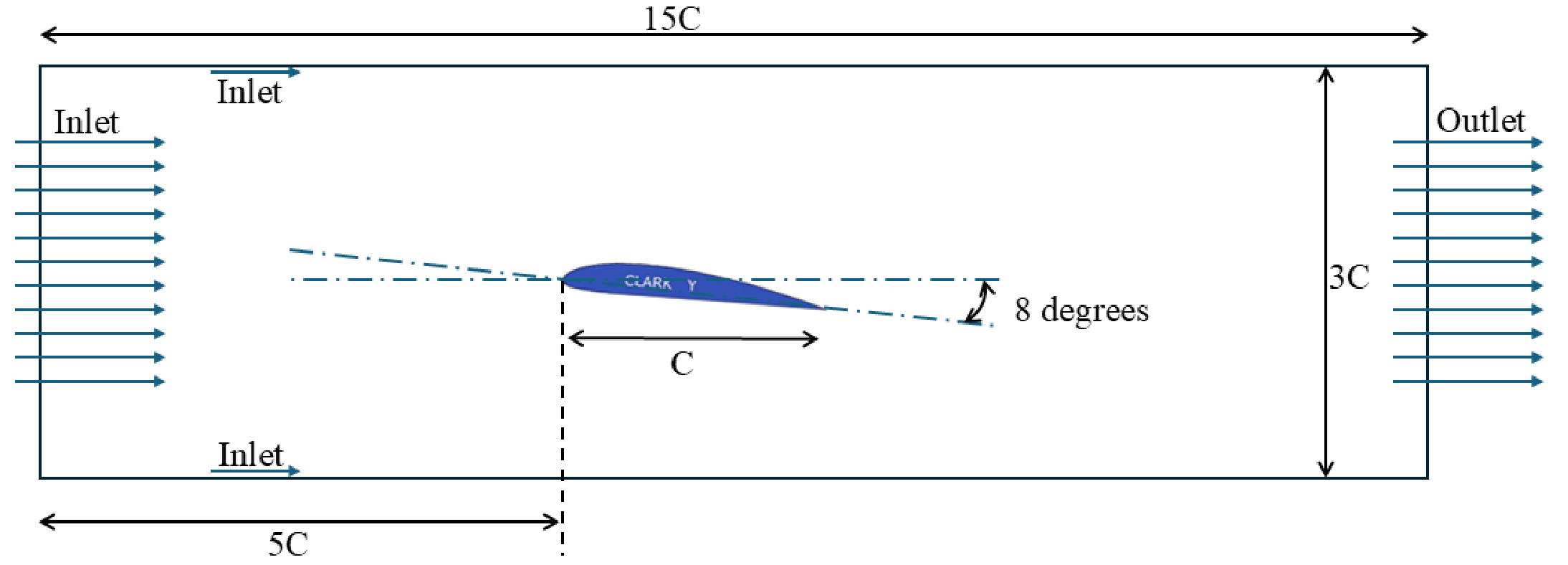}
    \caption{Schematic of the computational domain for cavitation simulation over a Clark-Y hydrofoil. The hydrofoil is placed at an angle of attack of 8 degrees, with inflow applied from the left boundary and outlet conditions at the right boundary. The domain size is \(15C \times 3C\), where \(C\) is the chord length.}
    \label{fig:domian}
\end{figure}

The computational domain used in this study is designed to simulate cavitation flow over a 2-D Clark-Y hydrofoil. As shown in Fig.~\ref{fig:domian}, the domain is a two-dimensional rectangular channel of total length \(15C\) and height \(3C\), where \(C\) denotes the chord length of the hydrofoil. The hydrofoil is positioned such that its leading edge is located \(5C\) downstream from the inlet and is inclined at an angle of attack of 8 degrees relative to the incoming flow direction.

A uniform velocity profile is applied at the left (inlet) boundary, with additional inflow specified at the top and bottom inlet segments to ensure a consistent upstream condition. The right boundary serves as the outlet with a fixed pressure condition to allow free outflow. This arrangement ensures adequate space upstream and downstream of the hydrofoil to accurately capture the development of cavitation structures, pressure recovery, and wake behavior.

Simulations are performed for cavitation numbers of 0.8 and 0.4 to represent different cavitation conditions ranging from incipient to fully developed cavity formation. The hydrofoil spans the entire domain width, assuming a two-dimensional approximation, and the chosen domain height avoids confinement effects, allowing the flow to develop naturally above and below the hydrofoil.

\section{Grid study}

To ensure the accuracy of the LES results, a comprehensive mesh independence study was conducted. Four different mesh resolutions were evaluated: very coarse, coarse, medium, and fine. All simulations were performed at a cavitation number of 0.4, and critical hydrodynamic parameters, including the lift coefficient ($C_l$) and cavity thickness at $x/C = 0.6$ ($\delta_{\text{0.6C}}$), were carefully monitored.

As presented in Table~\ref{tab:grid_independence}, the computational results exhibit a clear convergence as the mesh density increases. Specifically, the differences in both $C_l$ and $\delta_{\text{0.6C}}$ between the medium and fine meshes are negligible. This strong agreement provides confidence that the solution is largely mesh-independent at these resolutions. Therefore, the fine mesh, with approximately 6.5 million cells, was ultimately chosen for all subsequent simulations. This selection ensures sufficient resolution to accurately capture the intricate details of cavitation structures and their dynamic behavior.

\begin{figure}[H]
    \centering
    \includegraphics[width=0.6\linewidth]{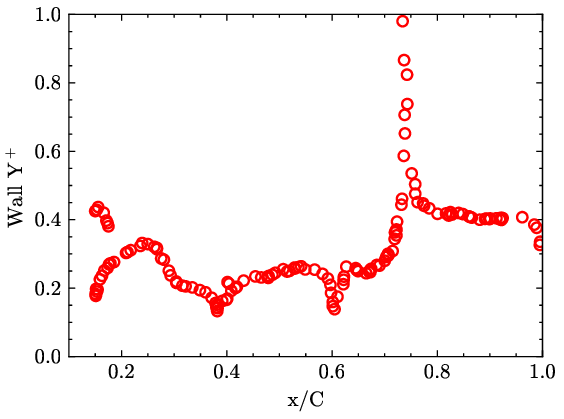}
\caption{Distribution of wall $Y^+$ values around the Clark Y hydrofoil at a cavitation number of 0.4 and angle of attack of 8 degrees. The results show that $Y^+$ remains below 1 across the hydrofoil surface, confirming that the near-wall resolution is adequate for LES and capable of capturing boundary layer dynamics without additional wall functions.}
\label{fig:y_plus}
\end{figure}

Furthermore, Fig.~\ref{fig:y_plus} presents the distribution of wall $Y^+$ values along the hydrofoil surface for the fine mesh. As shown in this figure, the results indicate that $Y^+$ remains consistently below 1 across the entire hydrofoil, including both the suction and pressure sides. This is particularly significant for LES because it ensures that the viscous sublayer of the boundary layer is fully resolved without relying on wall functions. The low $Y^+$ values validate the adequacy of the chosen mesh for capturing near-wall physics critical to cavitation inception, boundary layer development, and flow separation. The fine mesh therefore meets the LES requirements for high-fidelity resolution of wall-bounded turbulence and cavitation dynamics around the hydrofoil.

\begin{table}[h!]
\centering
\caption{Grid independence study at cavitation number = 0.4: Comparison of lift coefficient ($C_l$) and cavity thickness ($\delta_{\text{0.6C}}$) across four mesh resolutions.}
\label{tab:grid_independence}
\begin{tabular}{lccc}
\hline
Mesh Resolution     & Total Cells     & $C_l$ & Cavity Thickness ($\delta_{\text{0.6C}}$) \\
\hline
Very Coarse Mesh    & 1.5M            & 0.11  & 0.250 \\
Coarse Mesh         & 2.7M            & 0.46  & 0.100 \\
Medium Mesh         & 4.7M            & 0.28  & 0.122 \\
Fine Mesh           & 6.5M            & 0.286 & 0.121 \\
\hline
\end{tabular}
\end{table}

\section{Validation of Numerical Method}

To assess the accuracy of the present numerical approach in predicting cavitation behavior, simulation results were compared with available experimental data~\cite{wang2001dynamics}. 

\begin{figure}[H]
    \centering
    \begin{subfigure}[b]{0.47\linewidth}
        \centering
        \includegraphics[width=\linewidth]{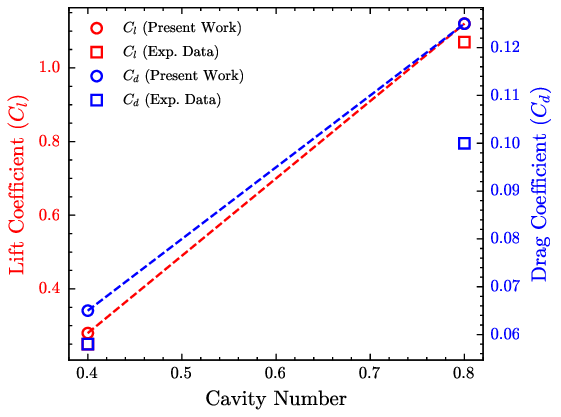}
        \subcaption{Lift ($C_l$) and drag ($C_d$) coefficients at two cavitation numbers.}
        \label{fig:validation_a}
    \end{subfigure}
    \hfill
    \begin{subfigure}[b]{0.48\linewidth}
        \centering
        \includegraphics[width=\linewidth]{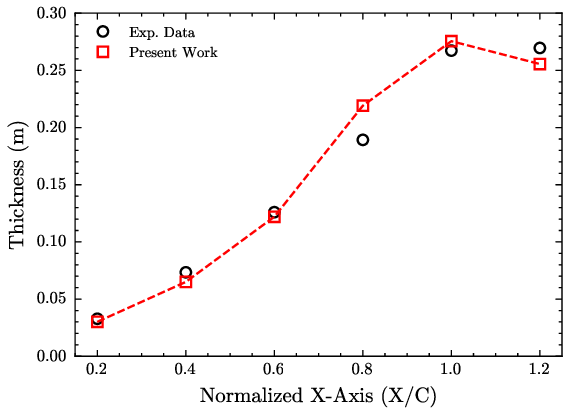}
        \subcaption{Cavity thickness along the normalized chordwise location ($X/C$) for cavitation number 0.4.}
        \label{fig:validation_b}
    \end{subfigure}
\caption{
Validation of the present numerical approach against experimental data~\cite{wang2001dynamics} for flow around a Clark Y hydrofoil at 8 degrees of angle of attack.
(a) Lift ($C_l$) and drag ($C_d$) coefficients at two cavitation numbers, showing excellent agreement with experimental results.
(b) Cavity thickness distribution along the normalized chordwise direction ($X/C$) for a cavitation number of 0.4, demonstrating the capability of the numerical method to capture cavity growth and the location of maximum thickness.
}
    \label{fig:validation}
\end{figure}

Figure~\ref{fig:validation}(a) presents the variation of lift ($C_l$) and drag ($C_d$) coefficients as functions of cavitation number. The present numerical results show excellent agreement with the experimental data ~\cite{wang2001dynamics}, accurately capturing both the magnitude and trend of the aerodynamic coefficients for varying cavitation intensities. The lift and drag coefficients increase with the cavitation number, and the numerical model successfully replicates this behavior, demonstrating its capability in resolving the effects of cavitation on the force coefficients.

Figure~\ref{fig:validation}(b) illustrates the cavity thickness distribution along the normalized chordwise location ($X/C$) for a cavitation number of 0.4. The present work closely matches the experimental measurements, particularly in capturing the growth of the cavity thickness towards the trailing edge and the location of maximum thickness. 

These comparisons validate that the current numerical methodology can reliably predict cavitation behavior under various operating conditions.

\section{Results and Discussion}

\subsection{Effect of Wall Contact Angle}

After verifying and validating the numerical method in the previous section, in this part, the influence of WCA on the hydrodynamic performance of a Clark Y hydrofoil is examined through time-averaged pressure distributions and corresponding vapor volume fraction fields. 

\begin{figure}[H]
    \centering
\caption{Time-averaged effects of WCA on flow around a Clark Y hydrofoil at 8 degrees of angle of attack and cavitation number of 0.8. a- Mean pressure distribution along the chord for various WCAs ($0^\circ$, $40^\circ$, $80^\circ$, $120^\circ$, $160^\circ$). b- Vapor volume fraction indicating vapor formation. Higher WCAs increase cavitation intensity and extent.}
    \label{fig:WCA_pressure_vapor}
    \begin{subfigure}[b]{0.47\linewidth}
        \centering
        \includegraphics[width=\linewidth]{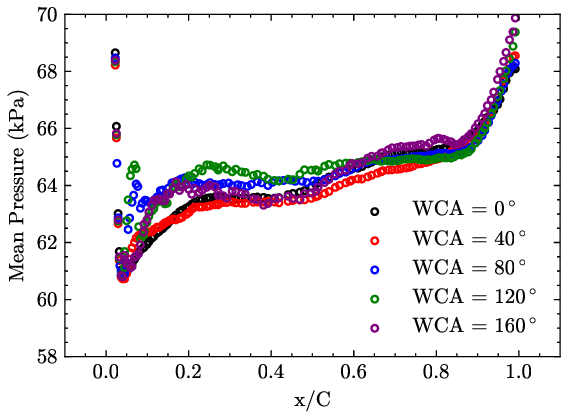}
            \subcaption{Time-averaged pressure distribution}
    \end{subfigure}
    \hfill
    \begin{subfigure}[b]{0.47\linewidth}
        \centering
        \includegraphics[width=\linewidth]{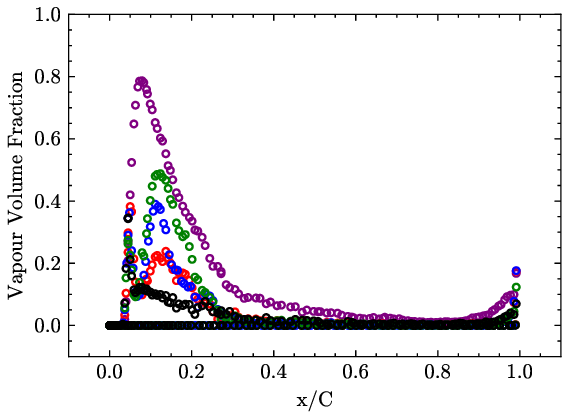}
            \subcaption{Time-averaged vapor volume fraction}
    \end{subfigure}

\end{figure}

Figure ~\ref{fig:WCA_pressure_vapor}a shows the pressure distribution along the suction side of the Clark Y hydrofoil at an angle of attack of $8^\circ$ and a cavitation number of $0.8$. Figure ~\ref{fig:WCA_pressure_vapor}b, on the other hand, illustrates the time-averaged vapor volume fraction around the hydrofoil, indicating the extent of cavitation.

Based on Fig. ~\ref{fig:WCA_pressure_vapor}a, distinct patterns in pressure distribution emerge depending on the WCA. For the more hydrophilic surfaces, specifically WCA = $0^\circ$ and $40^\circ$, the pressure distribution exhibits a relatively smooth drop on the suction side, maintaining a stable profile across the chord. As the surface wettability transitions towards intermediate hydrophobicity, particularly at WCA = $80^\circ$, the pressure profile begins to show noticeable fluctuations, especially in the pressure recovery region (around $x/C = 0.6$ to $0.8$). These fluctuations become even more pronounced and spatially extended for WCA = $120^\circ$, indicating a less stable and more oscillatory pressure field along a significant portion of the chord. For the superhydrophobic surface (WCA = $160^\circ$), while it exhibits the lowest minimum pressure on the suction side, the initial pressure drop to this minimum is relatively smooth. However, the subsequent pressure recovery phase also displays significant fluctuations, reflecting the complex flow dynamics associated with the extensive cavitation observed for this condition.

These pressure characteristics are directly corroborated by the time-averaged vapor volume fraction fields shown in Fig. ~\ref{fig:WCA_pressure_vapor}b, which vividly illustrate the extent and intensity of cavitation. For the highly hydrophilic cases (WCA = $0^\circ$ and $40^\circ$), where Fig. ~\ref{fig:WCA_pressure_vapor}a shows relatively high and smooth pressure distributions, Fig. ~\ref{fig:WCA_pressure_vapor}b confirms negligible vapor formation, with values consistently near zero. This clearly demonstrates that maintaining higher suction-side pressures effectively suppresses cavitation inception. As the WCA increases to $80^\circ$ and $120^\circ$, corresponding to the onset and intensification of pressure fluctuations and lower minimum pressures in Fig. ~\ref{fig:WCA_pressure_vapor}a, Fig. ~\ref{fig:WCA_pressure_vapor}b reveals increasingly significant vapor pockets. For WCA = $80^\circ$, distinct cavitation pockets with peak vapor volume fractions around $0.2$ to $0.3$ are observed in the fore-chord region (up to $x/C \approx 0.2$). The trend intensifies sharply for WCA = $120^\circ$, where a more pronounced and extended region of vapor forms, with peak values approaching $0.5$ around $x/C = 0.15$. The increased pressure depression and the growing pressure fluctuations observed in Fig. ~\ref{fig:WCA_pressure_vapor}a for these WCAs directly facilitate the inception and growth of these cavitation structures. Finally, for the superhydrophobic surface (WCA = $160^\circ$), which exhibits the lowest minimum pressure and significant fluctuations in its recovery region as shown in Fig. ~\ref{fig:WCA_pressure_vapor}a, Fig. ~\ref{fig:WCA_pressure_vapor}b displays the most extensive and intense vapor formation. Persistent vapor structures with peak values exceeding $0.7$ are distributed over a broad region from the leading edge to approximately $x/C = 0.3$. Furthermore, a secondary region of vapor reappears further downstream around $x/C = 0.9$, corresponding to the continued low pressure and fluctuations in that region of Fig. ~\ref{fig:WCA_pressure_vapor}a. This direct correlation underscores that the reduced pressure and the destabilizing effect of surface wettability on the local flow, as manifested by pronounced pressure drops and fluctuations, are primary drivers for enhanced cavitation intensity and extent.

\begin{figure}[H]
    \centering
    \includegraphics[width=0.6\linewidth]{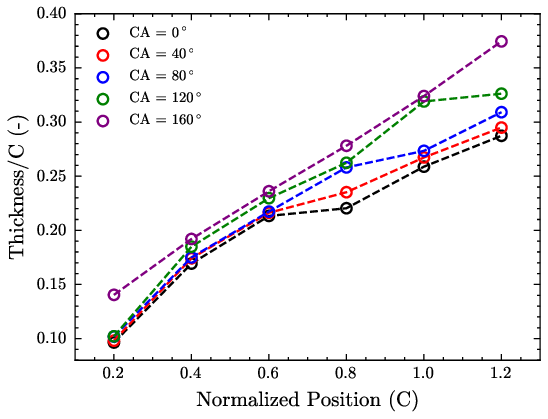}
\caption{Normalized cavity thickness as a function of chordwise position for different contact angles, for flow around a Clark Y hydrofoil at 8 degrees of angle of attack and cavitation number of 0.8.}
\label{fig:cavity_thickness}
\end{figure}

Figure~\ref{fig:cavity_thickness} illustrates the normalized cavity thickness ($\text{Thickness}/C$) as a function of chordwise position for various WCA. A clear trend of increasing cavity thickness with downstream progression is observed across all WCAs, which shows the growth of the vapor region along the hydrofoil.

As evident, superhydrophobic surfaces (WCA = $160^\circ$) consistently exhibit the largest cavity thicknesses at every chordwise position, notably so beyond the mid-chord region. This is attributed to the reduced liquid-solid adhesion on superhydrophobic surfaces, which effectively lowers the cavitation inception pressure and promotes larger nuclei formation. The presence of trapped air pockets within the surface texture of superhydrophobic materials also contributes to an effective slip velocity at the interface, reducing shear stress and potentially delaying flow reattachment, leading to more aggressive and extensive vapor accumulation.

Conversely, hydrophilic conditions, represented by WCA = $0^\circ$ and $40^\circ$, result in the smallest cavity thicknesses throughout the domain, suggesting suppressed cavitation. On hydrophilic surfaces, stronger adhesive forces between the liquid and solid stabilize the flow, raising the threshold for cavitation inception and promoting the rapid collapse of forming vapor bubbles due to increased viscous damping.

Intermediate WCAs, such as $80^\circ$ and $120^\circ$, show a progressive increase in cavity thickness as WCA rises, bridging the behavior between the hydrophilic and superhydrophobic extremes. This transitional behavior reflects the gradual weakening of liquid-solid interaction and the increasing propensity for heterogeneous cavitation nucleation. The distinct slopes of these profiles at different WCAs highlight how surface wettability directly impacts the rate at which the cavitation region expands, primarily by modulating the initial conditions for vapor bubble formation and their subsequent growth dynamics.

\begin{figure}[H]
    \centering
    \includegraphics[width=0.6\textwidth]{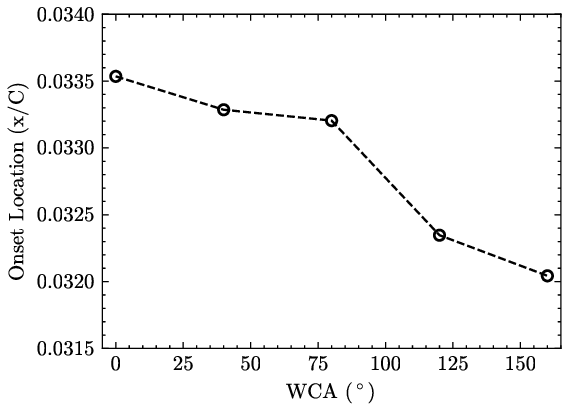}
\caption{Variation of cavitation onset location with WCA for flow around a Clark Y hydrofoil at 8 degrees of angle of attack nd cavitation number of 0.8. Higher WCAs result in upstream shift of vapor formation.}    \label{fig:Onset}
\end{figure}

The relationship between WCA and cavitation inception location is shown in Fig.~\ref{fig:Onset}. As the WCA increases, the onset of cavitation shifts upstream, indicating that reduced surface wettability accelerates vapor formation. While the differences in onset position are subtle in magnitude, the trend is consistent across all tested angles.

For low-WCA cases (e.g., $0^\circ$), cavitation begins further downstream at approximately $x/C = 0.0335$. This behavior aligns with the strong adhesive forces characteristic of hydrophilic surfaces, which promote liquid-solid interaction and stabilize the near-wall flow. Such surfaces effectively suppress the growth of any pre-existing gas nuclei or delay the formation of new ones by increasing the pressure required for phase change. The onset point at this WCA suggests that the flow must develop further along the hydrofoil to reach a sufficiently low-pressure minimum conducive to cavitation.

As WCA rises to intermediate values ($40^\circ$--$80^\circ$), a gradual upstream movement in the inception point is observed. For instance, at $40^\circ$, inception occurs around $x/C = 0.0332$, and at $80^\circ$, it moves slightly further upstream to about $x/C = 0.0326$. This progressive shift indicates a weakening of the liquid's adherence to the surface, making it marginally easier for cavitation to initiate. The transition from strong adhesion to a less adhesive state facilitates earlier nucleation as the local pressure drops approach the vapor pressure.

Beyond $100^\circ$, the shift becomes more pronounced, with the $160^\circ$ case exhibiting cavitation onset near $x/C = 0.0319$. This significant upstream shift for superhydrophobic surfaces is attributed to several factors. These surfaces can entrap stable micro- or nanobubbles within their surface texture, which act as readily available and easily activated heterogeneous nucleation sites, effectively lowering the cavitation threshold. Furthermore, the partial slip condition often observed on superhydrophobic surfaces can alter the velocity profile and pressure distribution within the boundary layer, potentially leading to the formation of more intense or spatially earlier local pressure minima near the leading edge. The combined effect of reduced wall shear resistance and enhanced capability to these pressure drops facilitates an earlier and more robust phase change. 

\begin{figure}[H]
    \centering
    \begin{subfigure}[b]{0.32\textwidth}
        \centering
        \includegraphics[width=\textwidth]{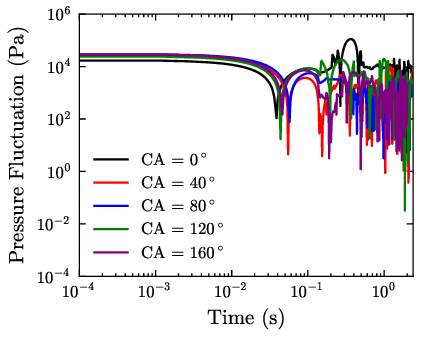}
        \subcaption{x=C}
    \end{subfigure}
    \hfill
    \begin{subfigure}[b]{0.32\textwidth}
        \centering
        \includegraphics[width=\textwidth]{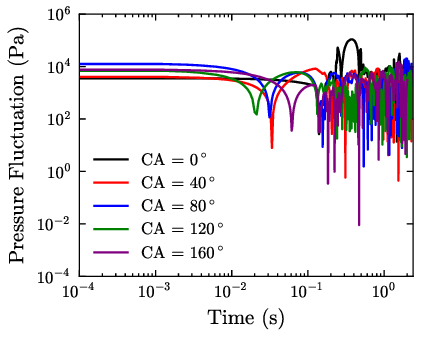}
        \subcaption{x=1.5C}
    \end{subfigure}
    \hfill
    \begin{subfigure}[b]{0.32\textwidth}
        \centering
        \includegraphics[width=\textwidth]{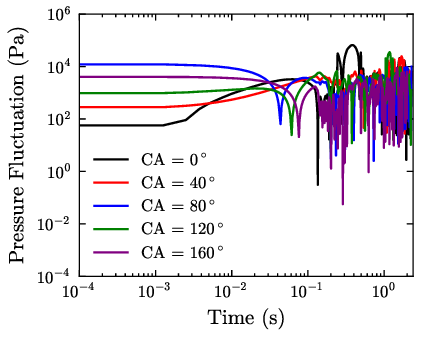}
        \subcaption{x=2C}
    \end{subfigure}
    \caption{Log-log plots of pressure fluctuations versus time at three downstream locations: \( x = C \), \( x = 1.5C \), and \( x = 2C \), for various WCAs. The angle of attack is 8 degrees and nd cavitation number of 0.8. Increased WCA intensifies early unsteadiness and high-frequency fluctuation content.}
    \label{fig:pressure_fluctuations}
\end{figure}

Figure~\ref{fig:pressure_fluctuations} displays the temporal evolution of pressure fluctuations at three downstream positions along the centerline: $x = C$, $x = 1.5C$, and $x = 2C$, for a range of WCAs. The data is presented in logarithmic scale, capturing both low-frequency trends and high-frequency cavitation-induced oscillations. The observed trends in pressure fluctuations can be fundamentally linked to the surface energy of the hydrofoil, which is directly modulated by WCA, as well as the altered near-wall flow dynamics.

At $x = C$, corresponding to the hydrofoil's trailing edge, the pressure spectra show relatively stable levels at early times. However, a sharp increase in fluctuation intensity emerges as unsteady cavity shedding commences. The onset of these fluctuations is notably accelerated for higher WCAs, particularly $120^\circ$ and $160^\circ$. This behavior is consistent with lower surface energy on hydrophobic surfaces, which promotes less stable vapor structures. Reduced liquid-solid interaction due to lower surface energy leads to weaker re-entrant jet development or more abrupt cavity detachment, resulting in earlier and more violent pressure oscillations near the shedding point. Furthermore, the presence of an effective slip velocity on superhydrophobic surfaces can influence the boundary layer, potentially leading to earlier flow separation or a more abrupt transition to turbulence, which would in turn promote earlier unsteady shedding. Conversely, lower WCAs such as $0^\circ$ and $40^\circ$, representing higher surface energy and stronger liquid adhesion, maintain lower amplitudes over a longer initial duration, indicative of more stable and coherent cavity dynamics and more robust flow attachment.

Moving to $x = 1.5C$, further downstream in the wake, the fluctuation magnitudes become more comparable across all WCAs. Nevertheless, the high-WCA cases continue to exhibit earlier and more intense broadband activity. At this position, the influence of secondary cavitation events or the impingement of re-entrant jets becomes more pronounced, particularly for $120^\circ$ and $160^\circ$ WCAs. The deeper valleys and sharper transitions in their mid-frequency regimes suggest more energetic and less damped bubble collapse events. Lower surface energy on hydrophobic surfaces can lead to the formation of larger, more unstable bubbles that collapse with greater intensity, generating stronger pressure waves that propagate into the downstream flow. The altered effective viscosity and surface tension effects at the interface due to varying wettability could also contribute to the different bubble collapse dynamics and pressure wave generation.

Finally, at $x = 2C$, the fluctuation field appears fully developed for all cases, characterized by densely packed high-frequency components dominating the spectra. While the difference between WCAs diminishes slightly in terms of peak amplitude, the spectral density remains richer for surfaces with reduced wettability. This implies that even further downstream, the history of more vigorous cavitation on hydrophobic surfaces contributes to a more turbulent and acoustically active wake. The low-WCA cases, linked to higher surface energy, display smoother spectral roll-off. This indicates that stronger liquid adhesion and more stable bubble dynamics on these surfaces lead to more coherent wake structures with less chaotic energy dissipation, perhaps due to a more ordered collapse process or reduced wake turbulence.

The inclusion of intermediate angles ($40^\circ$ and $120^\circ$) effectively bridges the response between highly wetting (high surface energy) and highly non-wetting (low surface energy) conditions. 

\begin{figure}[H]
\centering
\begin{tabular}{cccc}
\includegraphics[width=0.34\textwidth]{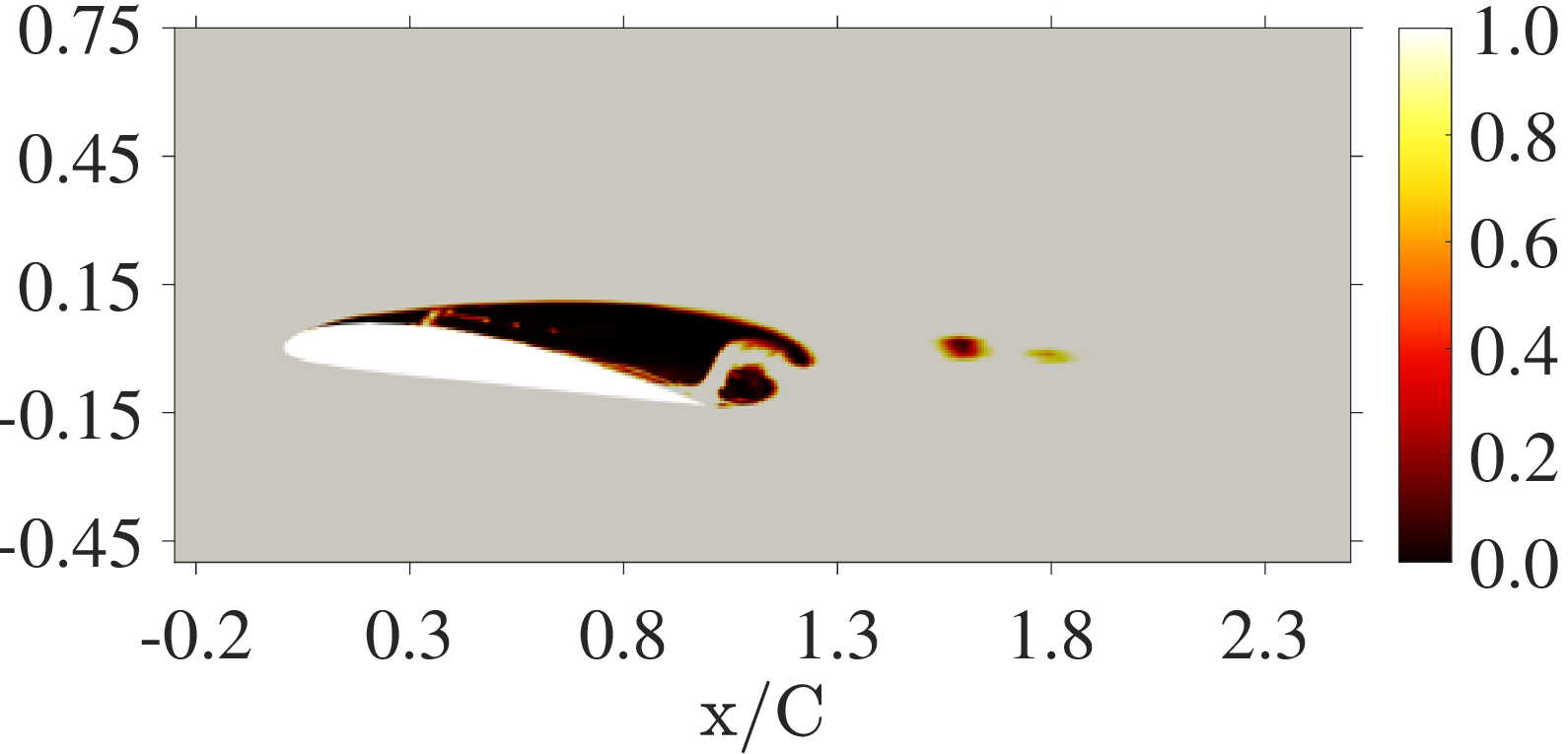} &
\includegraphics[width=0.34\textwidth]{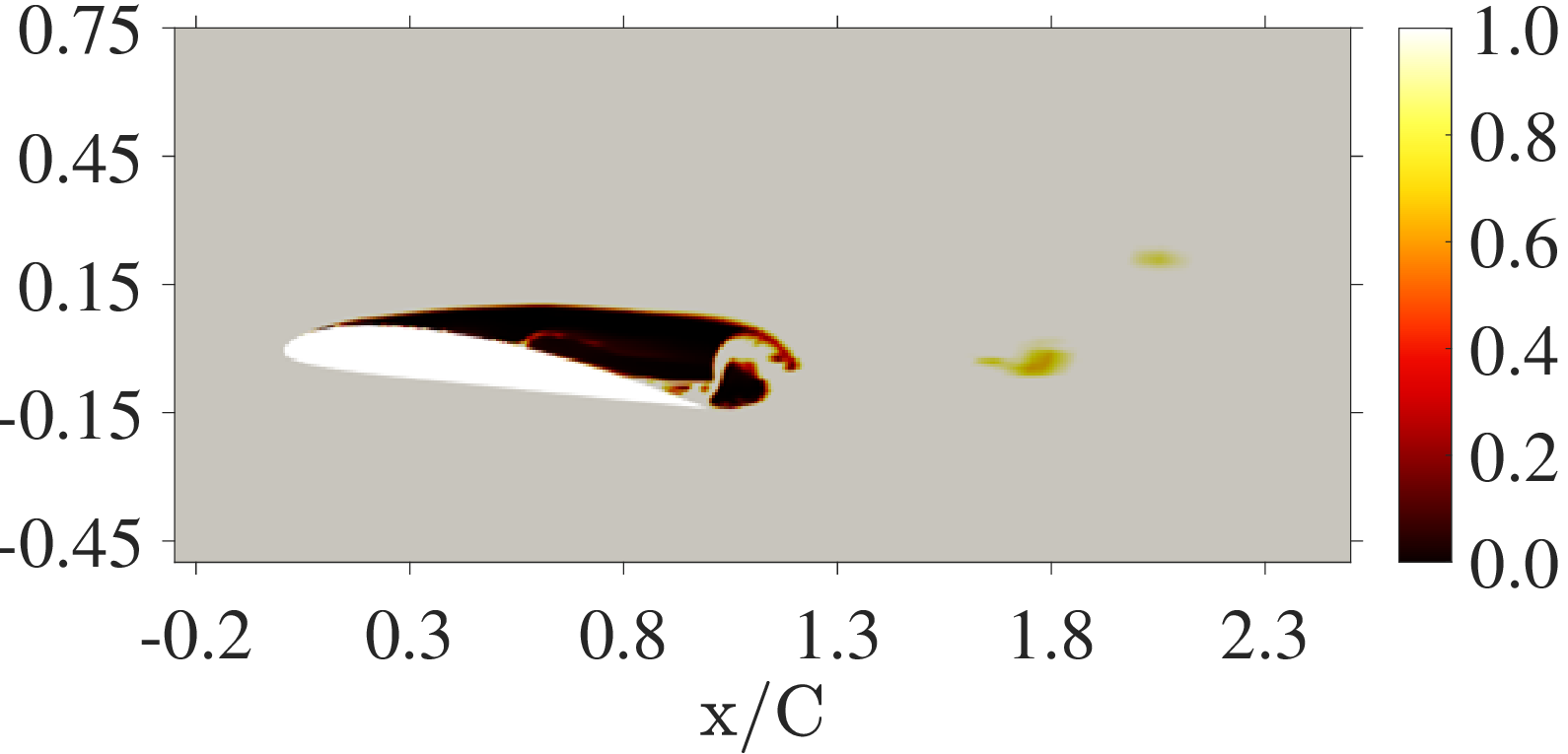} &
\includegraphics[width=0.34\textwidth]{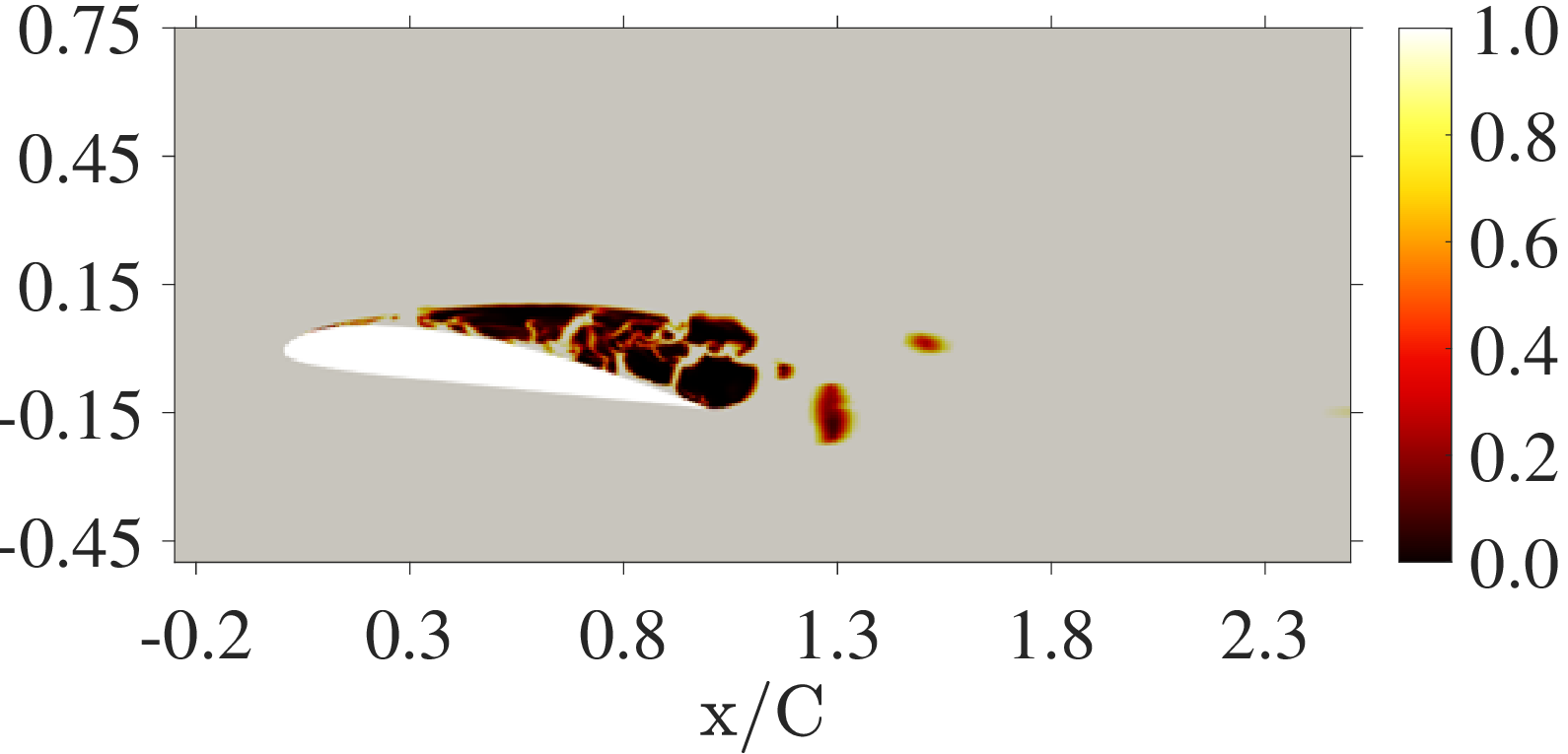} \\
\includegraphics[width=0.34\textwidth]{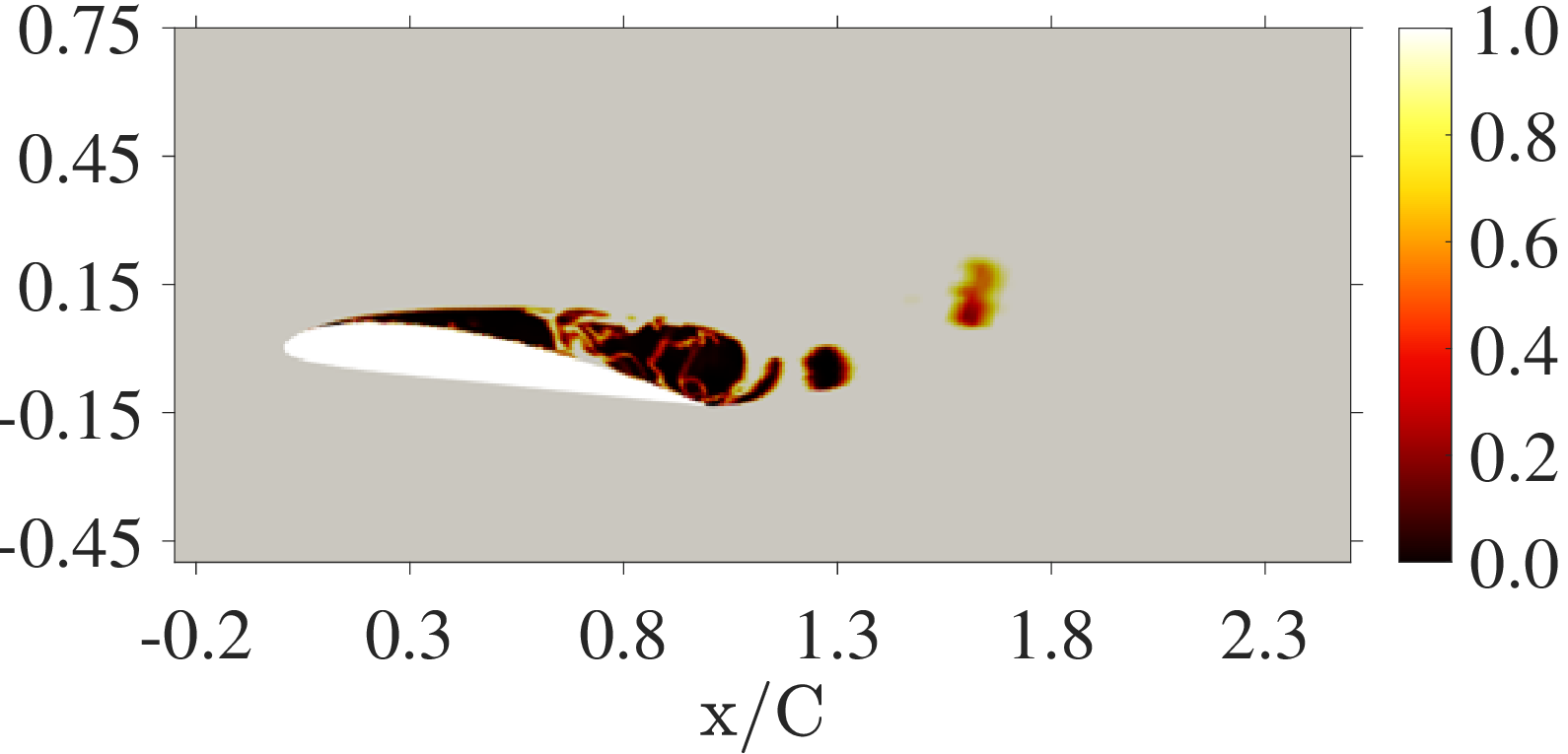} &
\includegraphics[width=0.34\textwidth]{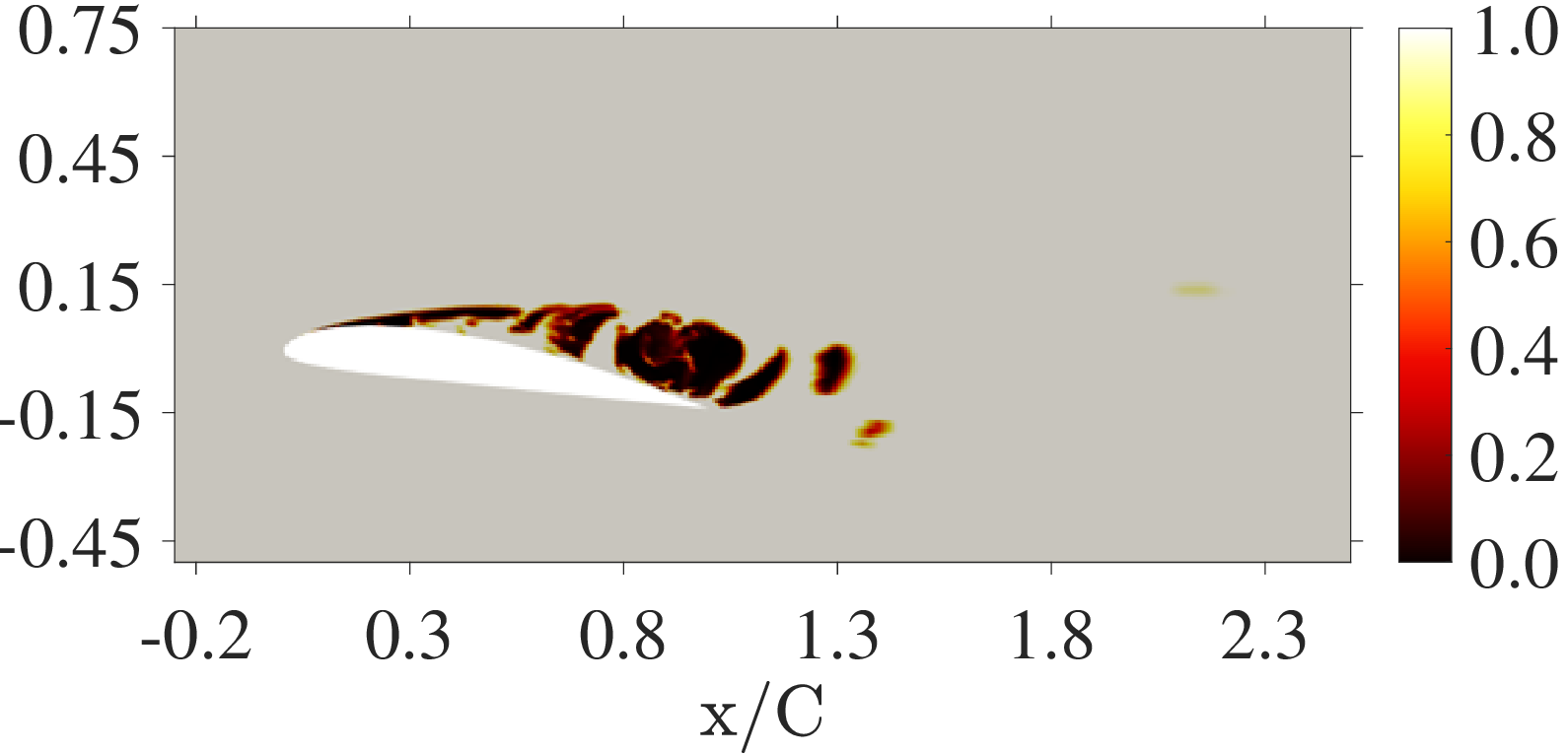} &
\includegraphics[width=0.34\textwidth]{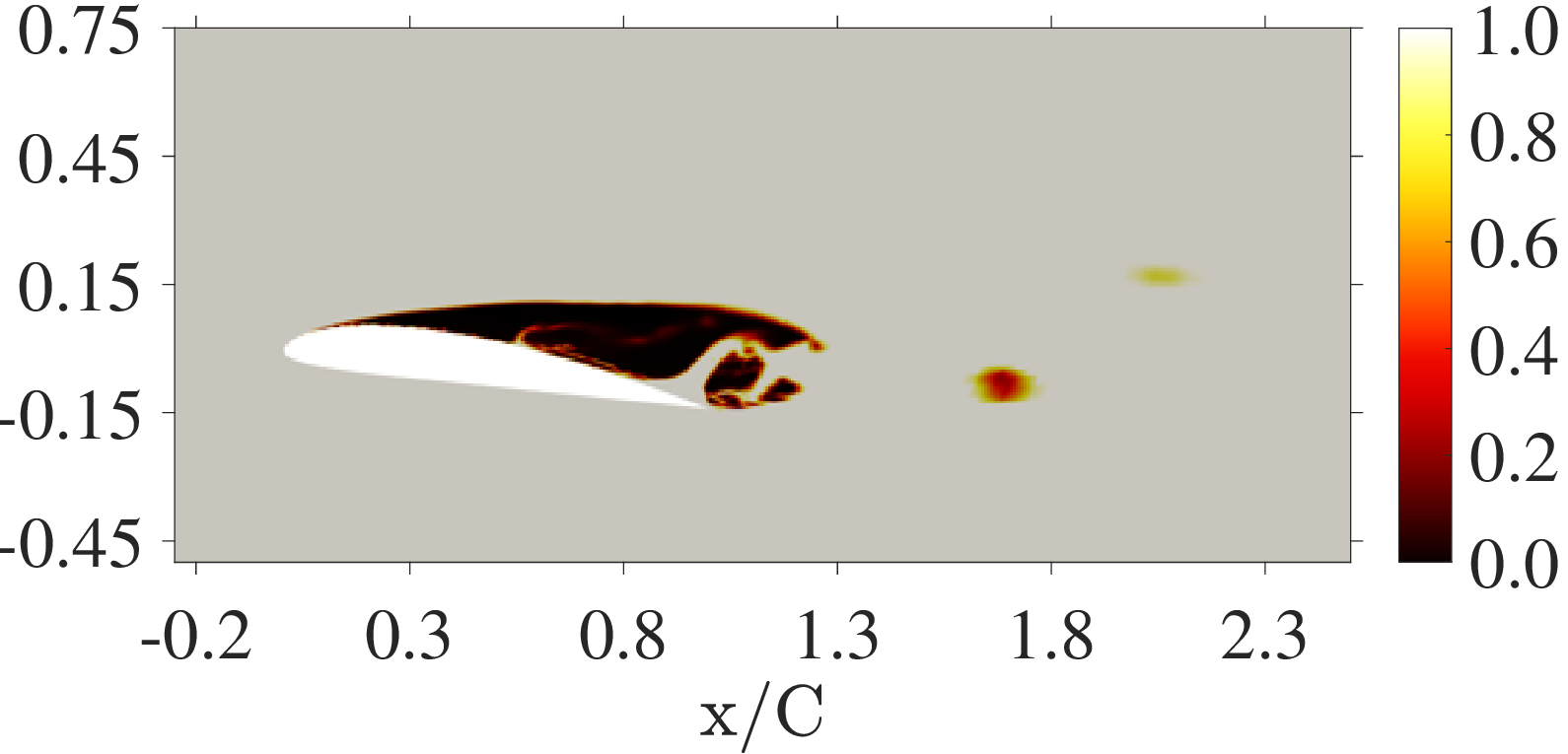} \\
\includegraphics[width=0.34\textwidth]{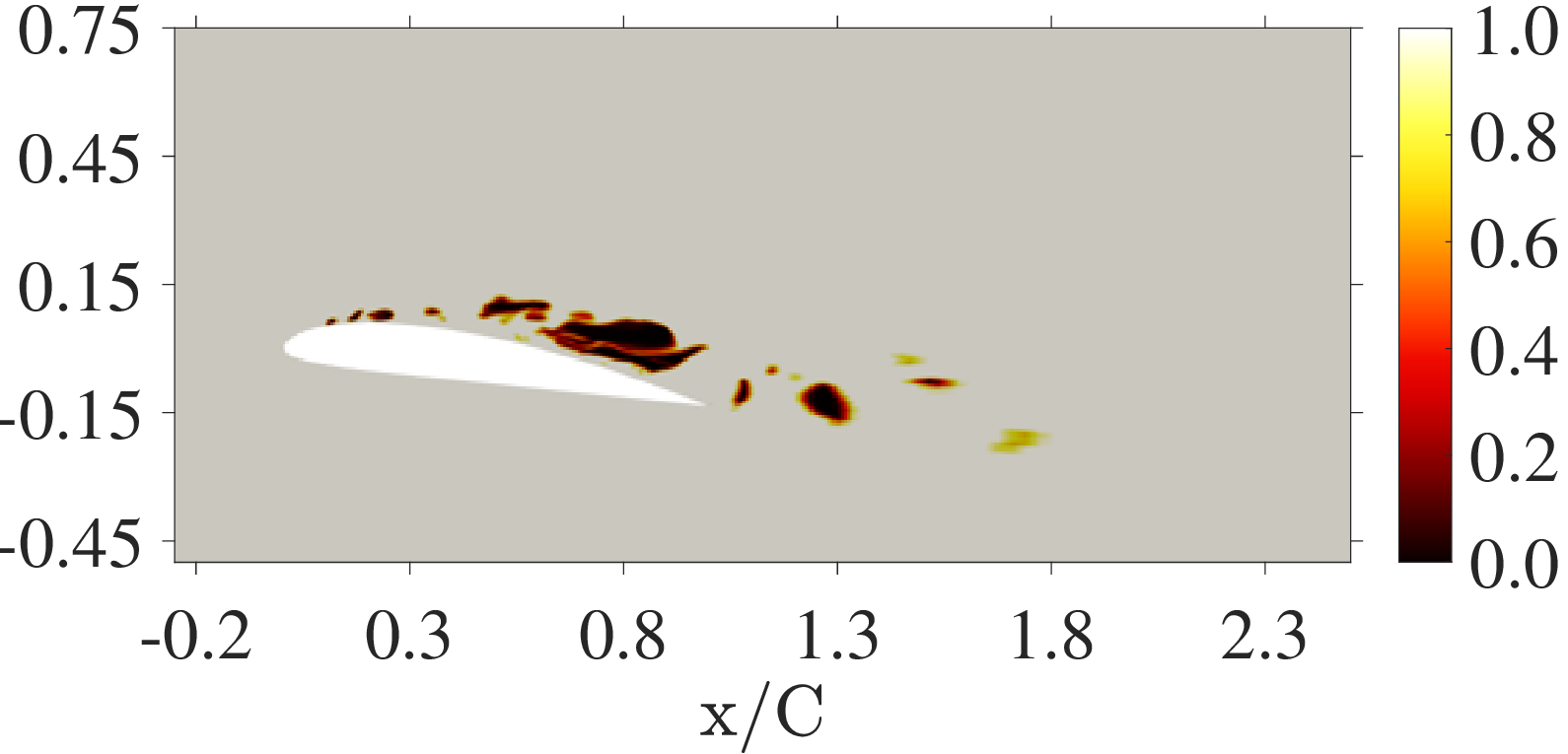}  &
\includegraphics[width=0.34\textwidth]{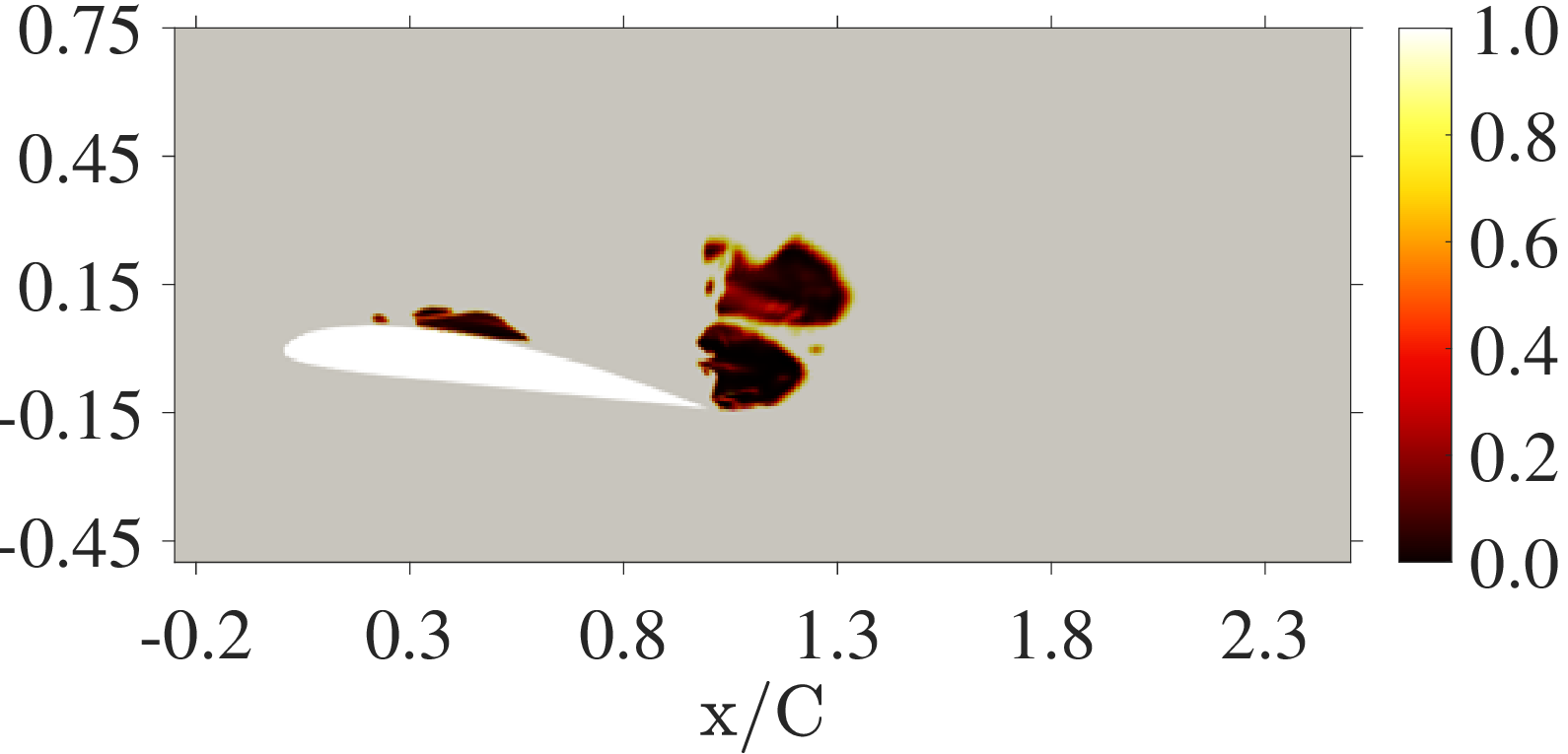} &
\includegraphics[width=0.34\textwidth]{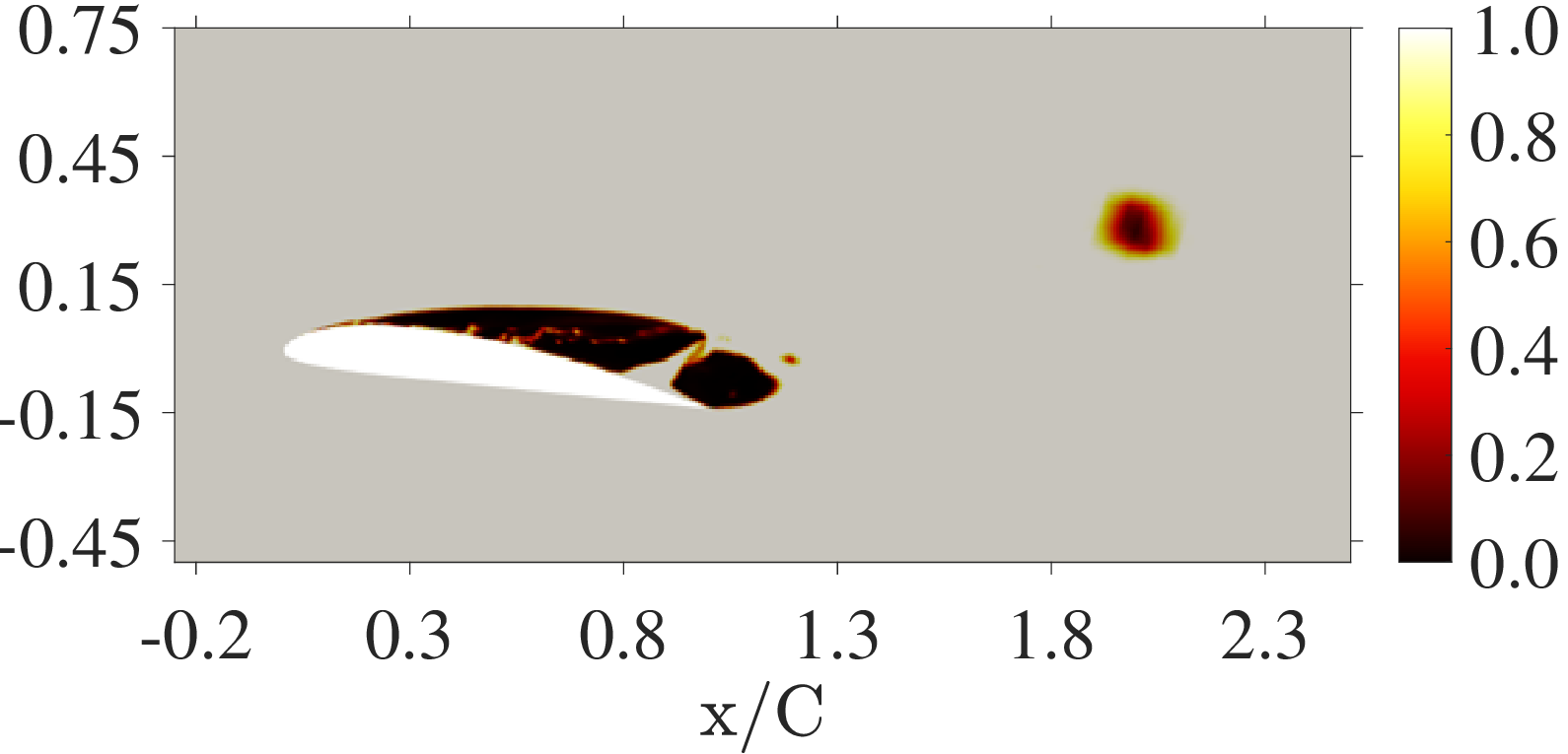} \\
\includegraphics[width=0.34\textwidth]{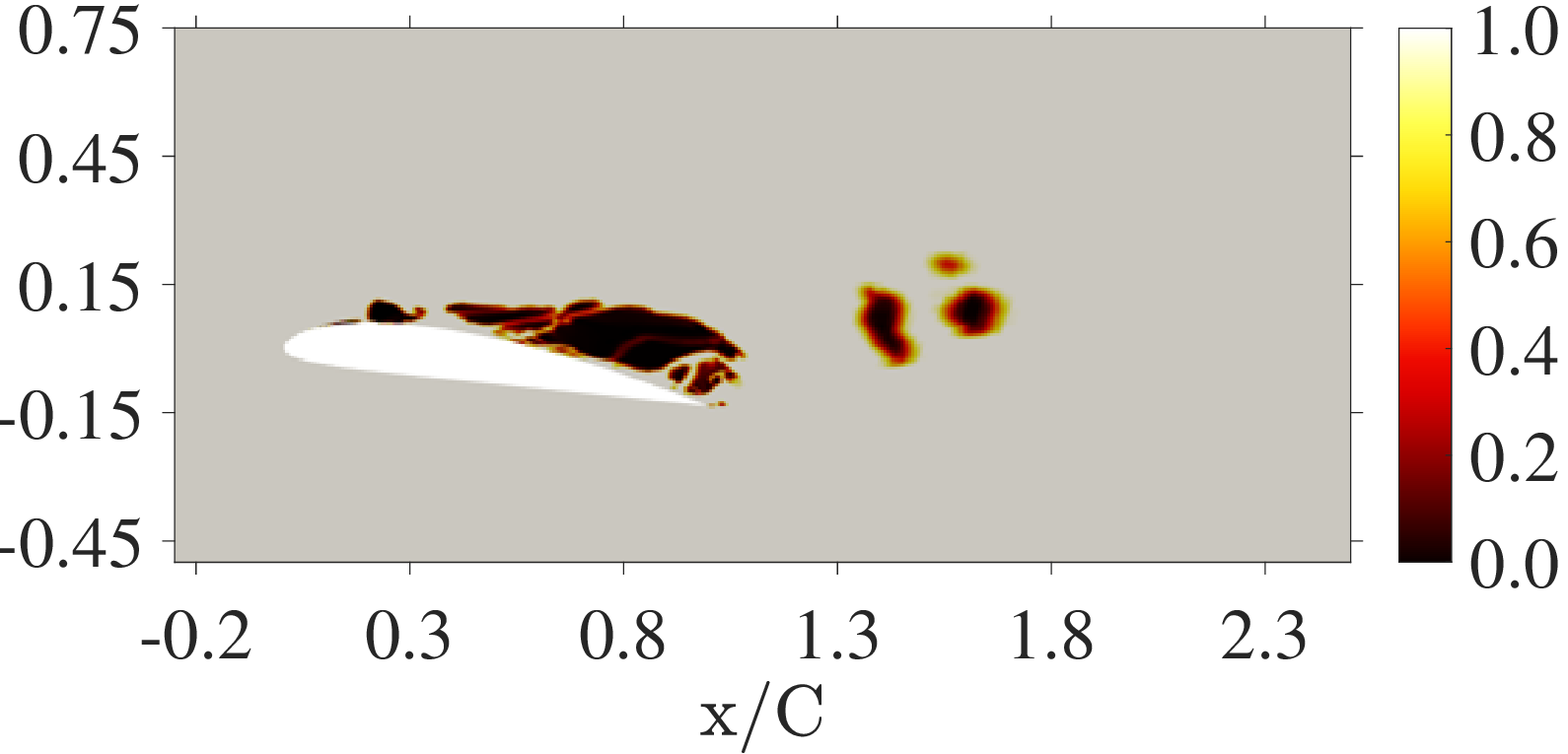} &
\includegraphics[width=0.34\textwidth]{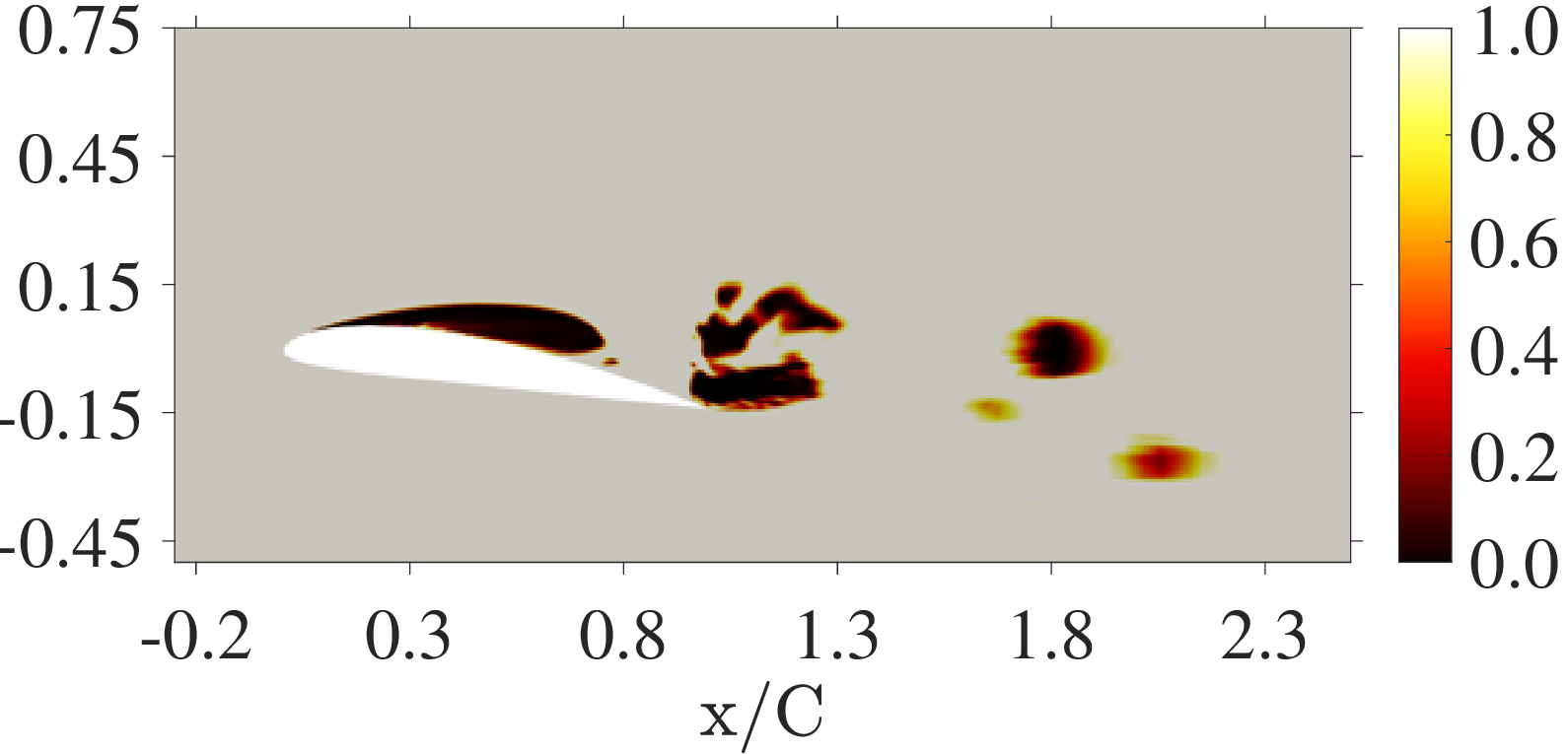} &
\includegraphics[width=0.34\textwidth]{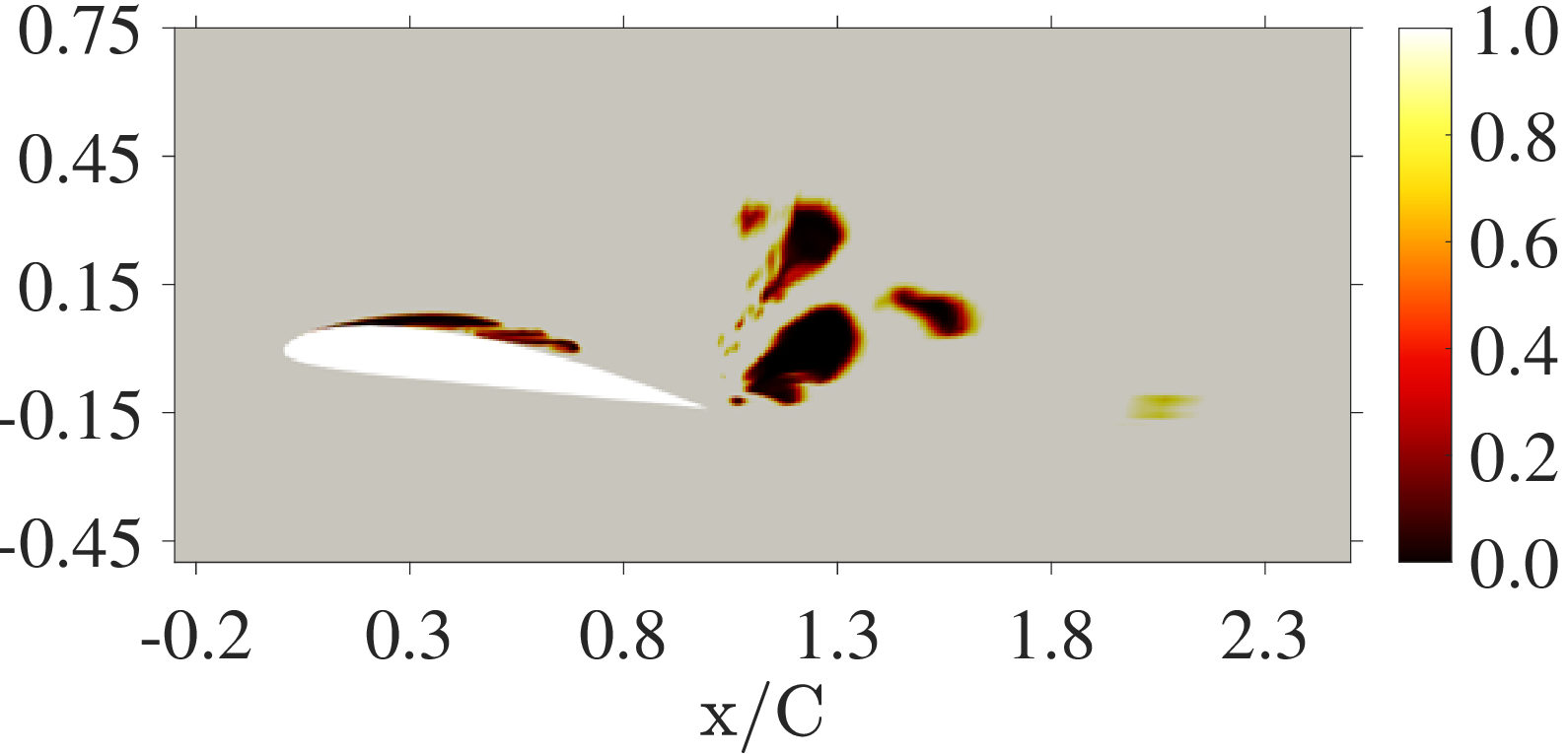} \\
\includegraphics[width=0.34\textwidth]{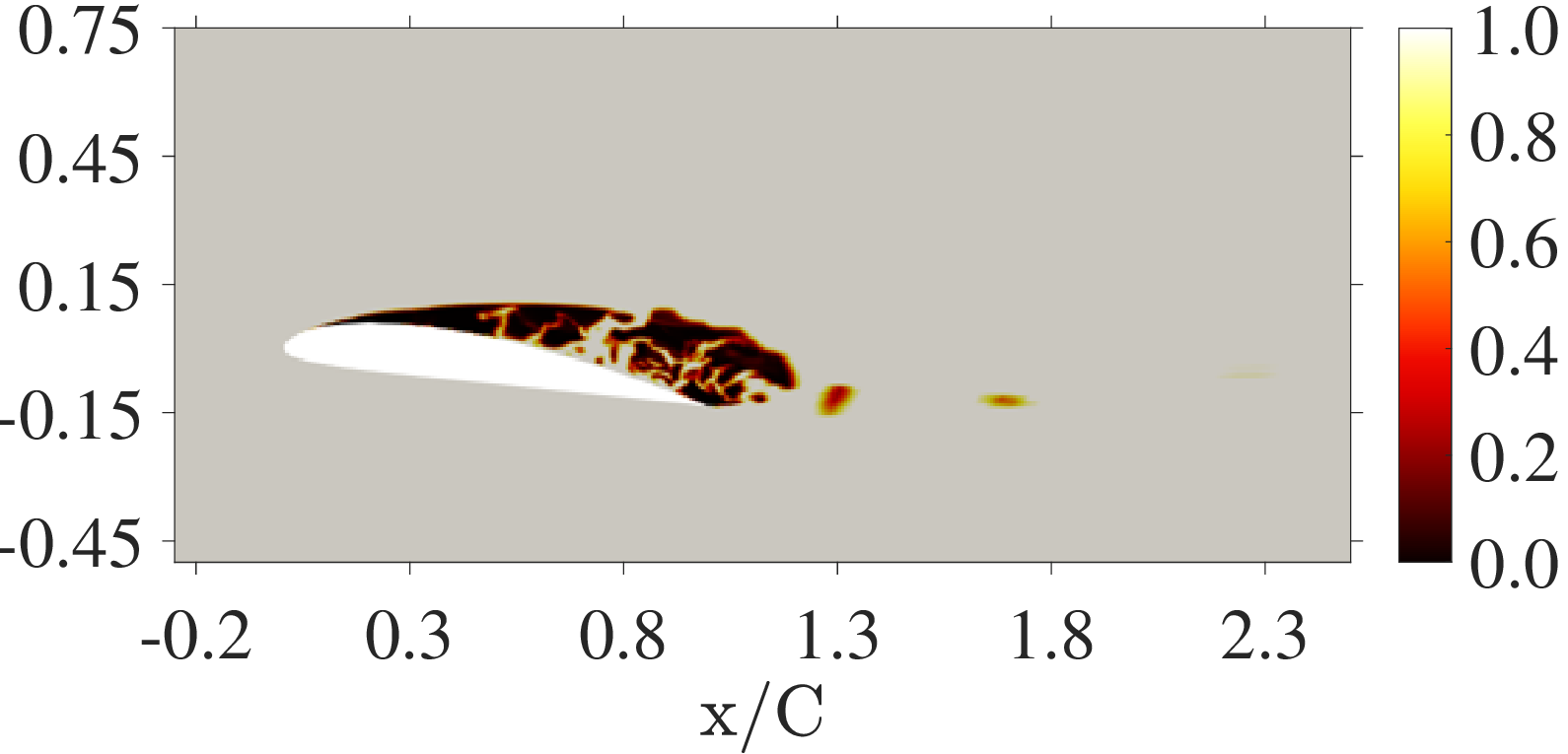}  &
\includegraphics[width=0.34\textwidth]{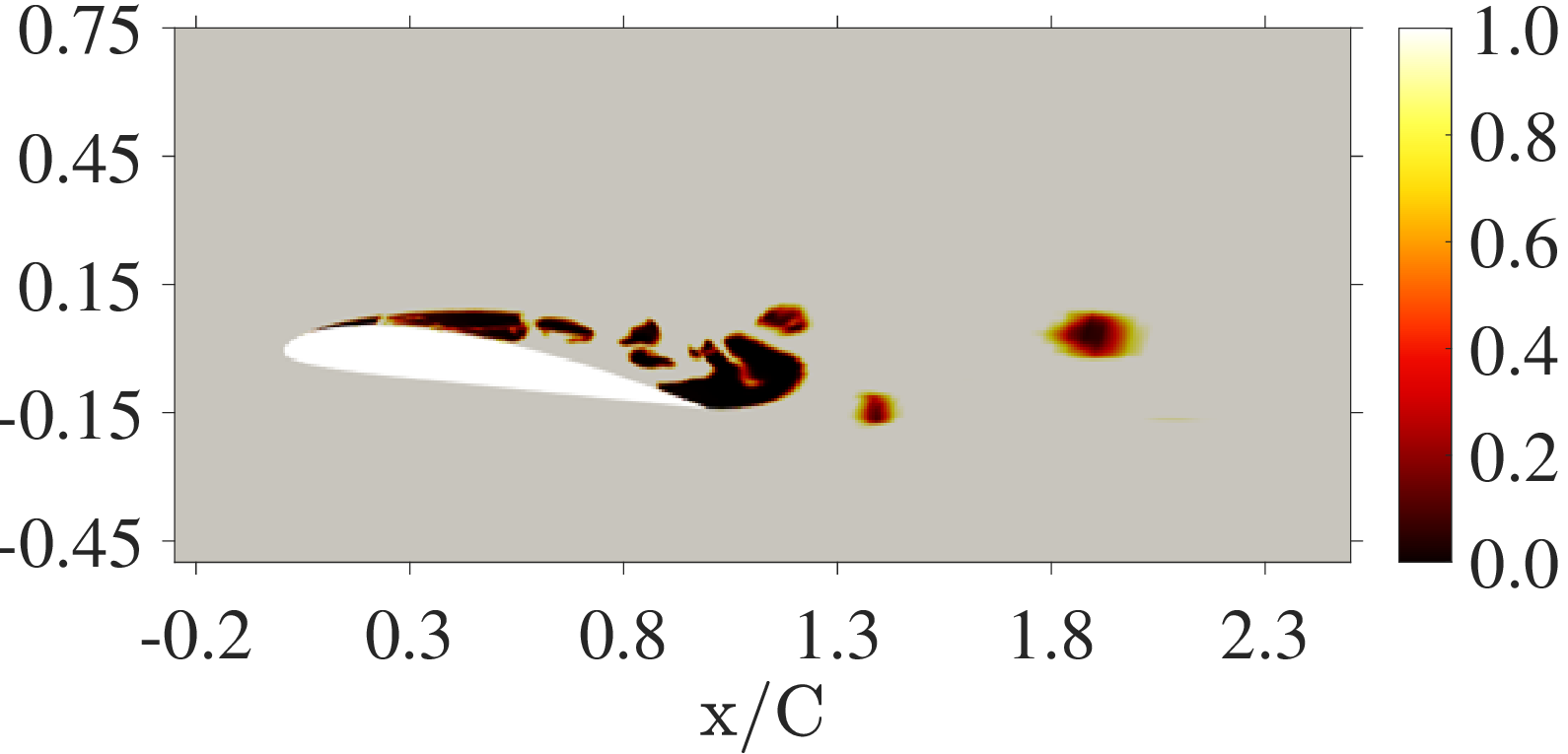} &
\includegraphics[width=0.34\textwidth]{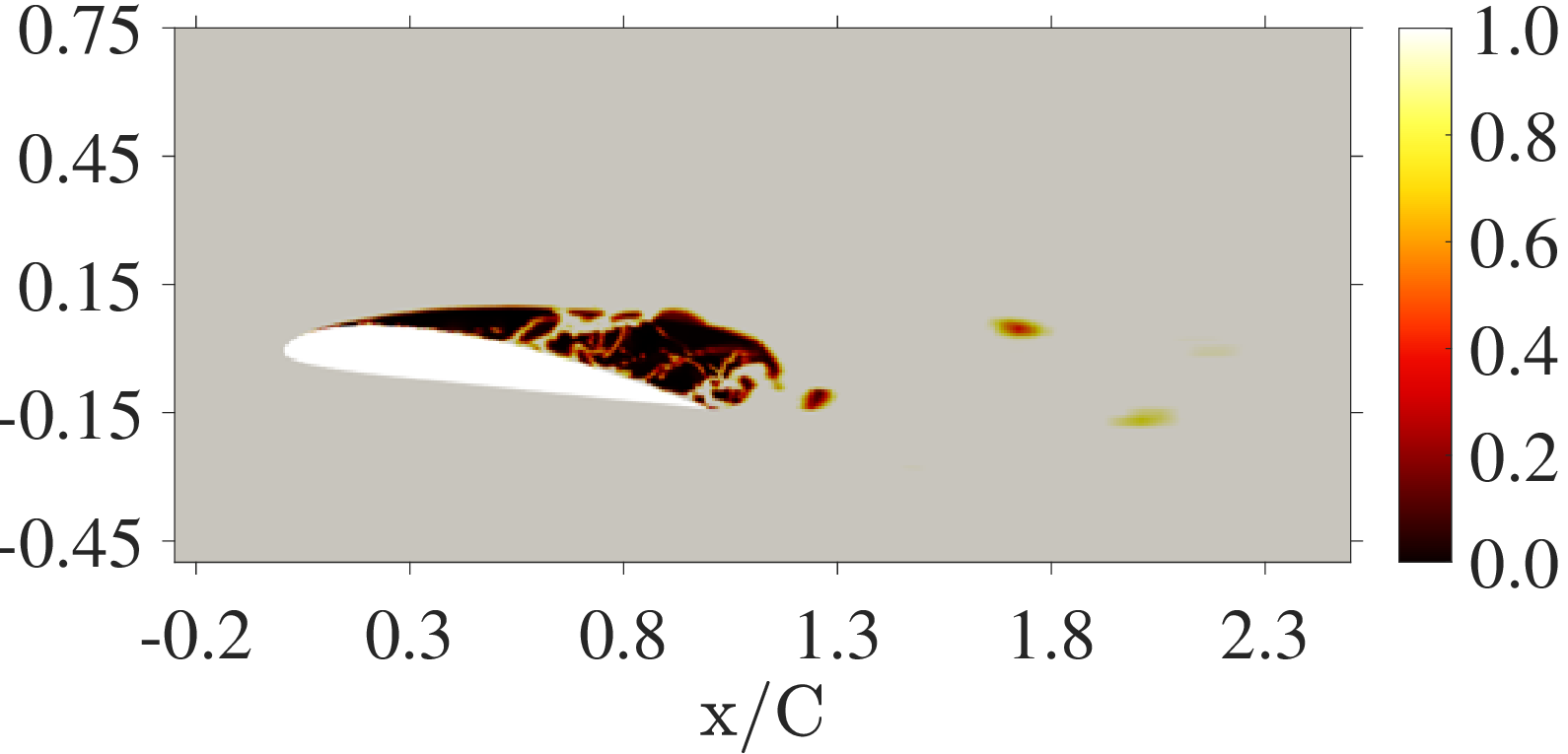} \\
\text{WCA=$0^\circ$} & \text{WCA=$80^\circ$} & \text{WCA=$160^\circ$} \\
\end{tabular}
\caption{Snapshots of water volume fraction ($\alpha_{\text{water}}$) contours at five flow times: $t = 0.80$, 0.85, 0.90, 0.95, and 1.00 s (top to bottom) for flow around a Clark Y hydrofoil at 8 degrees of angle of attack and cavitation number of 0.8. Columns represent WCA of $0^\circ$, $80^\circ$, and $160^\circ$, respectively. Higher WCAs lead to larger, more detached vapor structures and greater wake interaction.}
    \label{fig:vapor_snapshots}
\end{figure}

Figure~\ref{fig:vapor_snapshots} illustrates the instantaneous contours of the water volume fraction ($\alpha_{\text{water}}$) at five consecutive time instants: $t = 0.80$, 0.85, 0.90, 0.95, and 1.00 s, presented from top to bottom. Each column corresponds to a different WCA: $0^\circ$, $80^\circ$, and $160^\circ$, displayed in Figs.~\ref{fig:vapor_snapshots}(a), (b), and (c), respectively.

In Fig.~\ref{fig:vapor_snapshots}(a), representing WCA = $0^\circ$, cavitation is minimal and localized solely near the suction side of the hydrofoil with negligible downstream extension. The vapor structures are small and exhibit strong surface adhesion, with only occasional, transient detachment events that quickly vanish. This indicates a highly stable flow regime where the hydrophilic surface actively suppresses extensive vapor formation and growth, likely due to strong liquid-solid attractive forces.

As WCA increases to $80^\circ$ (Fig.~\ref{fig:vapor_snapshots}(b)), a marked shift in cavitation behavior becomes apparent. Vapor pockets become more prevalent and detach further upstream along the hydrofoil's chord, evolving into larger cloud cavitation structures that propagate into the wake. These vapor clouds display increased spatial intermittency and reduced structural coherence across the time sequence, signifying a rise in flow unsteadiness compared to the $0^\circ$ case. This intermediate wettability allows for more robust bubble nucleation and growth, leading to a more dynamic shedding process.

For WCA = $160^\circ$ (Fig.~\ref{fig:vapor_snapshots}(c)), the most profound changes are observed. The vapor regions expand dramatically, covering a significant portion of the hydrofoil's chord and extending deeply into the wake. Large-scale cavity detachment events dominate the flow, with voluminous vapor structures persisting and actively interacting with the downstream shear layers. The temporal snapshots reveal significantly broader spatial coverage of vapor, with clear indications of secondary cavitation forming near the trailing edge, further exacerbating the vapor extent. This extreme hydrophobicity promotes a highly destabilized cavitation regime, where the reduced liquid-solid interaction facilitates early and widespread vapor generation, leading to an exceptionally turbulent and active wake.

\begin{figure}[h]
\centering
\begin{tabular}{cccc}
\includegraphics[width=0.34\textwidth]{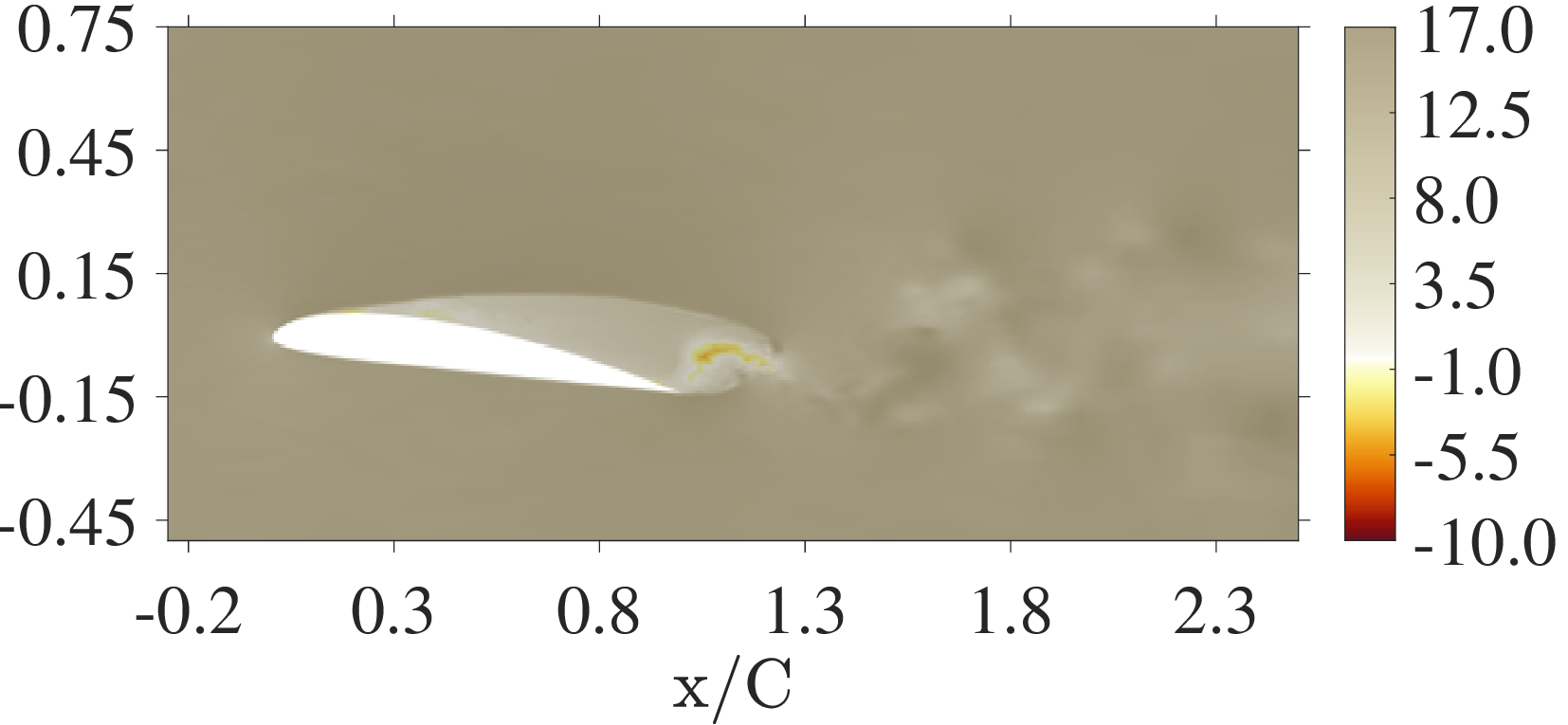} &
\includegraphics[width=0.34\textwidth]{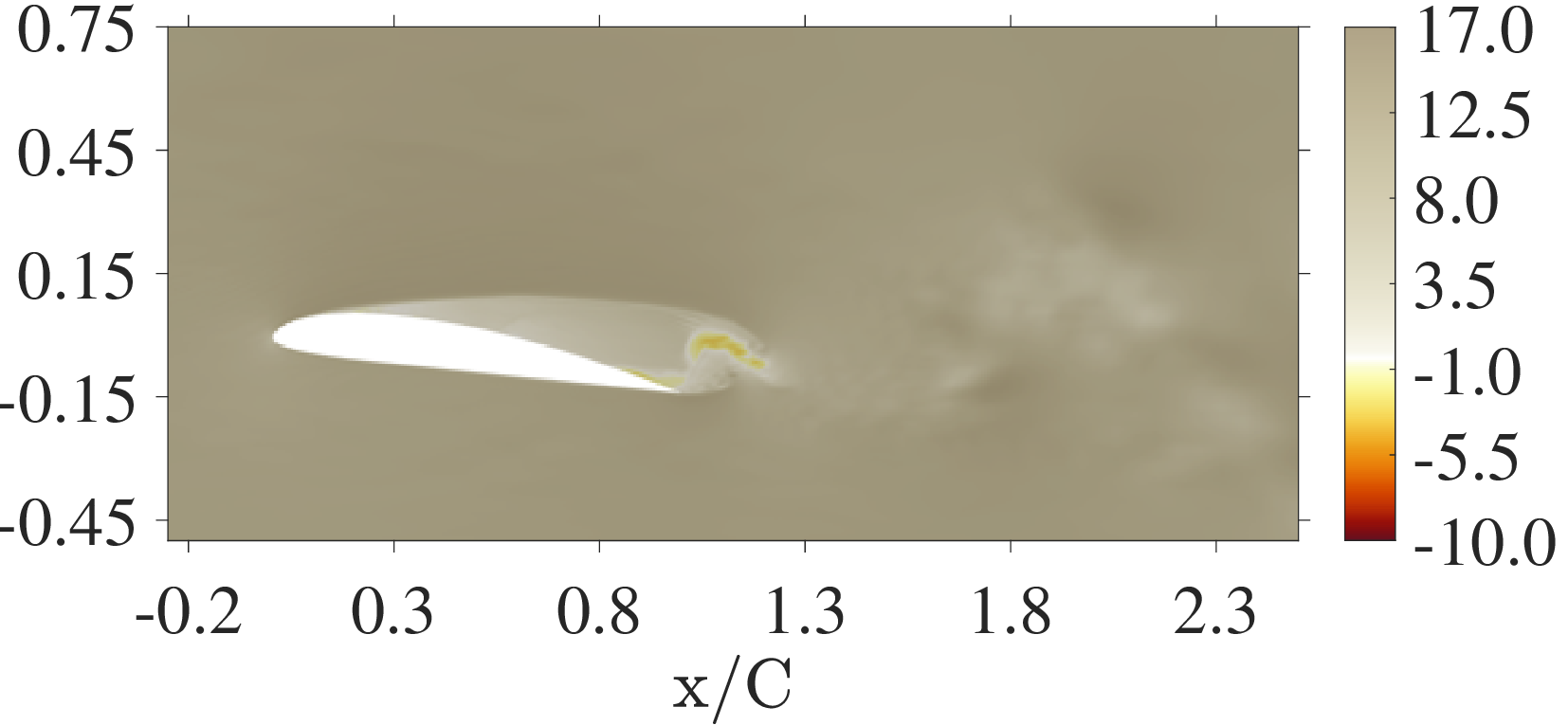} &
\includegraphics[width=0.34\textwidth]{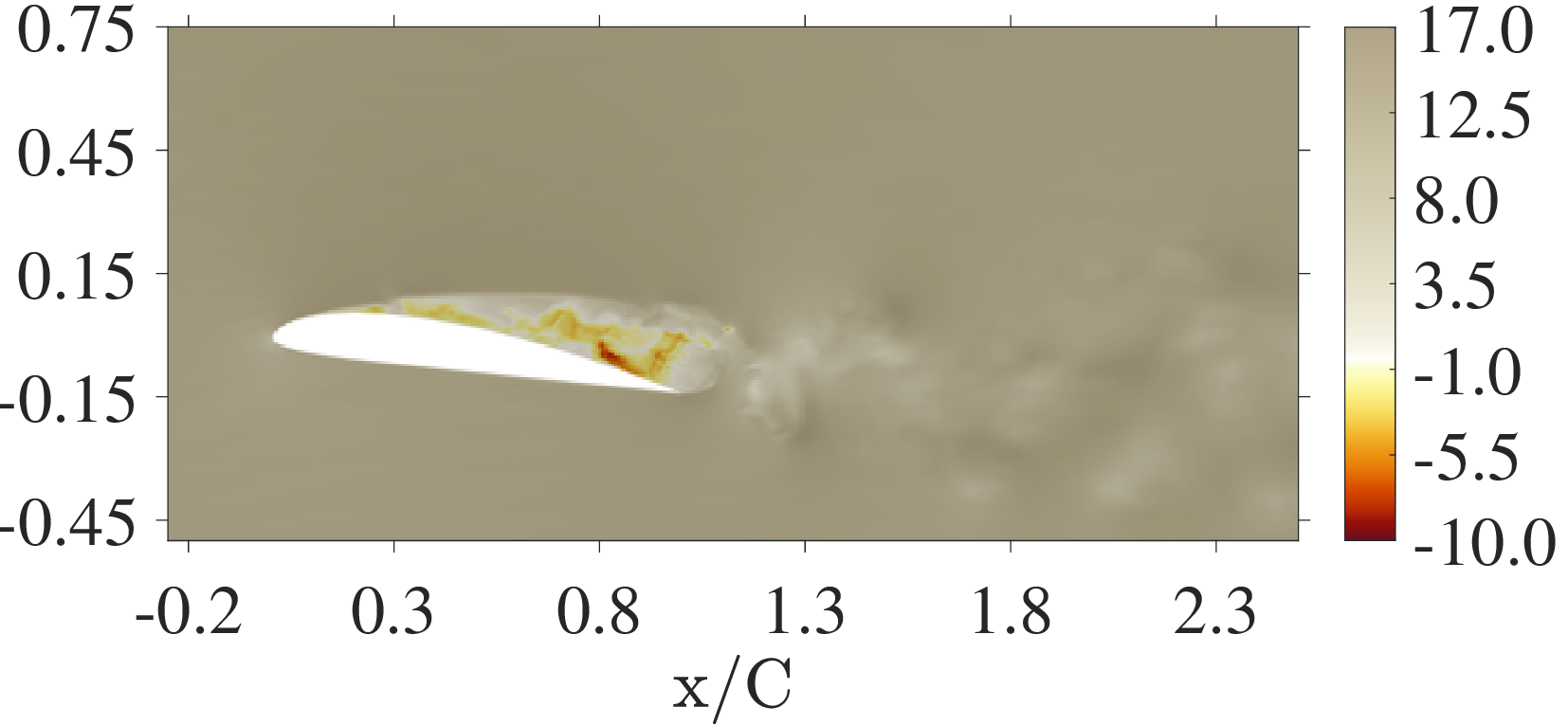} \\

\includegraphics[width=0.34\textwidth]{figures/zero_V10/Ux.0102.eps} &
\includegraphics[width=0.34\textwidth]{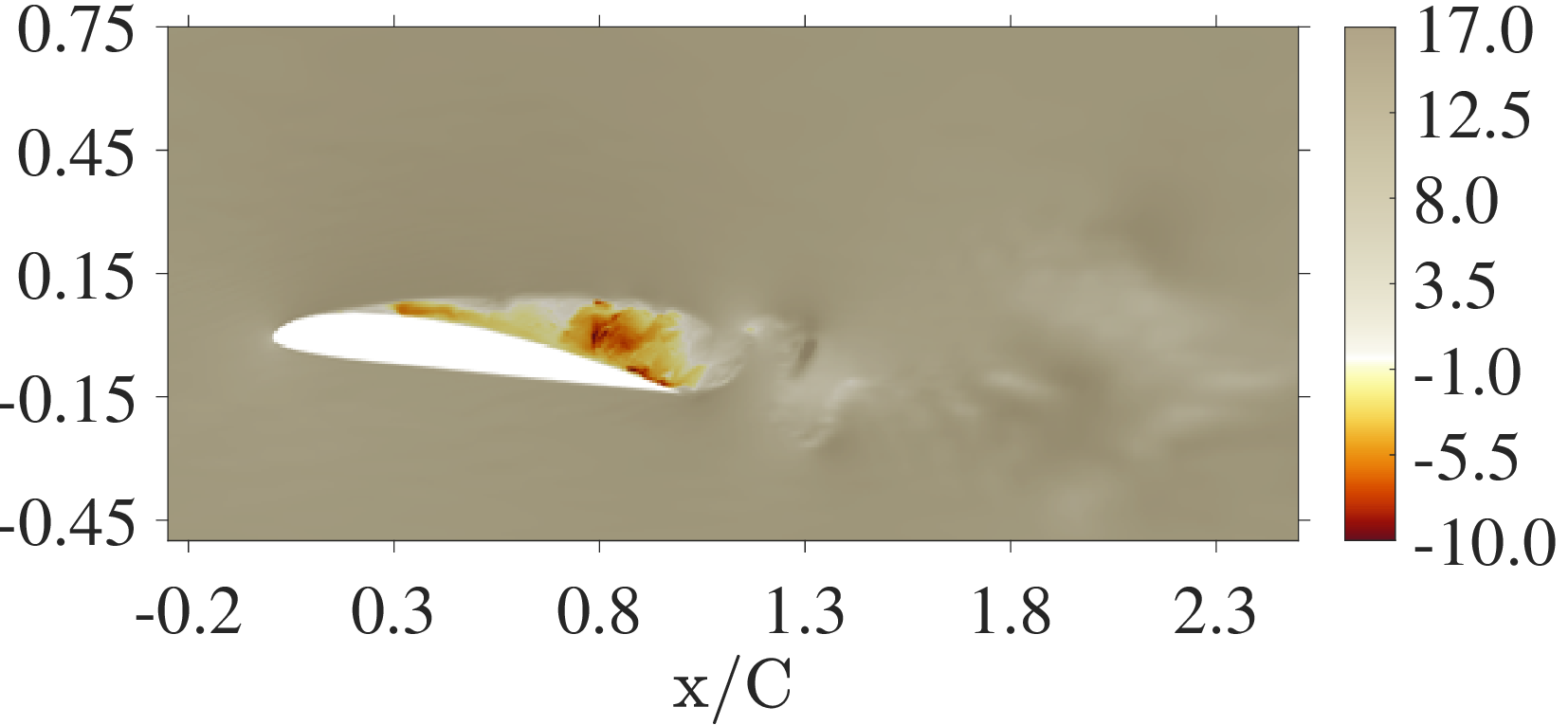} &
\includegraphics[width=0.34\textwidth]{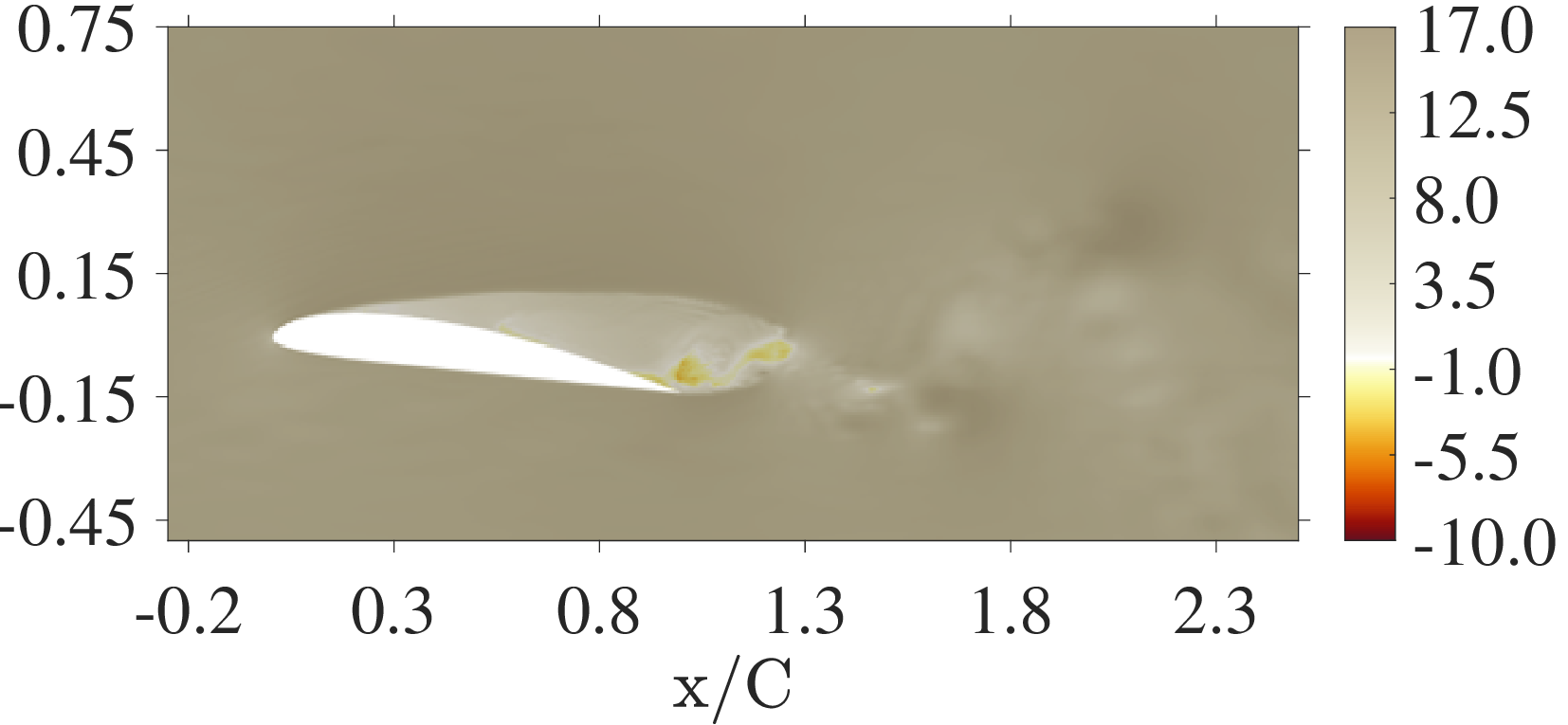} \\
\includegraphics[width=0.34\textwidth]{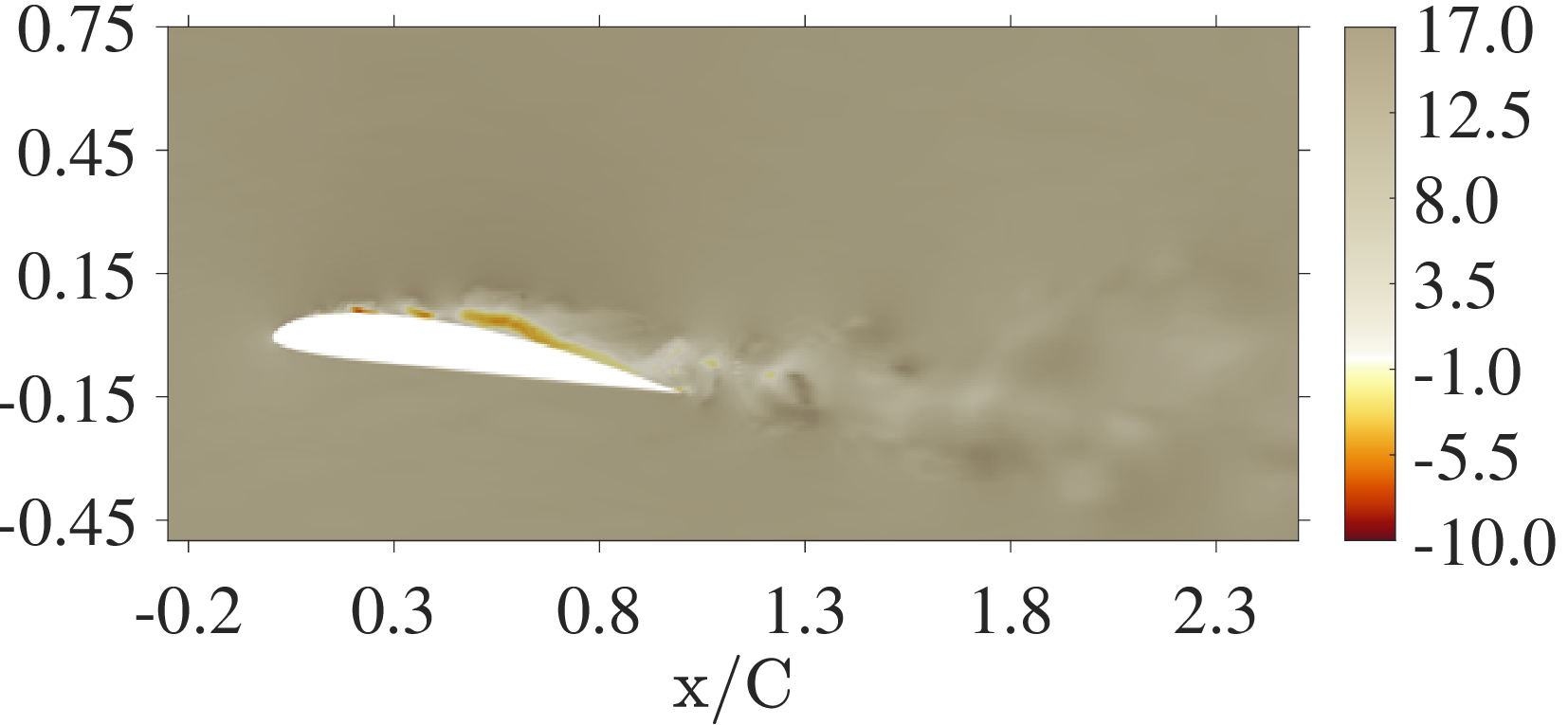}  &
\includegraphics[width=0.34\textwidth]{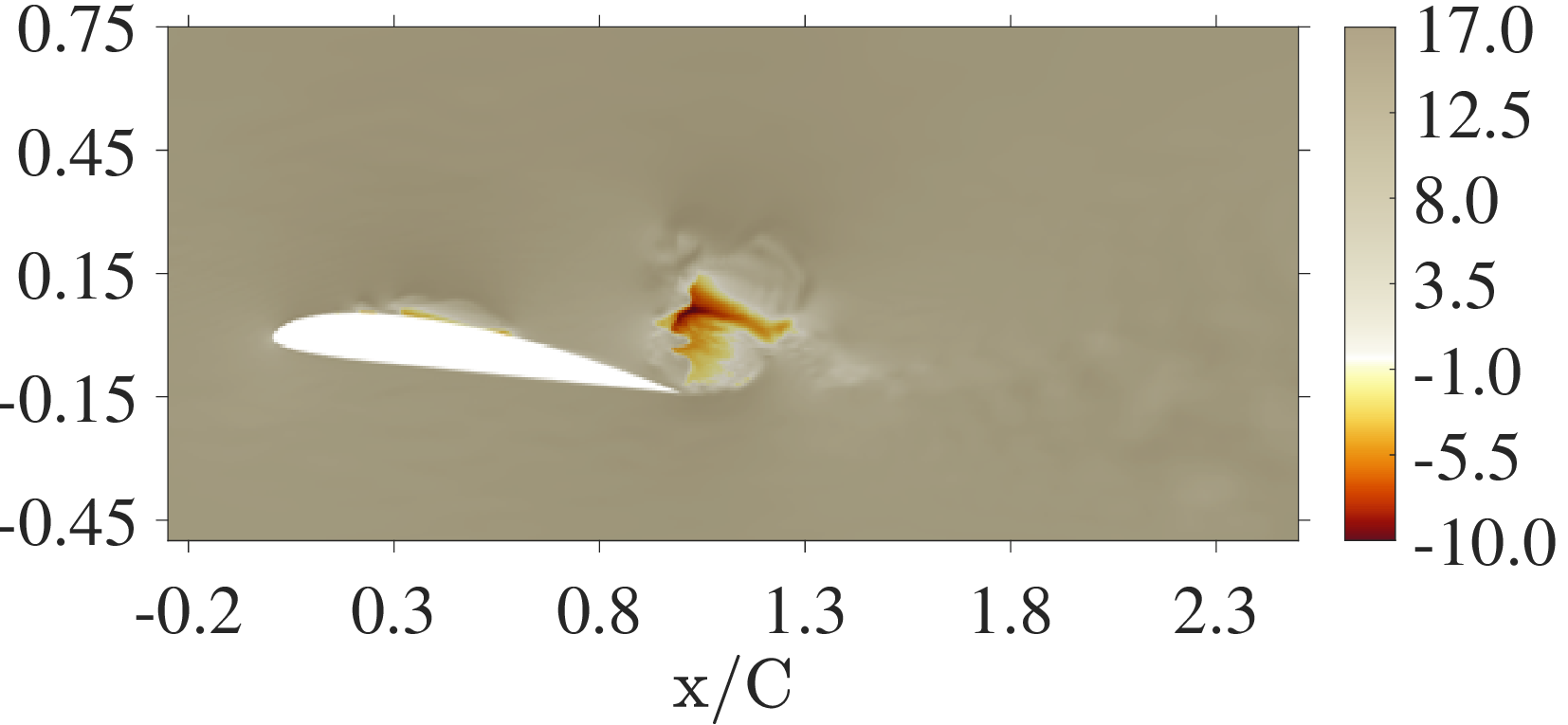} &
\includegraphics[width=0.34\textwidth]{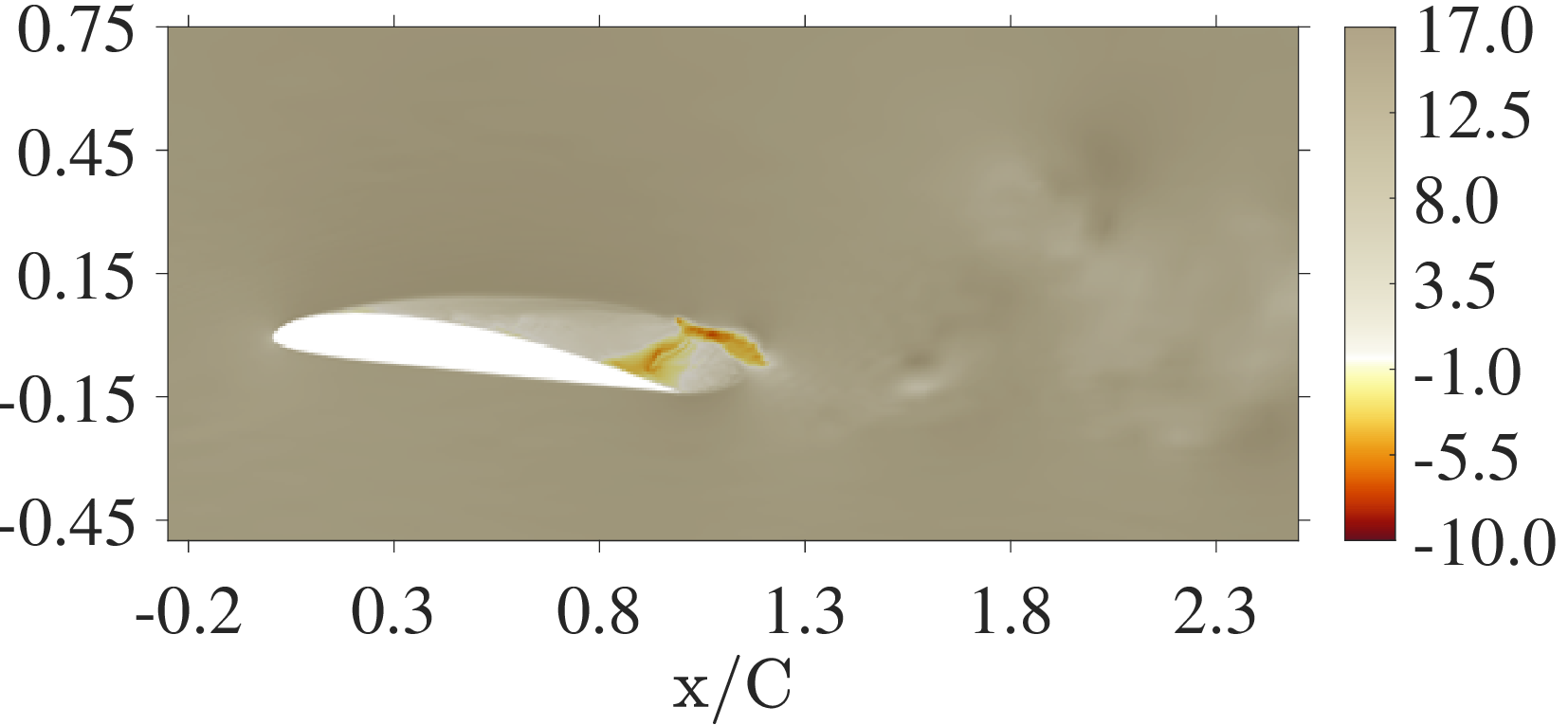} \\

\includegraphics[width=0.34\textwidth]{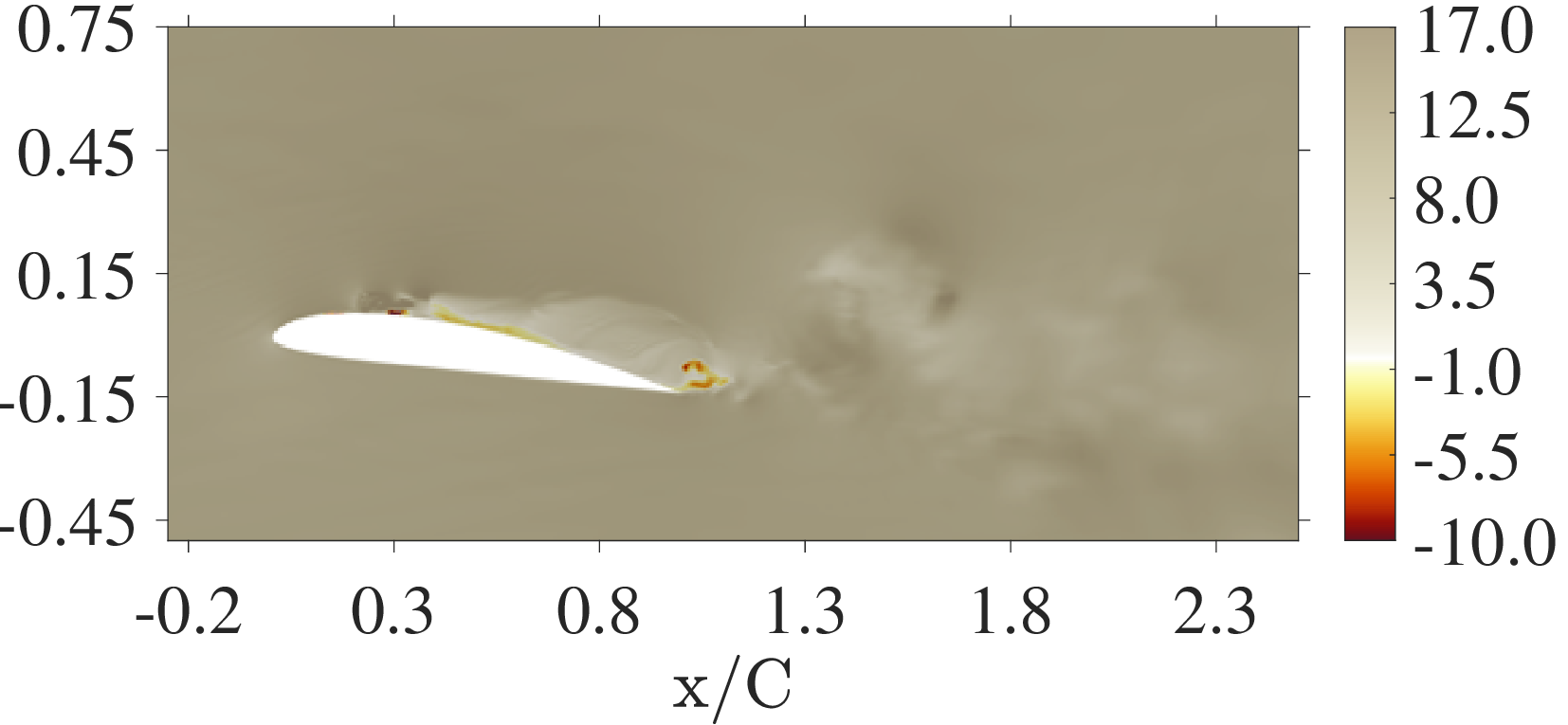} &
\includegraphics[width=0.34\textwidth]{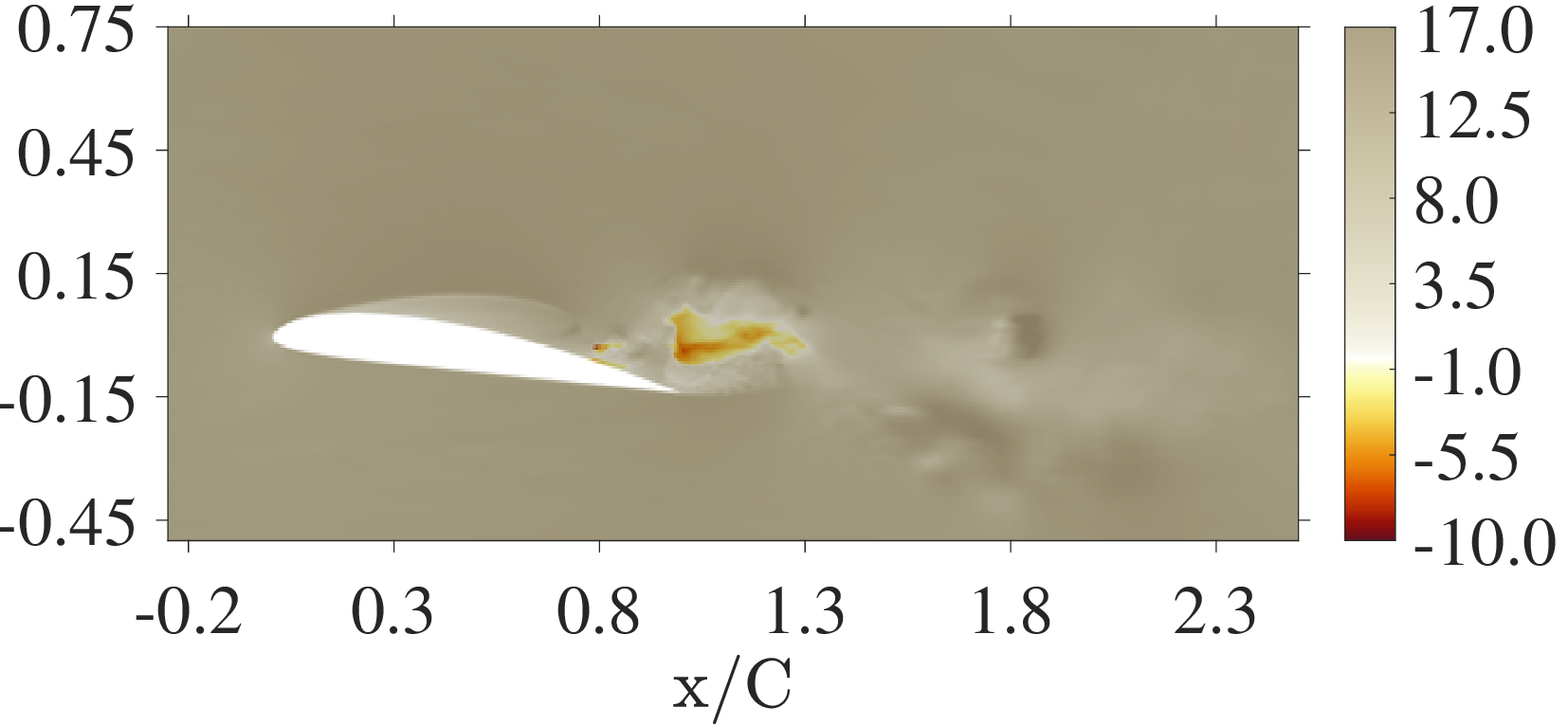} &
\includegraphics[width=0.34\textwidth]{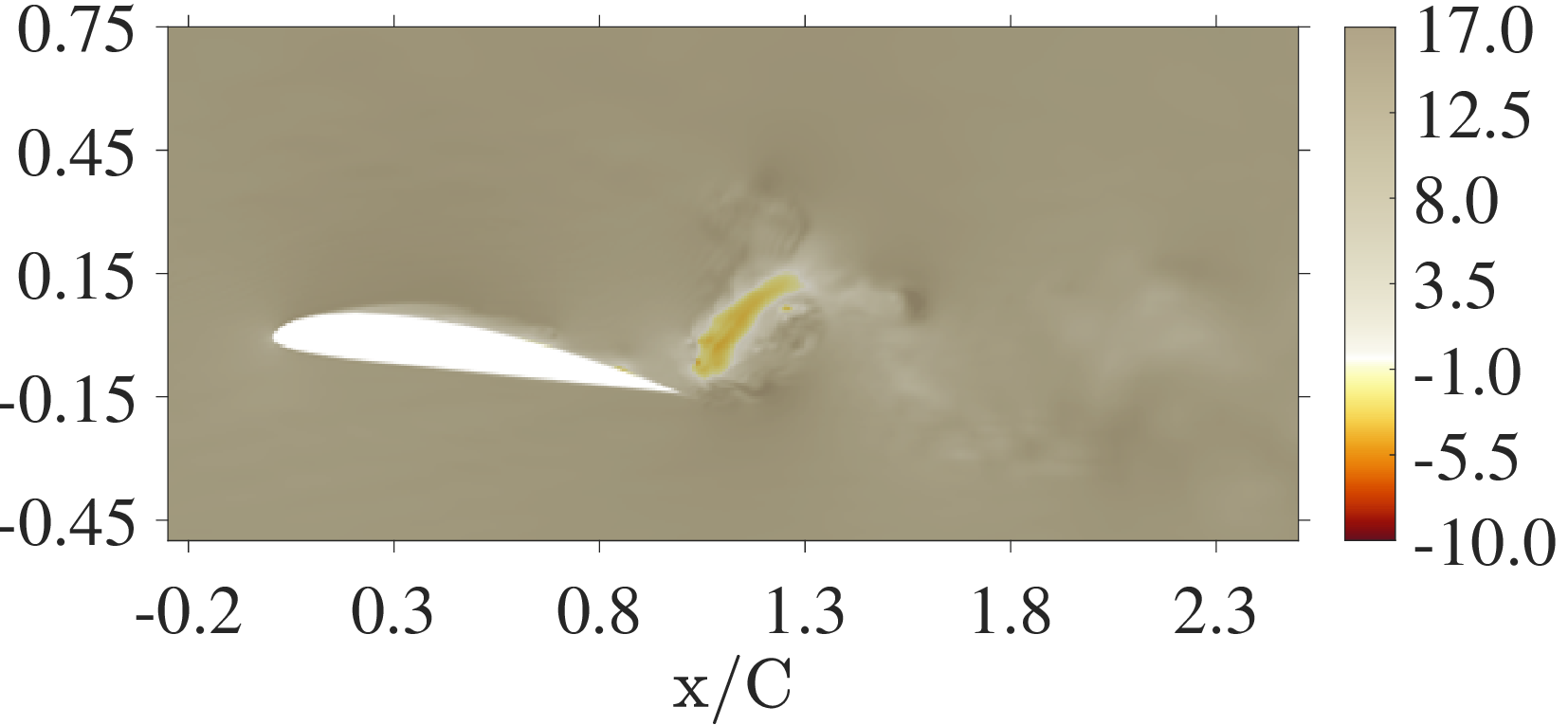} \\
\includegraphics[width=0.34\textwidth]{figures/zero_V10/Ux.0162.eps}  &
\includegraphics[width=0.34\textwidth]{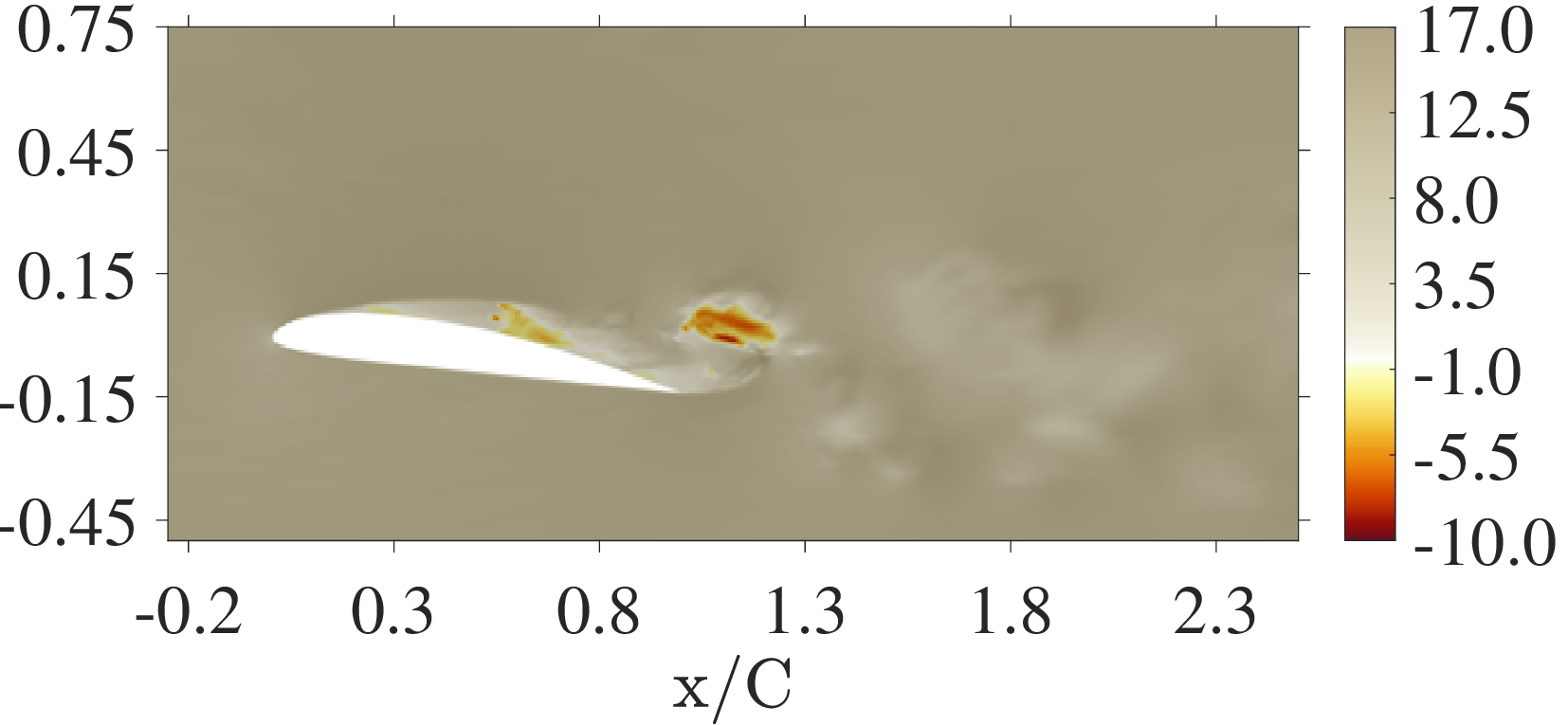} &
\includegraphics[width=0.34\textwidth]{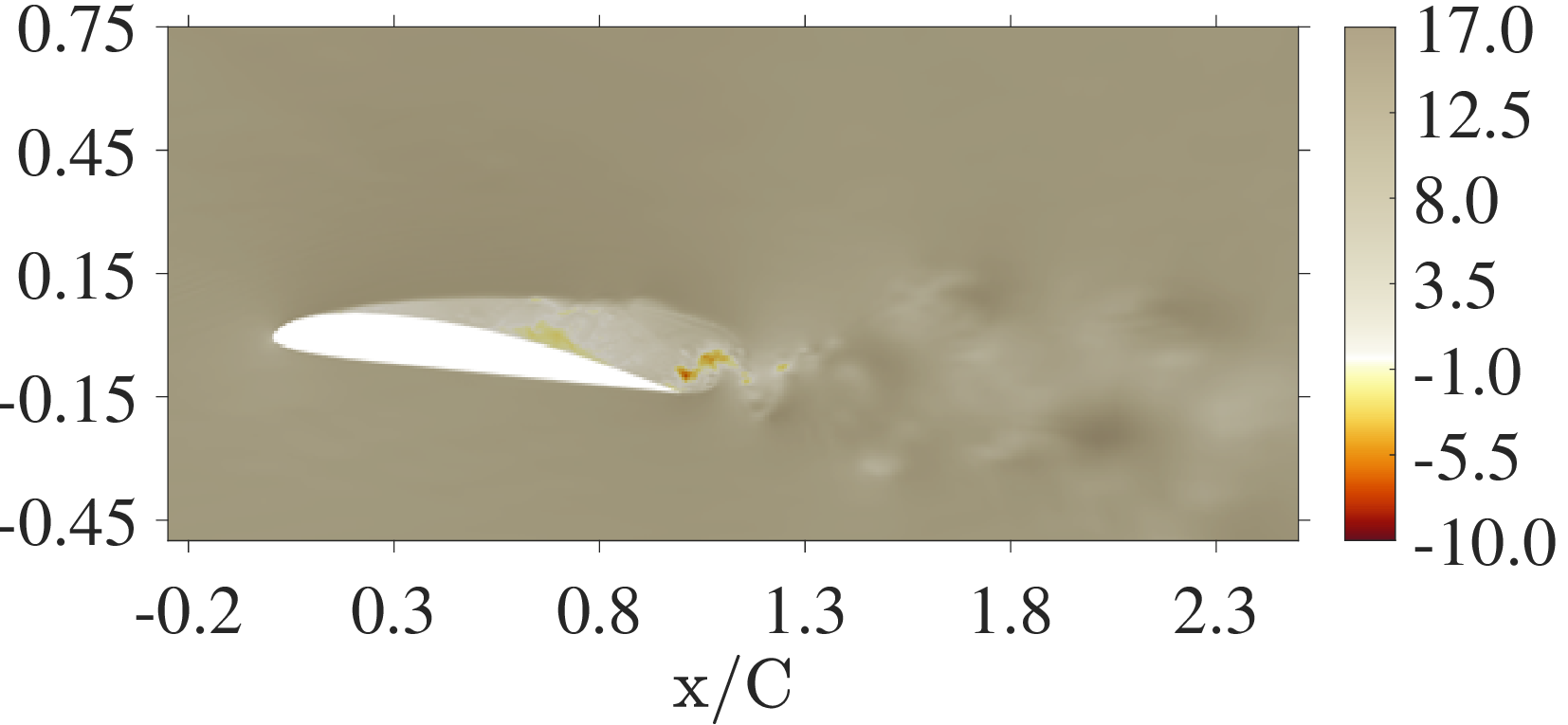} \\
\text{WCA=$0^\circ$} & \text{WCA=$80^\circ$} & \text{WCA=$160^\circ$} \\
\end{tabular}
\caption{Snapshots of axial velocity ($U_{\text{x}}$) contours at five flow times: $t = 0.80$, 0.85, 0.90, 0.95, and 1.00 s (top to bottom) for flow around a Clark Y hydrofoil at 8 degrees of angle of attack and cavitation number of 0.8. Columns represent WCA of $0^\circ$, $80^\circ$, and $160^\circ$, respectively.}    \label{fig:xVelocity}
\end{figure}

Figure~\ref{fig:xVelocity} presents instantaneous contours of the axial velocity component ($U_x$) at five time instances ranging from $t = 0.80$ to 1.00 s (top to bottom), for WCAs of $0^\circ$, $80^\circ$, and $160^\circ$, shown in Figs.~\ref{fig:xVelocity}(a)--(c), respectively. These visualizations show the evolution of the near-wake structure and the  influence of wall wettability on momentum transport and flow stability.

In Fig.~\ref{fig:xVelocity}(a), corresponding to the hydrophilic $0^\circ$ WCA, the flow remains stable and largely attached downstream of the hydrofoil. Only negligible velocity gradients are discernible near the trailing edge, and the wake maintains a narrow, coherent profile. The recirculation zone is notably weak, consistent with the suppressed cavitation activity and minimal vapor detachment observed in the corresponding water volume fraction snapshots (Fig.~\ref{fig:vapor_snapshots}(a)). This demonstrates that strong liquid-solid adhesion effectively preserves the momentum of the mean flow.

Figure~\ref{fig:xVelocity}(b) for $80^\circ$ WCA reveals a transition towards increased flow disturbance. Localized velocity deficits begin to form near the suction surface and extend into the wake, indicating the onset of stronger flow unsteadiness and cavitation-induced separation. The wake exhibits increasing asymmetry and clear signs of momentum loss downstream, mirroring the more frequent and larger vapor pockets seen in Fig.~\ref{fig:vapor_snapshots}(b). This intermediate wettability facilitates a more active interaction between the vapor structures and the surrounding liquid, leading to greater velocity fluctuations and disrupted wake recovery.

In Fig.~\ref{fig:xVelocity}(c), representing the superhydrophobic surface with $160^\circ$ WCA, the flow field is dramatically altered. High-velocity gradients and pronounced asymmetric structures dominate the trailing edge and near-wake region. This is a direct consequence of the extensive and dynamic vapor cavities observed in Fig.~\ref{fig:vapor_snapshots}(c), where large-scale detachment and interactions with the shear layer induce significant flow distortion. Large-scale vortex shedding and recirculating zones appear, with clear detachment of low-speed fluid packets into the far wake. The flow field is markedly disturbed, with strong spatial variations indicating significant influence from vigorous vapor cavity collapse and the powerful re-entrant jet dynamics characteristic of superhydrophobic surfaces. These high-WCA conditions lead to a substantial redistribution of momentum in the wake, manifesting as broader and more turbulent velocity fluctuations compared to less hydrophobic cases.

\begin{figure}[H]
\caption{Normalized streamwise velocity (\( U/U_\infty \)) profiles at transverse sections: 
(a) \(x/C = 0.6\), (b) \(x/C = 0.8\), (c) \(x/C = 1.0\), and (d) \(x/C = 1.2\), for different WCAs and flow around a Clark Y hydrofoil at 8 degrees of angle of attack and cavitation number of 0.8.}

    \label{fig:velocity_profiles}
    \begin{subfigure}[b]{0.48\linewidth}
        \centering
        \includegraphics[width=\linewidth]{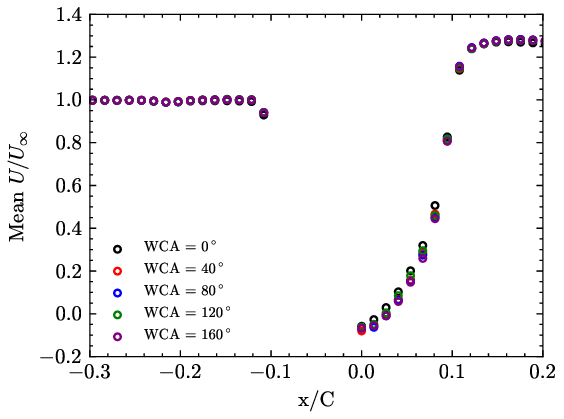}
            \subcaption{x/C=0.6}

    \end{subfigure}
    \hfill
    \begin{subfigure}[b]{0.48\linewidth}
        \centering
        \includegraphics[width=\linewidth]{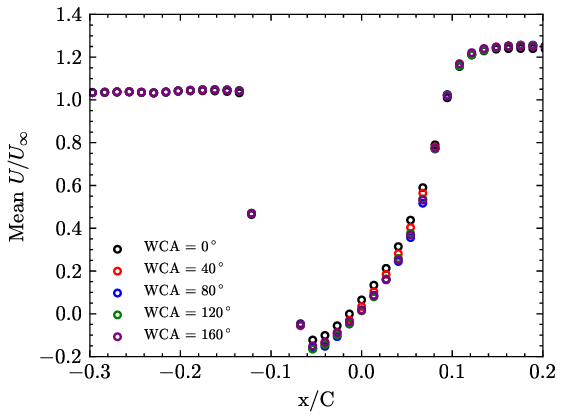}
            \subcaption{x/C=0.8}
    \end{subfigure}
    \vspace{0.5em}
    \begin{subfigure}[b]{0.475\linewidth}
        \centering
        \includegraphics[width=\linewidth]{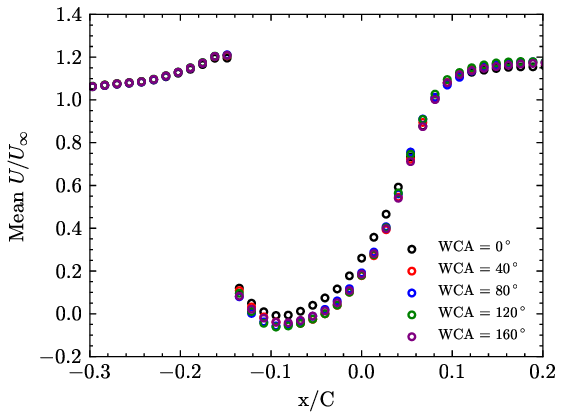}
            \subcaption{x/C=1.0}

    \end{subfigure}
    \hfill
    \begin{subfigure}[b]{0.475\linewidth}
        \centering
        \includegraphics[width=\linewidth]{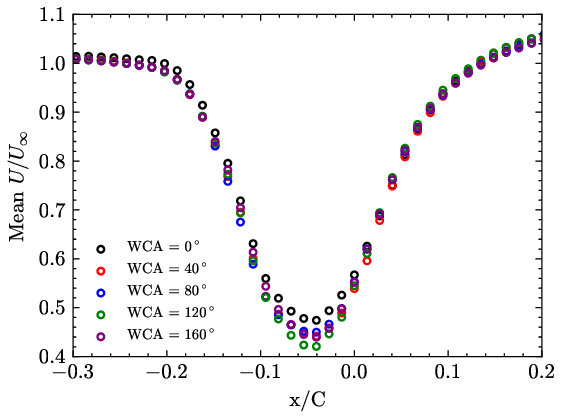}
            \subcaption{x/C=1.2}

    \end{subfigure}
\end{figure}

Figure~\ref{fig:velocity_profiles} displays the mean streamwise velocity normalized by the free-stream velocity ($U/U_\infty$) across the transverse direction at four distinct streamwise positions: $x/C = 0.6$, 0.8, 1.0, and 1.2, labeled as Figs.~\ref{fig:velocity_profiles}(a)--(d), respectively. The plots compare the influence of five different WCAs: $0^\circ$, $40^\circ$, $80^\circ$, $120^\circ$, and $160^\circ$, which show the momentum transport characteristics in the wake.

In Fig.~\ref{fig:velocity_profiles}(a), at $x/C = 0.6$, all WCA cases exhibit a pronounced velocity deficit near the centerline, indicative of the evolving cavity. While the profiles closely align away from the immediate shear layer, negligible differences in the boundary layer region hint at the influence of surface wettability, with the $160^\circ$ WCA case showing marginally greater asymmetry. This initial deviation suggests an earlier impact of reduced surface adhesion on the flow structure.

Further downstream at $x/C = 0.8$ (Fig.~\ref{fig:velocity_profiles}(b)), the core velocity deficit becomes more pronounced and spatially wider for all conditions. The intensification of the velocity gradient across the shear layer is evident, and a clearer separation emerges between the profiles. Higher WCAs, particularly $120^\circ$ and $160^\circ$, demonstrate slower velocity recovery compared to lower WCAs, indicating increased momentum loss in the boundary layer. This corresponds to the larger and more unstable vapor structures observed at these higher WCAs (Fig.~\ref{fig:vapor_snapshots}).

At $x/C = 1.0$ (Fig.~\ref{fig:velocity_profiles}(c)), the wake continues its evolution, characterized by a substantial low-speed region centered near the hydrofoil's trajectory. The distinctions among the different WCA conditions become more prominent here. The higher WCA angles consistently exhibit deeper velocity troughs and a more protracted re-acceleration of the flow, which is directly linked to the sustained influence of larger and more persistently shed vapor structures. This signifies that reduced surface wettability leads to greater wake thickness and diminished kinetic energy recovery.

Finally, at $x/C = 1.2$ (Fig.~\ref{fig:velocity_profiles}(d)), some degree of velocity recovery begins to occur more uniformly across the cases. However, residual wake asymmetry persists. The velocity profiles for lower WCAs show better alignment, indicating a quicker return to undisturbed flow. Conversely, the higher WCAs maintain a broader region of depressed velocity, consistent with the more extensive and turbulent wake regions previously observed (Fig.~\ref{fig:xVelocity}). This comprehensive view of the velocity profiles reinforces that higher WCAs, by promoting more aggressive and persistent cavitation, significantly alter the momentum distribution in the wake, leading to increased drag and altered hydrodynamic performance.

\subsection{Hydrodynamic Behavior at Low Cavitation Number ($\sigma = 0.4$)}

\begin{figure}[H]
\centering
\begin{tabular}{cccc}
\includegraphics[width=0.475\textwidth]{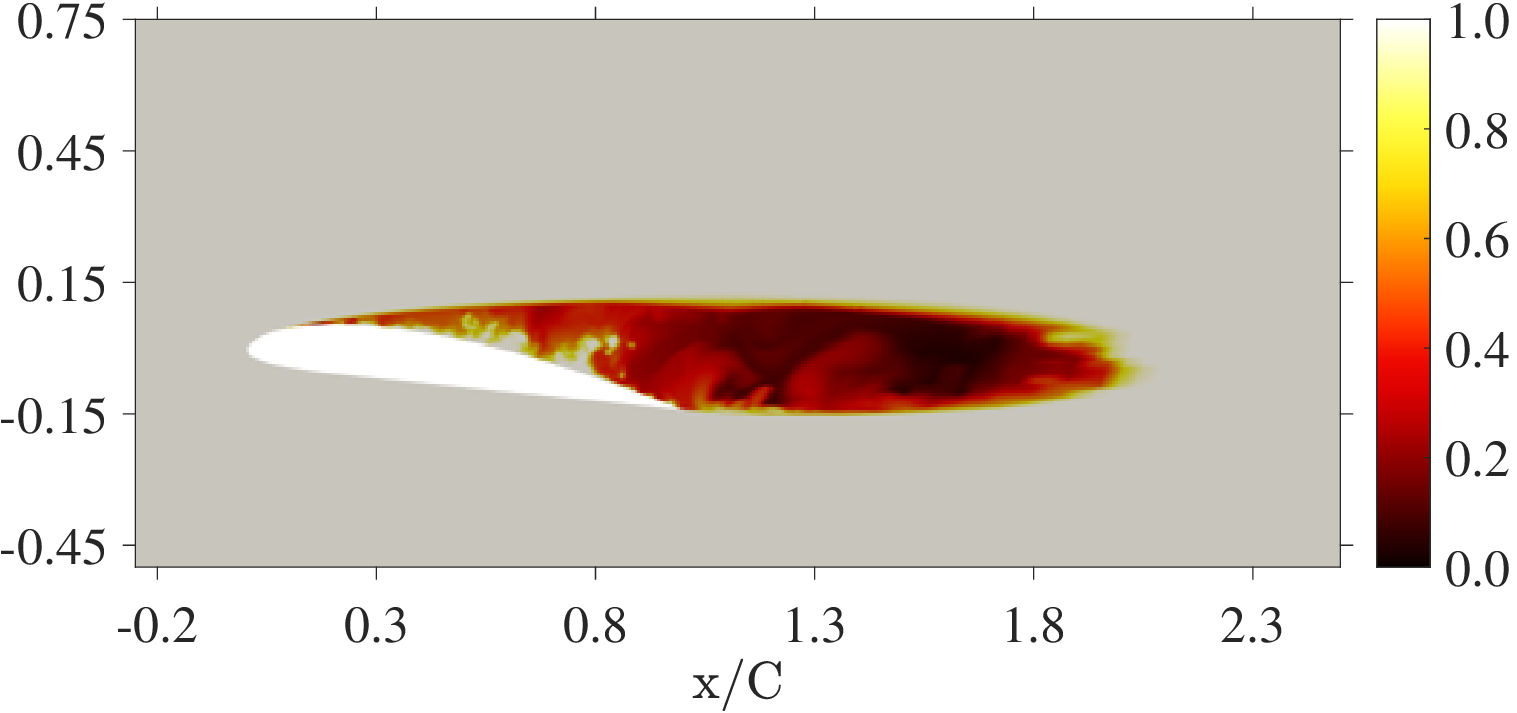} &

\includegraphics[width=0.475\textwidth]{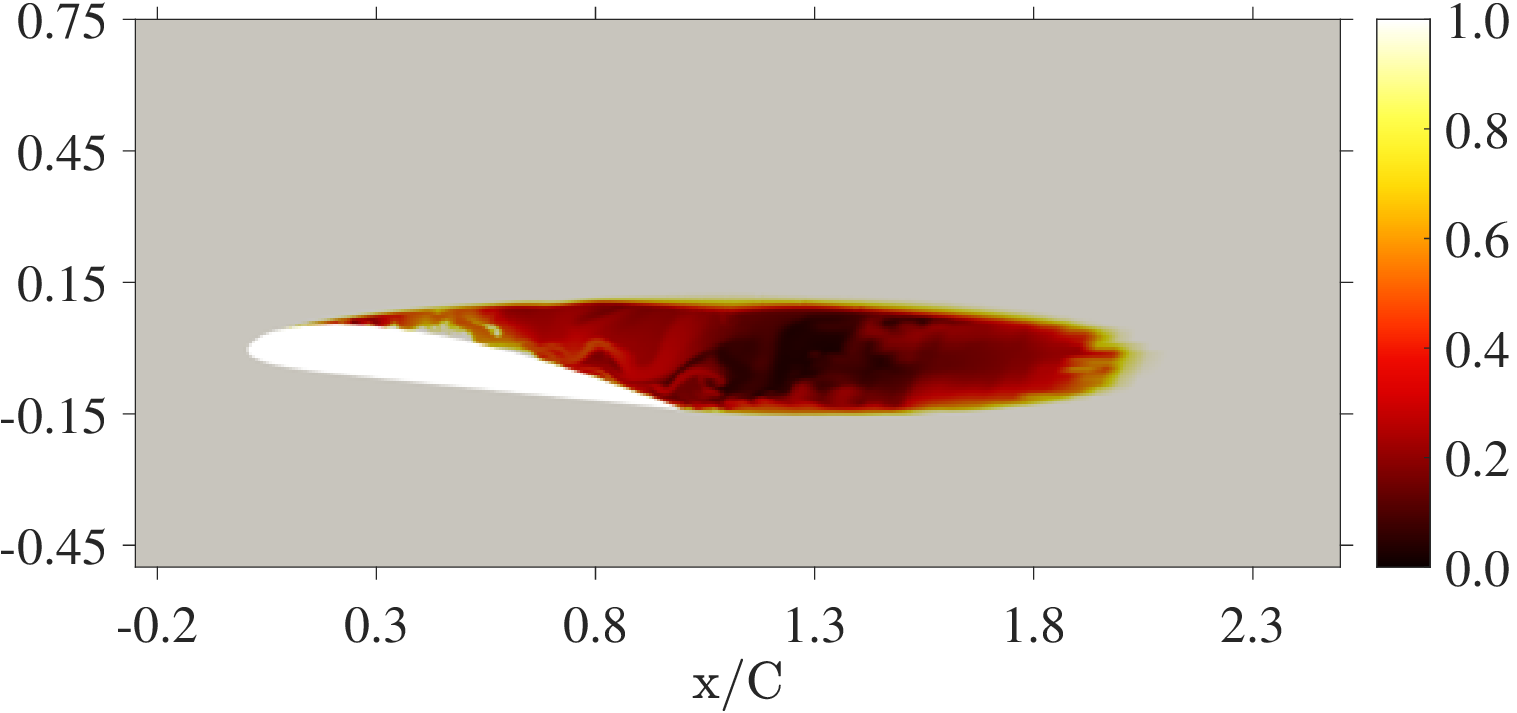} \\

\includegraphics[width=0.475\textwidth]{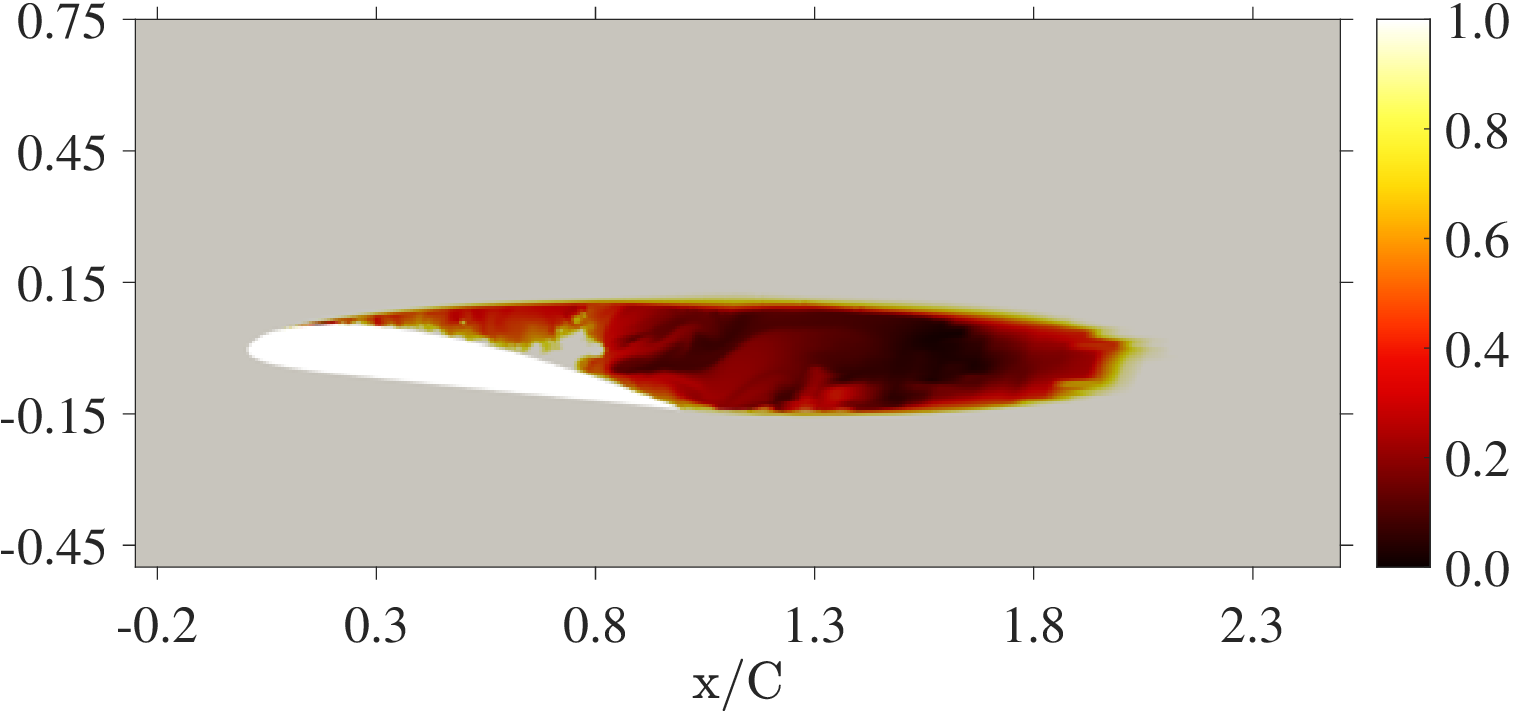} &

\includegraphics[width=0.475\textwidth]{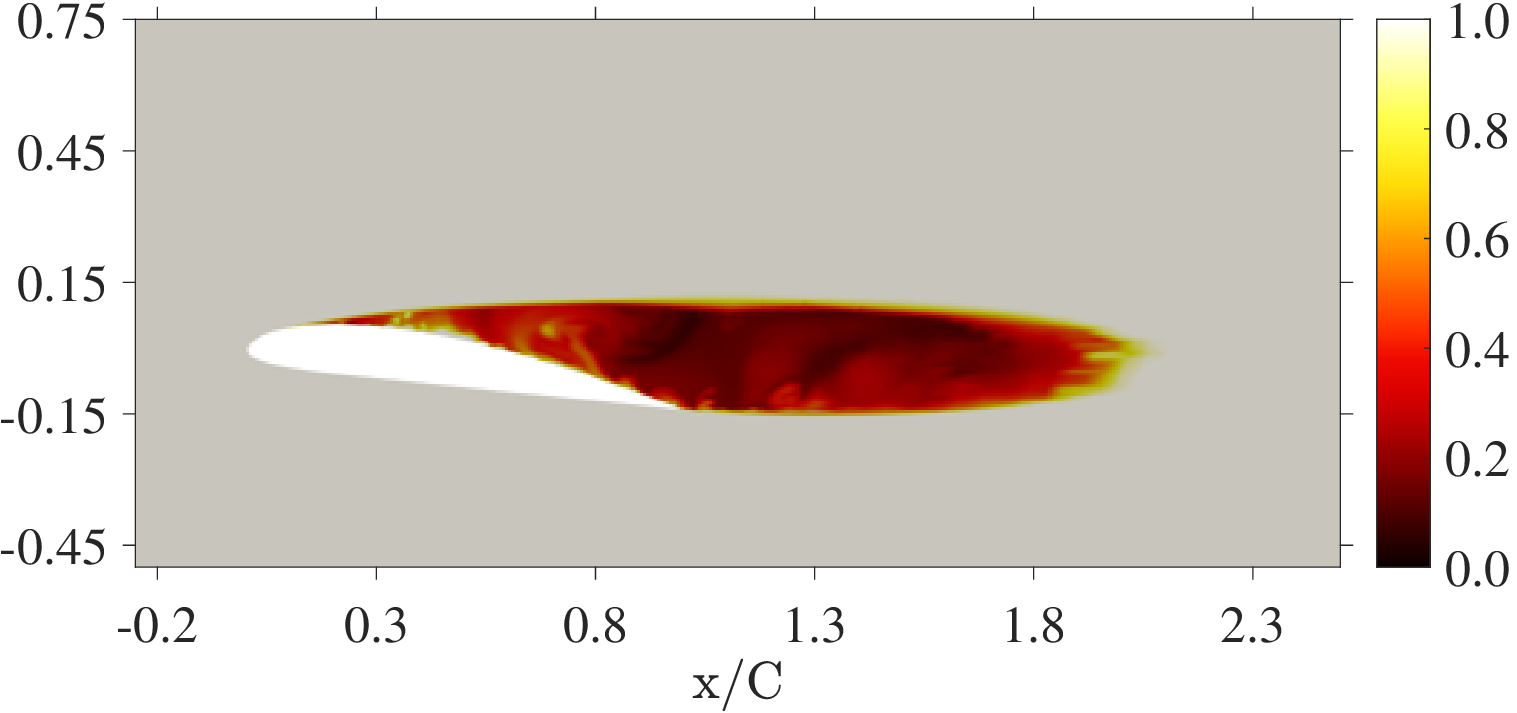} \\
\includegraphics[width=0.475\textwidth]{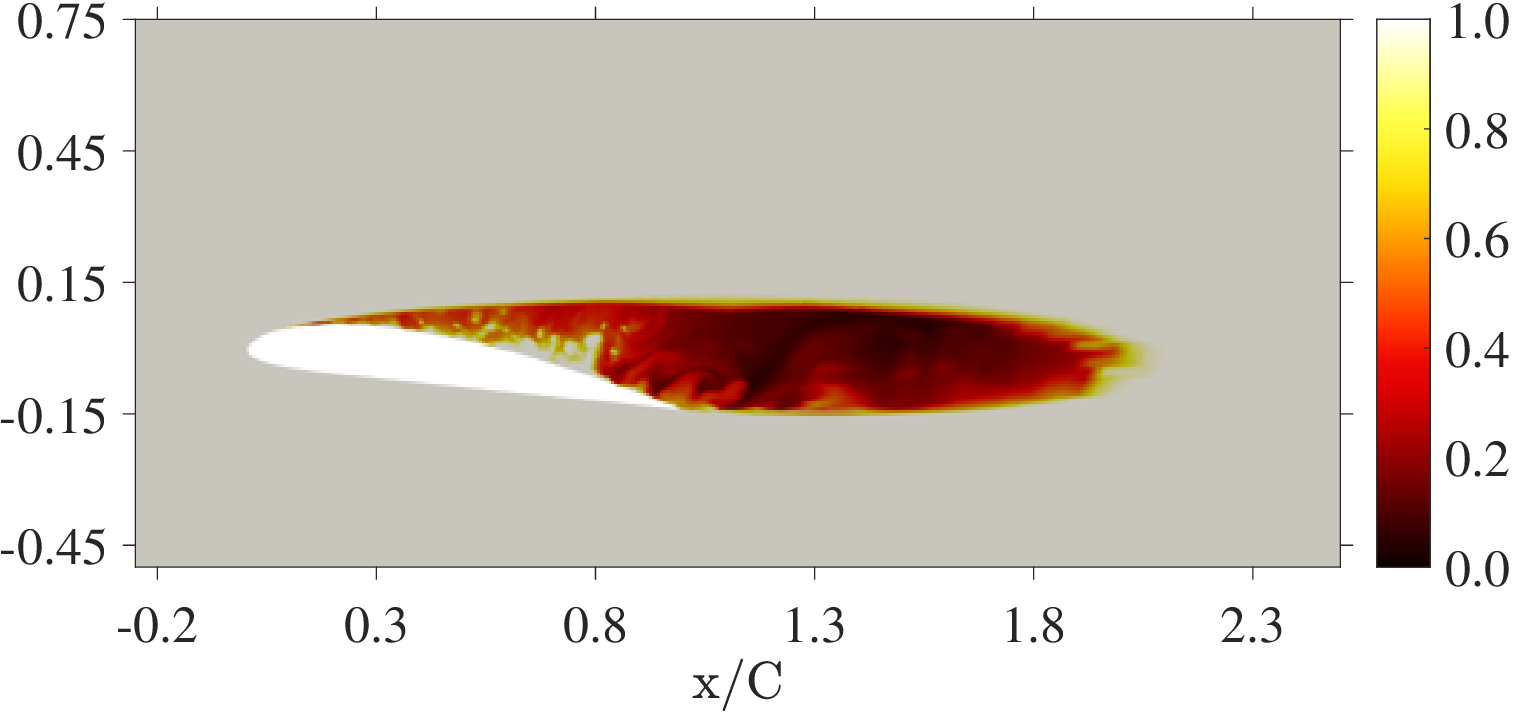}  &

\includegraphics[width=0.475\textwidth]{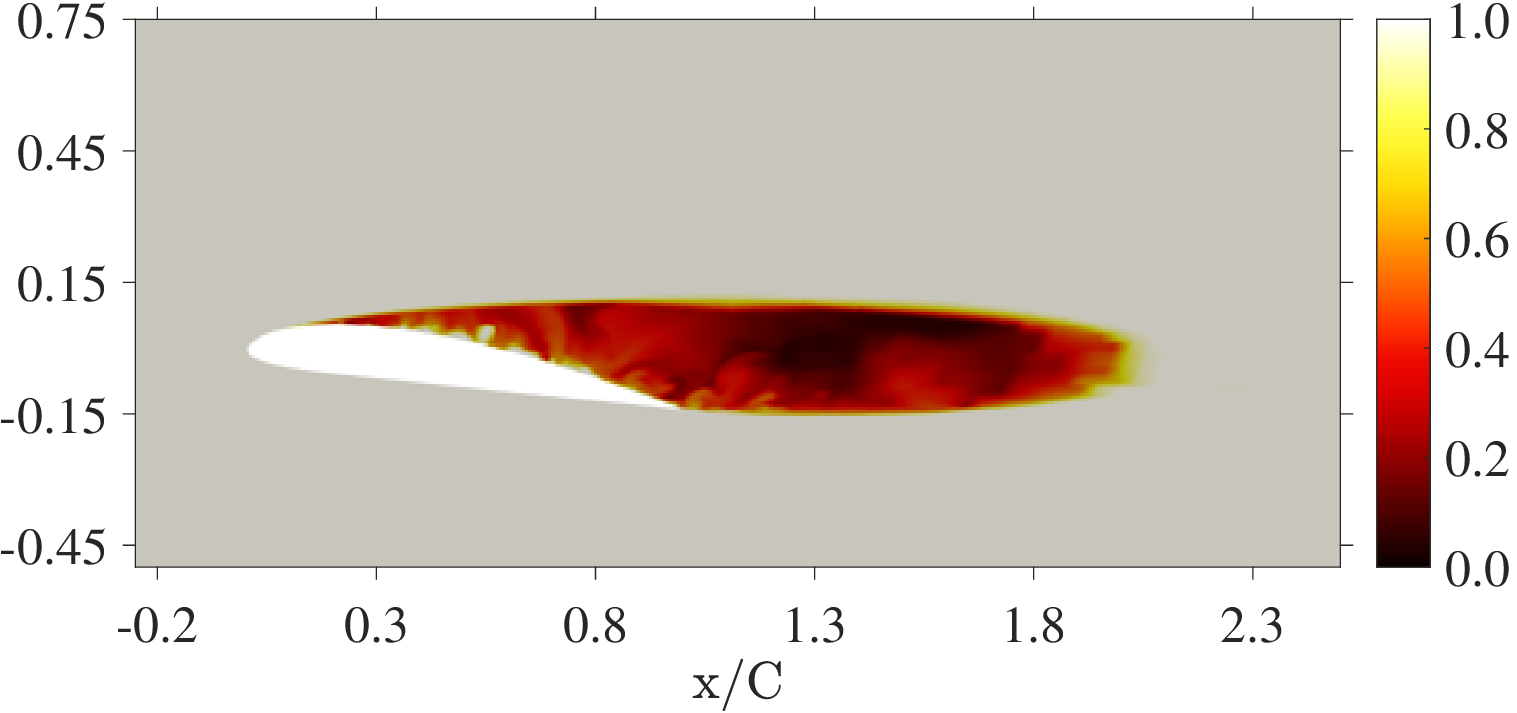} \\

\includegraphics[width=0.475\textwidth]{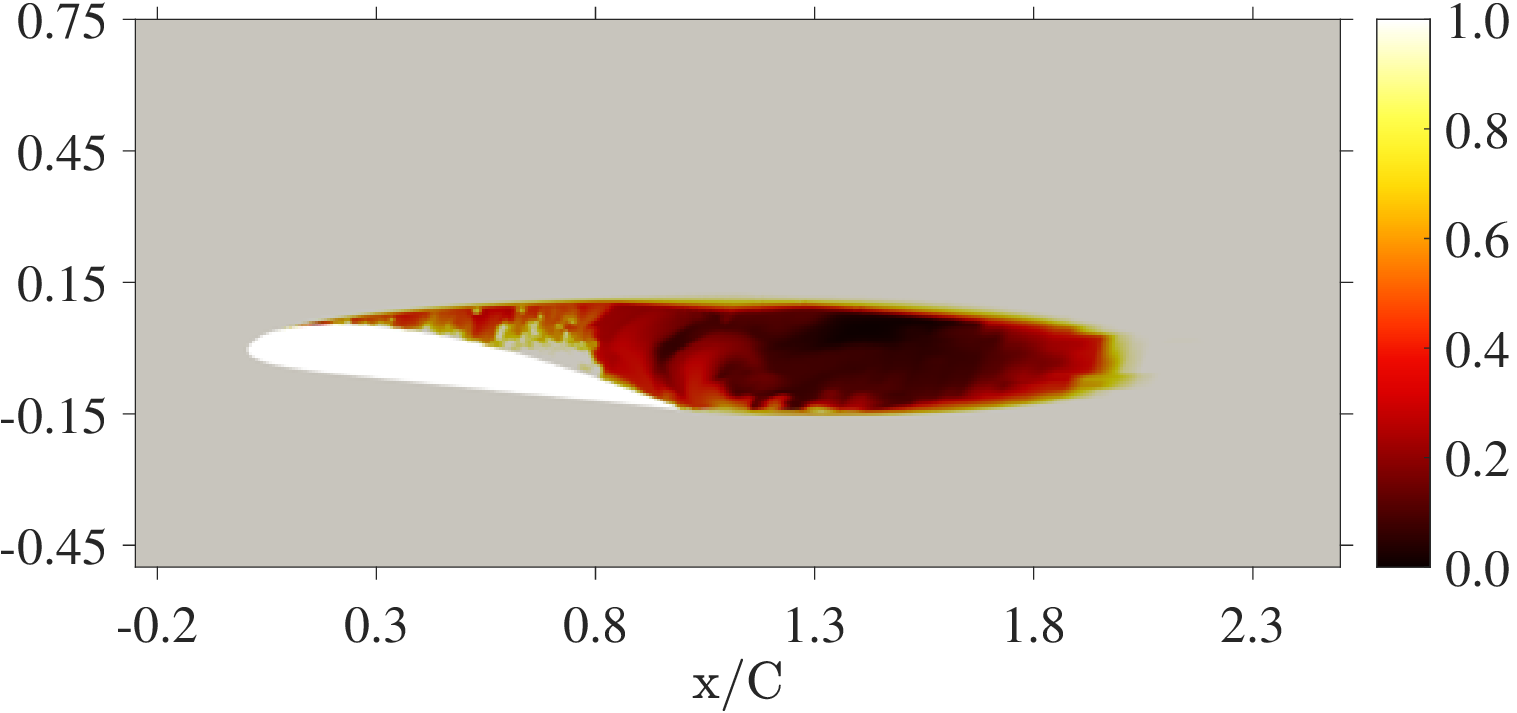} &

\includegraphics[width=0.475\textwidth]{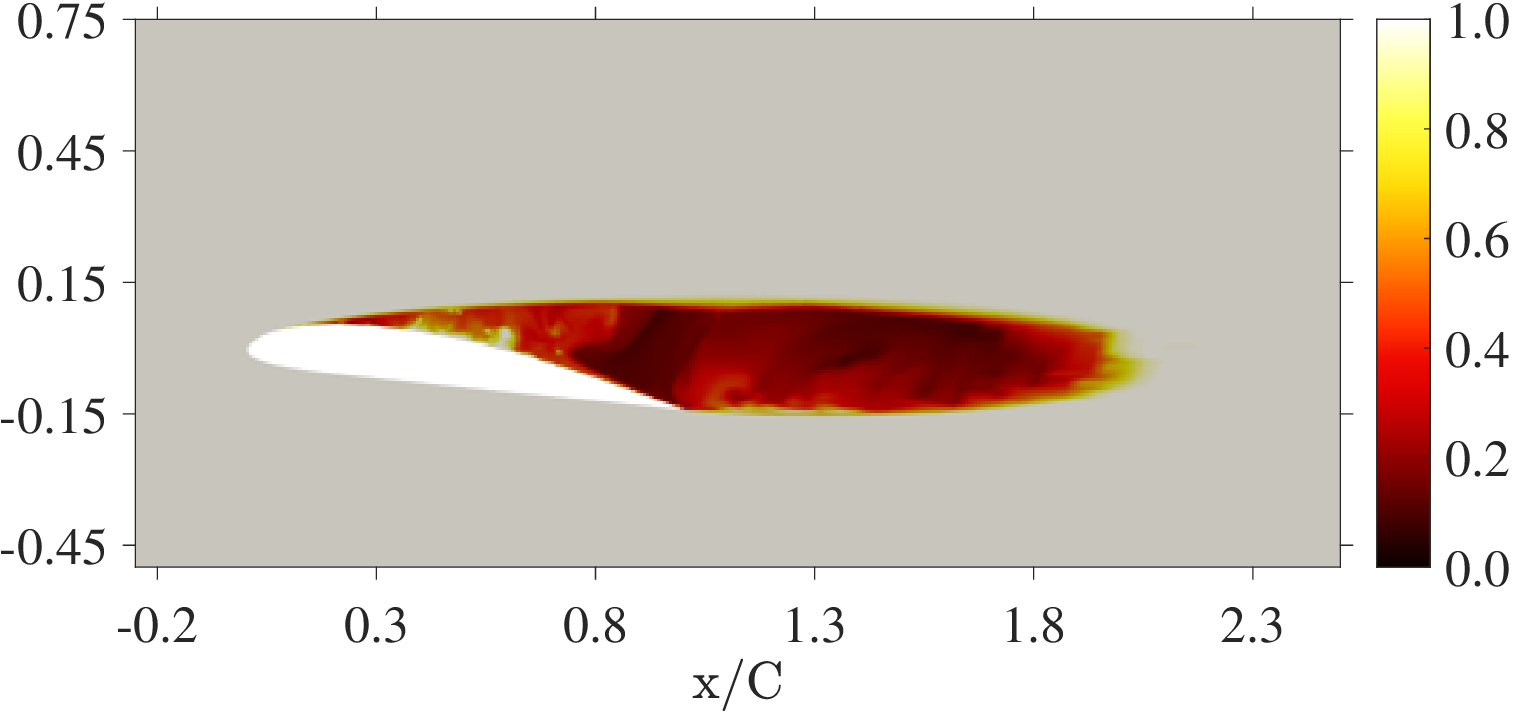} \\
\includegraphics[width=0.475\textwidth]{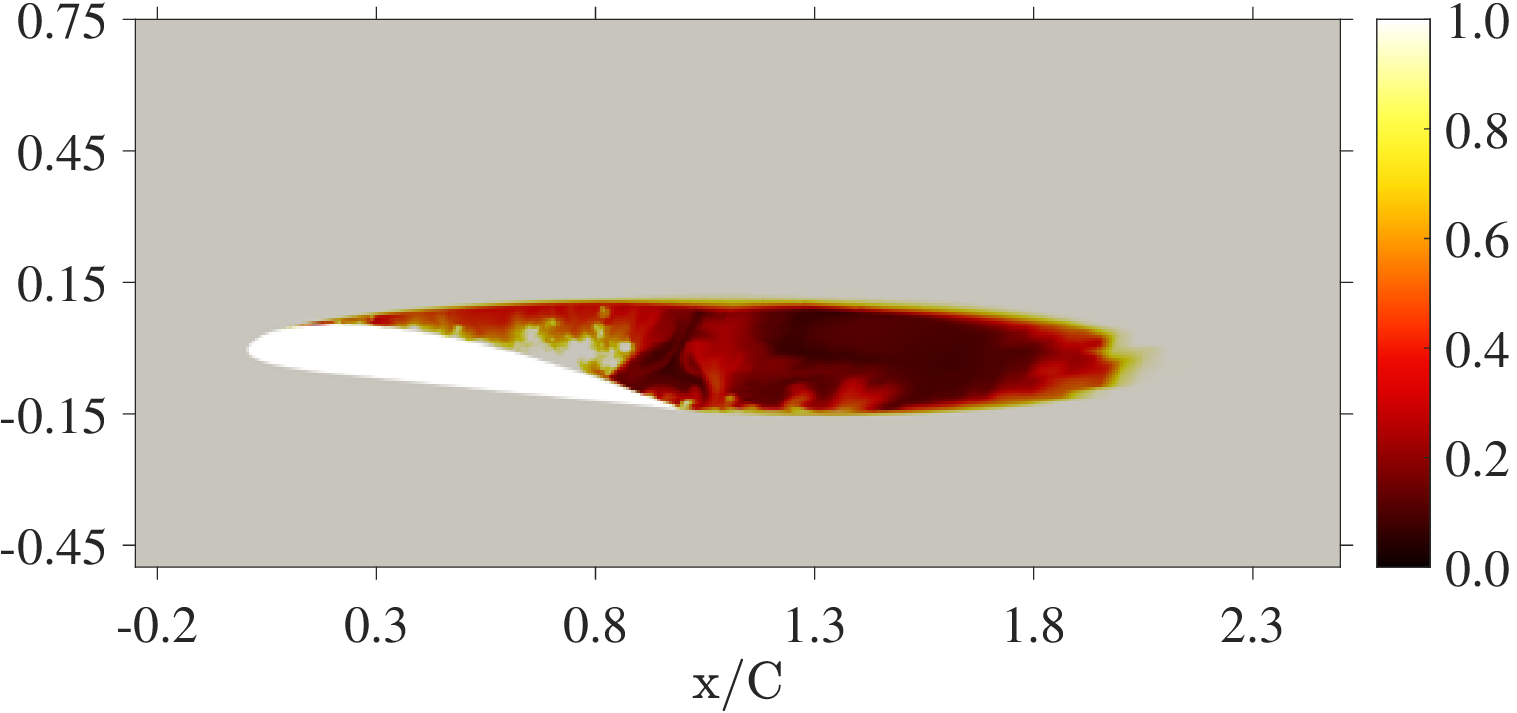}  &

\includegraphics[width=0.475\textwidth]{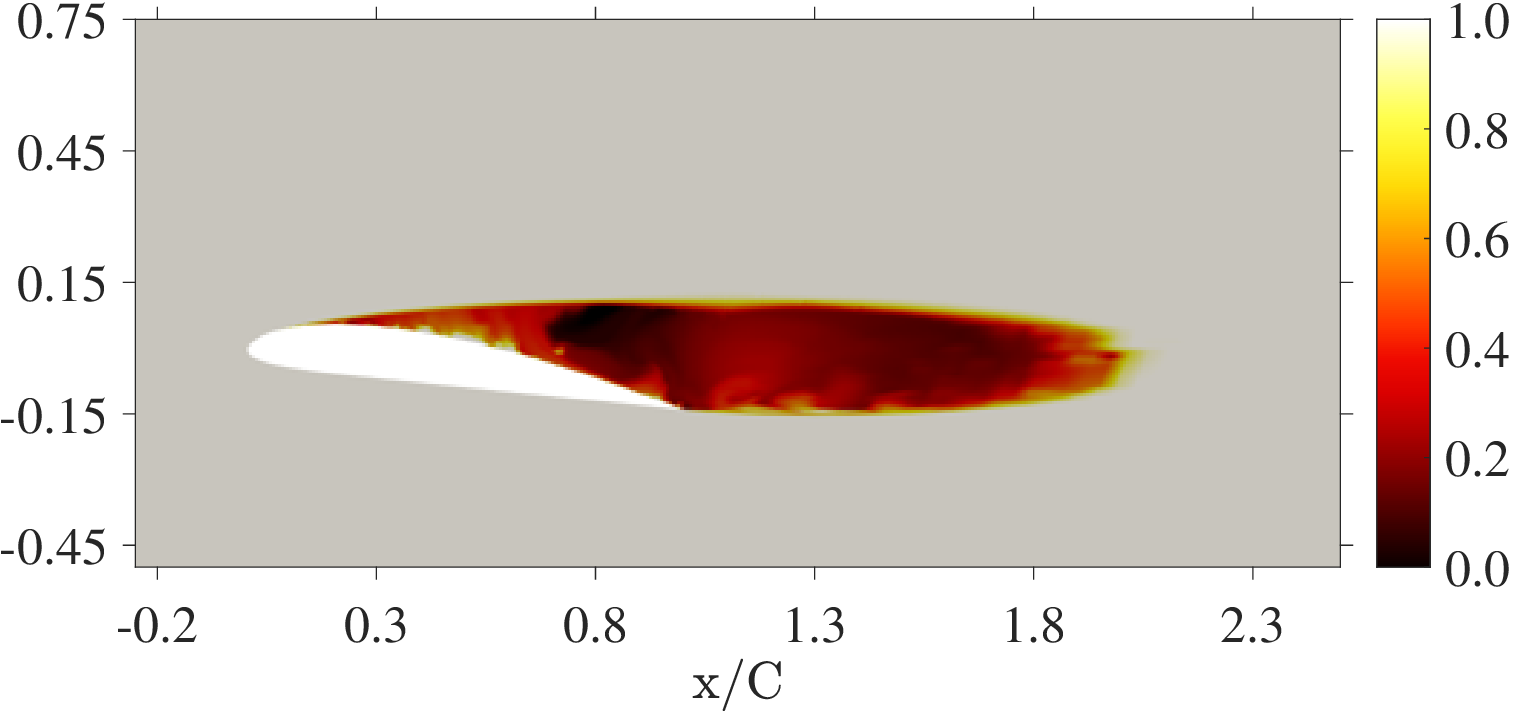} \\
\\
\text{WCA=$0^\circ$} & \text{WCA=$160^\circ$} \\

\end{tabular}
\caption{Instantaneous contours of water volume fraction ($\alpha_{\text{water}}$) at five flow times ($t = 0.80$ to 1.00 s, top to bottom) for $\sigma = 0.4$ for flow around a Clark Y hydrofoil at 8 degrees of angle of attack. Left: $\text{WCA} = 0^\circ$; Right: $\text{WCA} = 160^\circ$. Higher WCA results in closer wall-adhered vapor and reduced interface dynamics.}    \label{fig:alpha_snapshots_sigma04}
\end{figure}

To investigate the effect of wall wettability in a more cavitation-prone regime, flow fields for a cavitation number of $\sigma = 0.4$ were analyzed for WCA values of $0^\circ$ and $160^\circ$. Figure~\ref{fig:alpha_snapshots_sigma04} shows instantaneous contours of the water volume fraction ($\alpha_{\text{water}}$) at five time steps ranging from $t = 0.80$ to 1.00 s. These snapshots illustrate the spatial vapor development and the degree of wall attachment between the cavity and the hydrofoil.

At $\text{WCA} = 0^\circ$, the vapor layer exhibits significant temporal variability. The contours in Fig.~\ref{fig:alpha_snapshots_sigma04} (Left Column) reveal that the vapor structures intermittently detach and reattach to the suction surface, with frequent changes in shape and position. This behavior is indicative of a dynamic shedding process driven by re-entrant jets, where the strong liquid-solid adhesion on the hydrophilic surface (high surface energy) promotes vigorous re-wetting and subsequent re-cavitation. The consistent observation of minimal spacing between the hydrofoil surface and the vapor interface for the hydrophilic case suggests a tendency for the liquid to strongly adhere, even during cavity development, leading to a more localized and oscillating vapor region.

In contrast, the superhydrophobic surface ($160^\circ$) maintains a consistently thin vapor layer that remains closely attached to the wall along the entire chord, as seen in Fig.~\ref{fig:alpha_snapshots_sigma04} (Right Column). The minimal spacing between the hydrofoil surface and vapor interface across all times demonstrates persistent cavity coverage and a pronounced suppression of re-entrant jet effects, which are typically responsible for cavity shedding. This stability is directly linked to the low surface energy and the presence of a trapped air layer (plastron) on superhydrophobic surfaces. The reduced liquid-solid interaction allows the vapor cavity to spread smoothly along the surface with less interference from the surrounding liquid, maintaining a more continuous vapor film.

\begin{figure}[H]
    \centering
    \caption{Time evolution of liquid volume fraction for \( \sigma = 0.4 \) from \( t = 1.5 \) to 3.0 s. 
    (a) WCA = \(0^\circ\): strong temporal fluctuations in vapor structure and wall reattachment. 
    (b) WCA = \(160^\circ\): consistent vapor coverage with minimal changes in layer thickness.}
    \label{fig:volume_fraction_time_series}
    \begin{subfigure}[b]{0.47\linewidth}
        \centering
        \includegraphics[width=\linewidth]{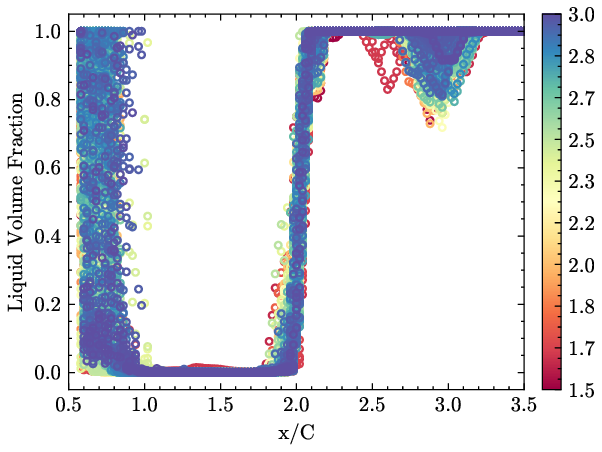}
            \subcaption{WCA=$0^\circ$}
    \end{subfigure}
    \hfill
    \begin{subfigure}[b]{0.47\linewidth}
        \centering
        \includegraphics[width=\linewidth]{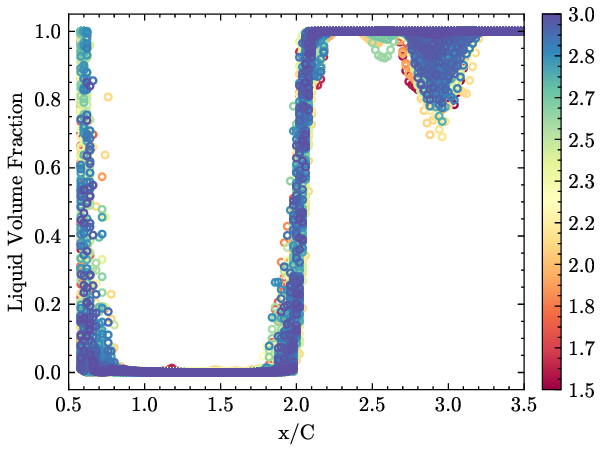}
            \subcaption{WCA=$160^\circ$}
    \end{subfigure}

\end{figure}

These trends are further supported by the temporal evolution of the liquid volume fraction in Figure~\ref{fig:volume_fraction_time_series}, spanning the time interval $t = 1.5$ to 3.0 s. For $\text{WCA} = 0^\circ$, significant temporal fluctuations in volume fraction are observed between $x/C = 0.5$ and 1.0 (Fig.~\ref{fig:volume_fraction_time_series}(a)), indicating dynamic re-filling of the vapor region by the surrounding liquid. This oscillatory behavior directly reflects unsteady cavity shedding and partial flow reattachment, leading to substantial changes in the effective hydrofoil shape. For $160^\circ$, the liquid fraction remains relatively stable across the same spatial window (Fig.~\ref{fig:volume_fraction_time_series}(b)), highlighting the sustained presence of vapor and significantly reduced interface instability. This consistency underscores the ability of superhydrophobic surfaces to maintain a stable vapor-liquid interface, minimizing temporal variations in the flow.

\begin{figure}[H]
    \centering
    \includegraphics[width=0.6\textwidth]{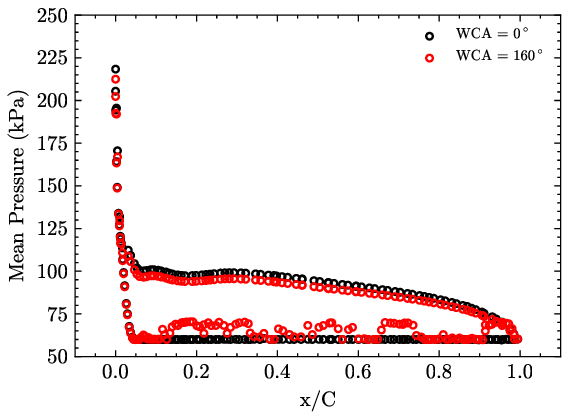}
\caption{Mean pressure distribution along the chord for WCA = $0^\circ$ and $160^\circ$ at $\sigma = 0.4$ for flow around a Clark Y hydrofoil at 8 degrees of angle of attack. The hydrophilic surface exhibits larger pressure fluctuation ranges due to unsteady cavity shedding and re-entrant jet activity.}
    \label{fig:mean_pressure_sigma04}
\end{figure}

The corresponding mean pressure distribution plotted in Fig.~\ref{fig:mean_pressure_sigma04} confirms the presence of stronger unsteadiness and broader pressure fluctuations in the hydrophilic case. While both cases show low-pressure zones near the leading edge, the pressure field for $160^\circ$ is more uniform along the chord, with a sustained low-pressure region extending further downstream, characteristic of a stable attached cavity. Notably, pressure fluctuations are more intense and span a wider range in the lower-WCA case, consistent with periodic cavity collapse and redevelopment, which generate strong pressure waves. This distinction in pressure behavior further illustrates the stabilizing influence of superhydrophobic surfaces on the attached cavitation morphology in low-$\sigma$ cavitating flows, contrasting with the highly unsteady behavior observed on hydrophilic surfaces. The lower surface energy of the superhydrophobic surface effectively stabilizes the vapor layer, leading to more consistent hydrodynamic loading.

\subsection{Effect of Wall Wettability at High Cavitation Number ($\sigma = 1.6$)}
\begin{figure}[H]
\centering
\begin{tabular}{cc}
\includegraphics[width=0.35\textwidth]{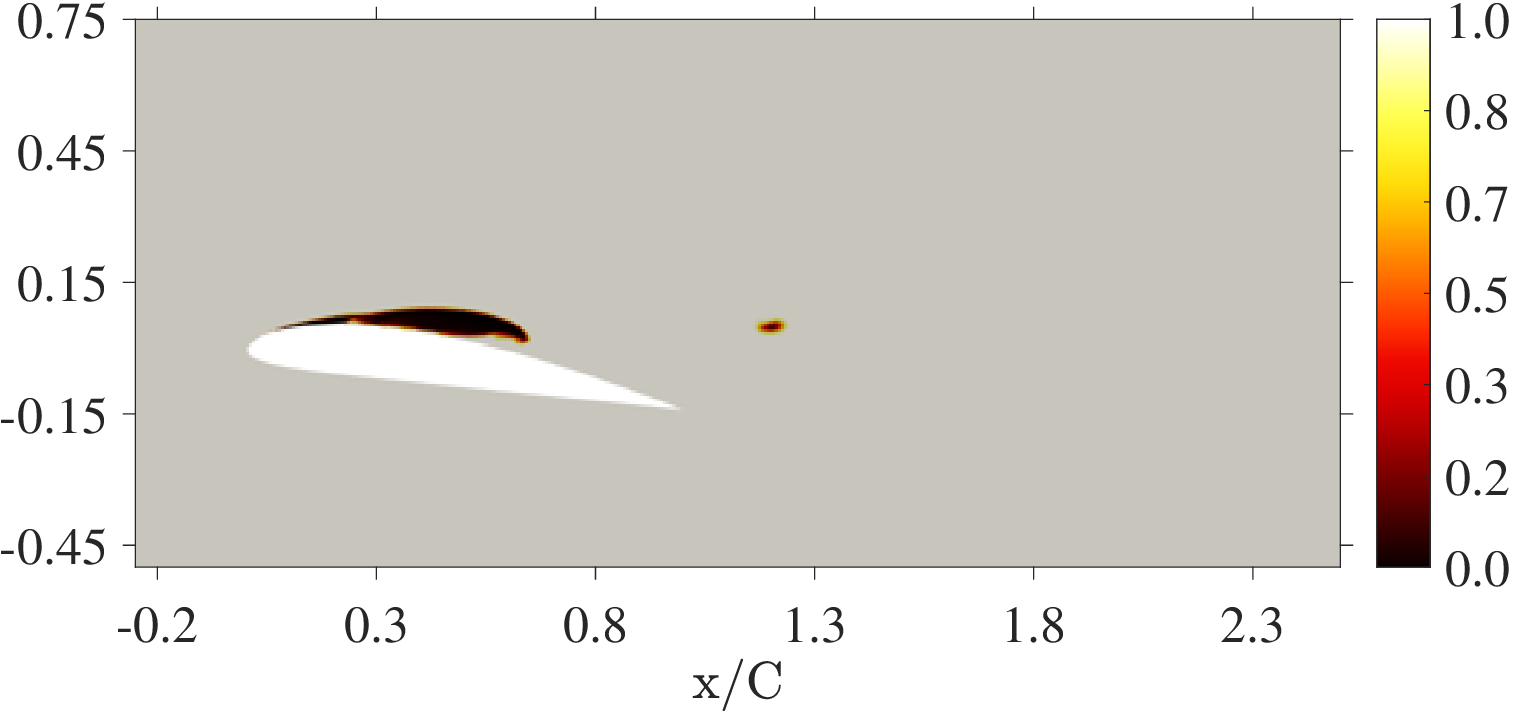} & \includegraphics[width=0.35\textwidth]{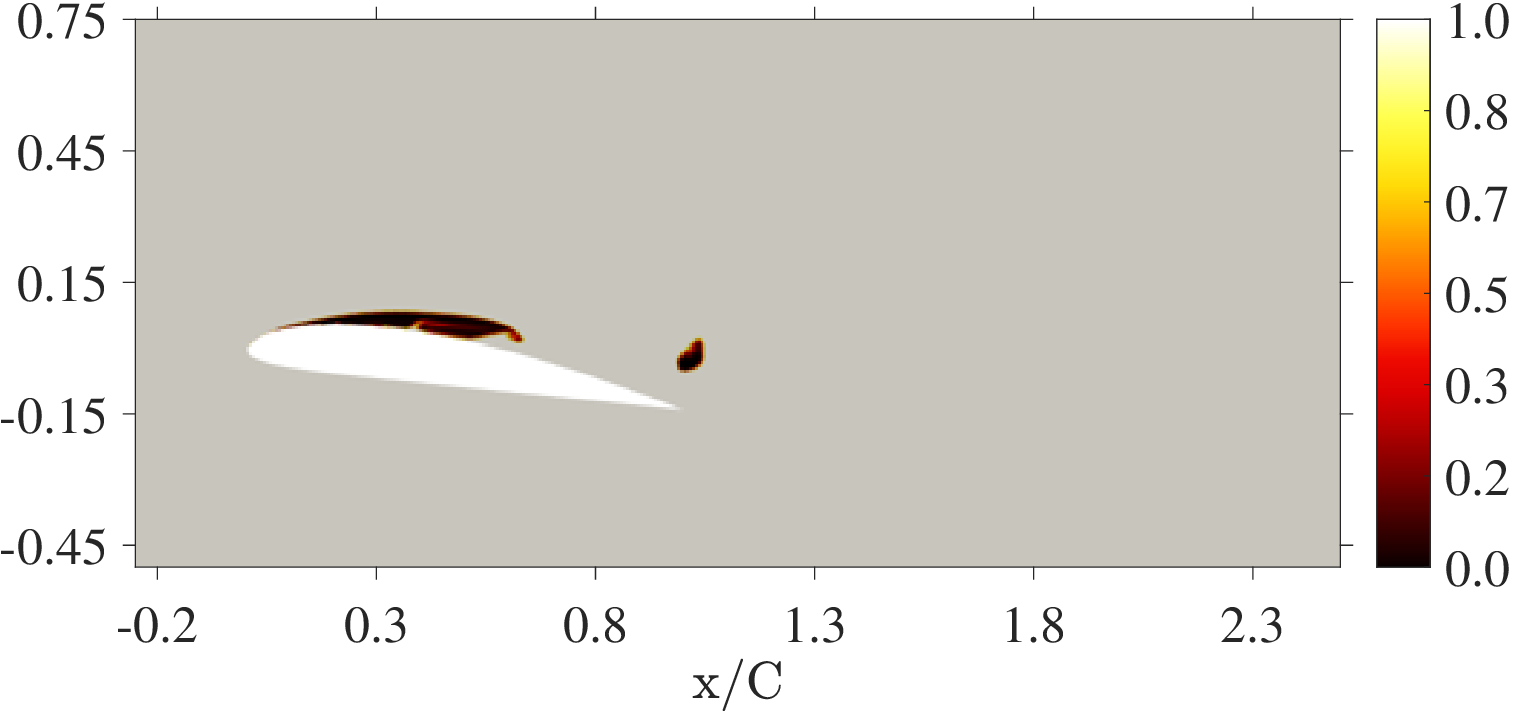} \\
\includegraphics[width=0.35\textwidth]{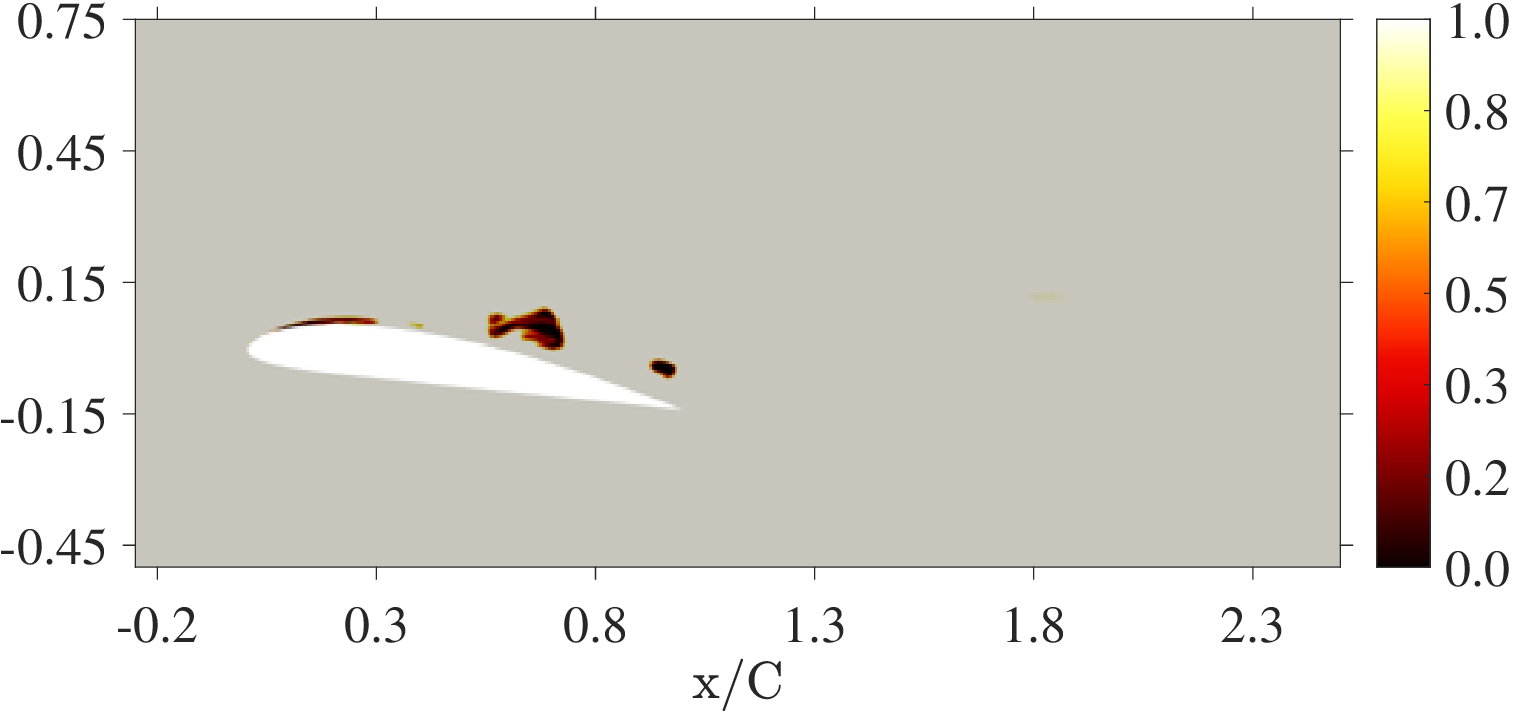} & \includegraphics[width=0.35\textwidth]{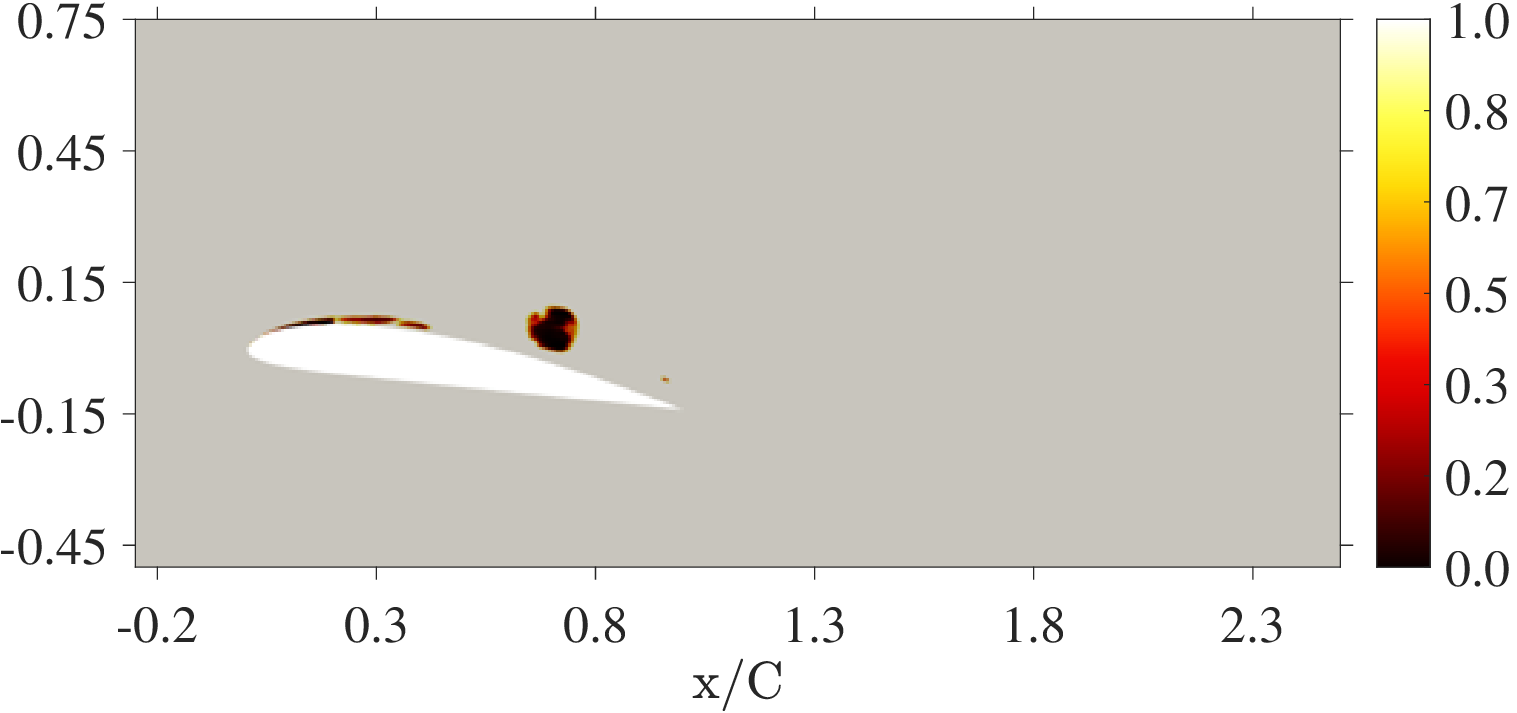} \\
\includegraphics[width=0.35\textwidth]{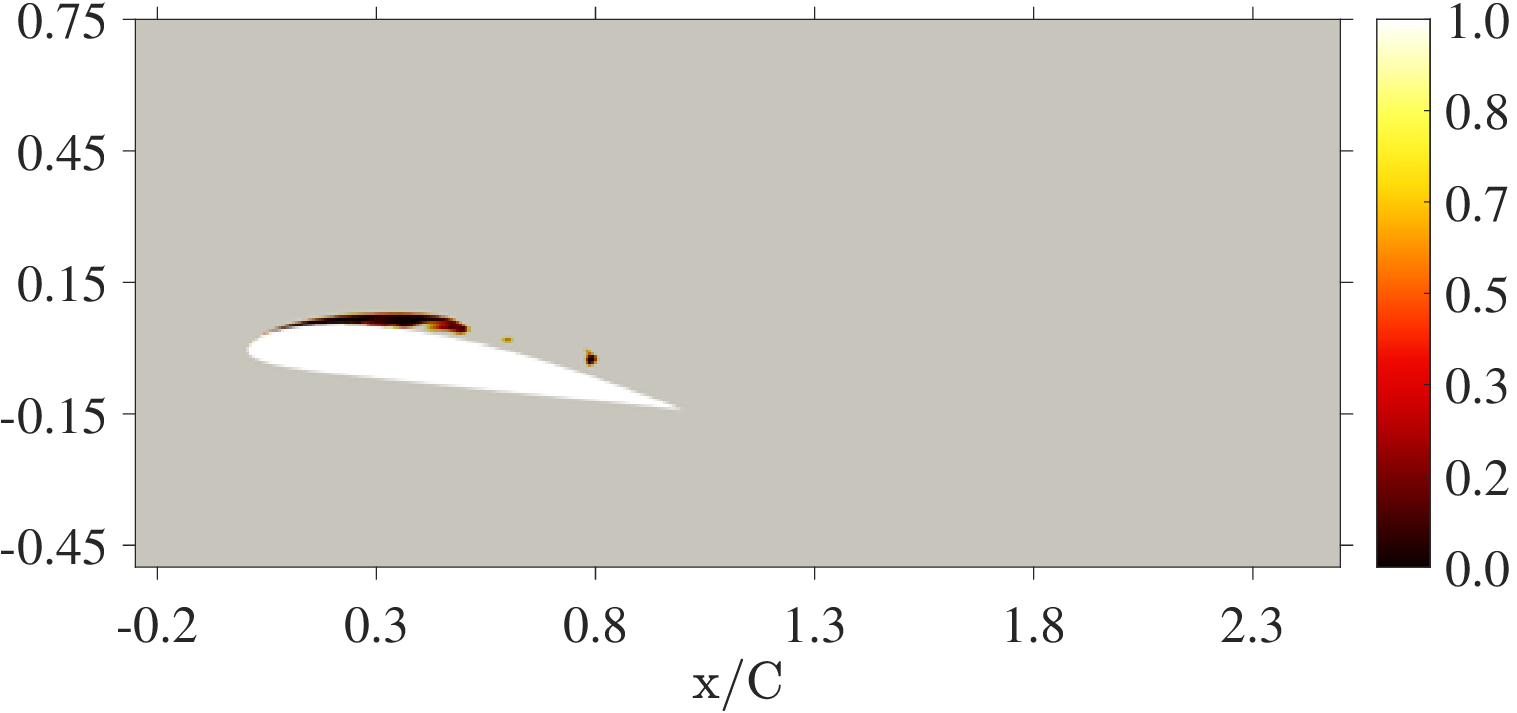} & \includegraphics[width=0.35\textwidth]{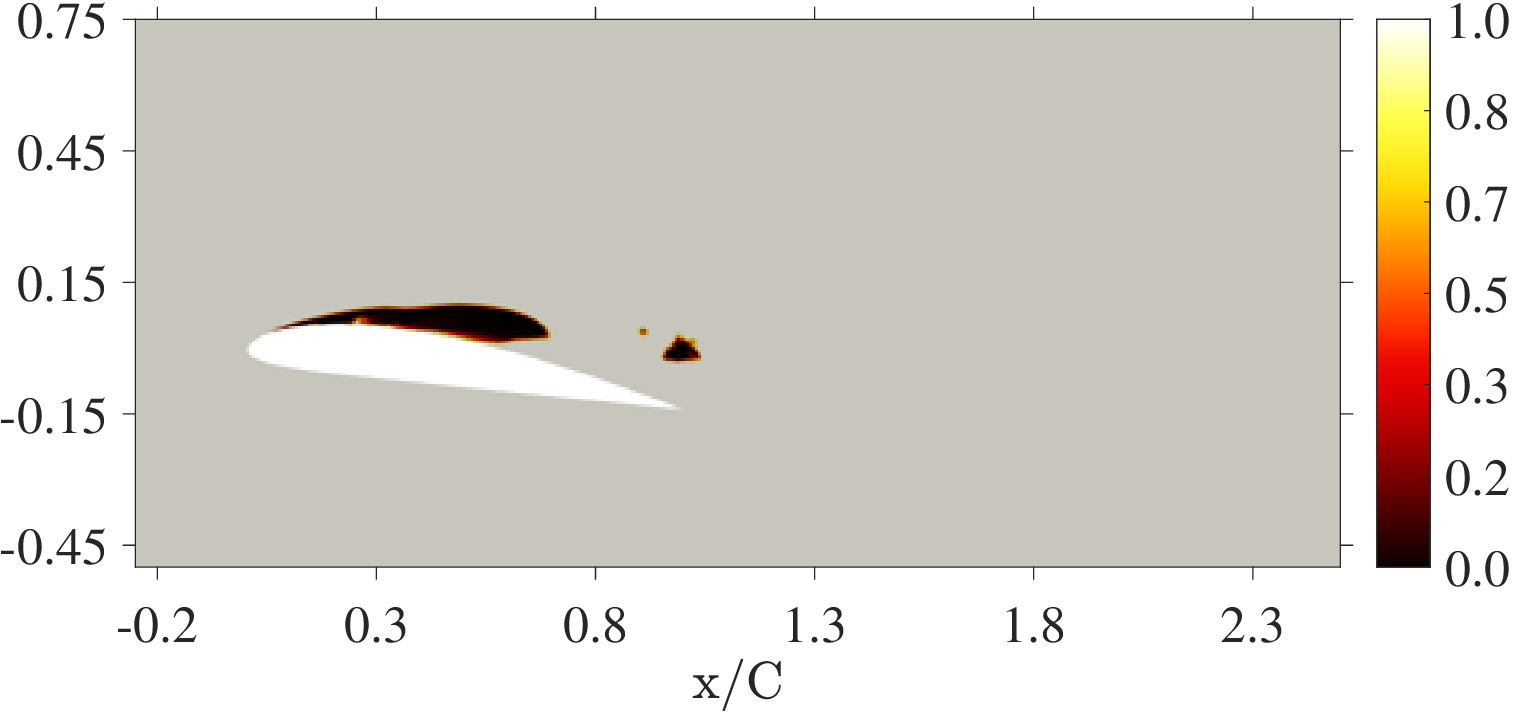} \\
\includegraphics[width=0.35\textwidth]{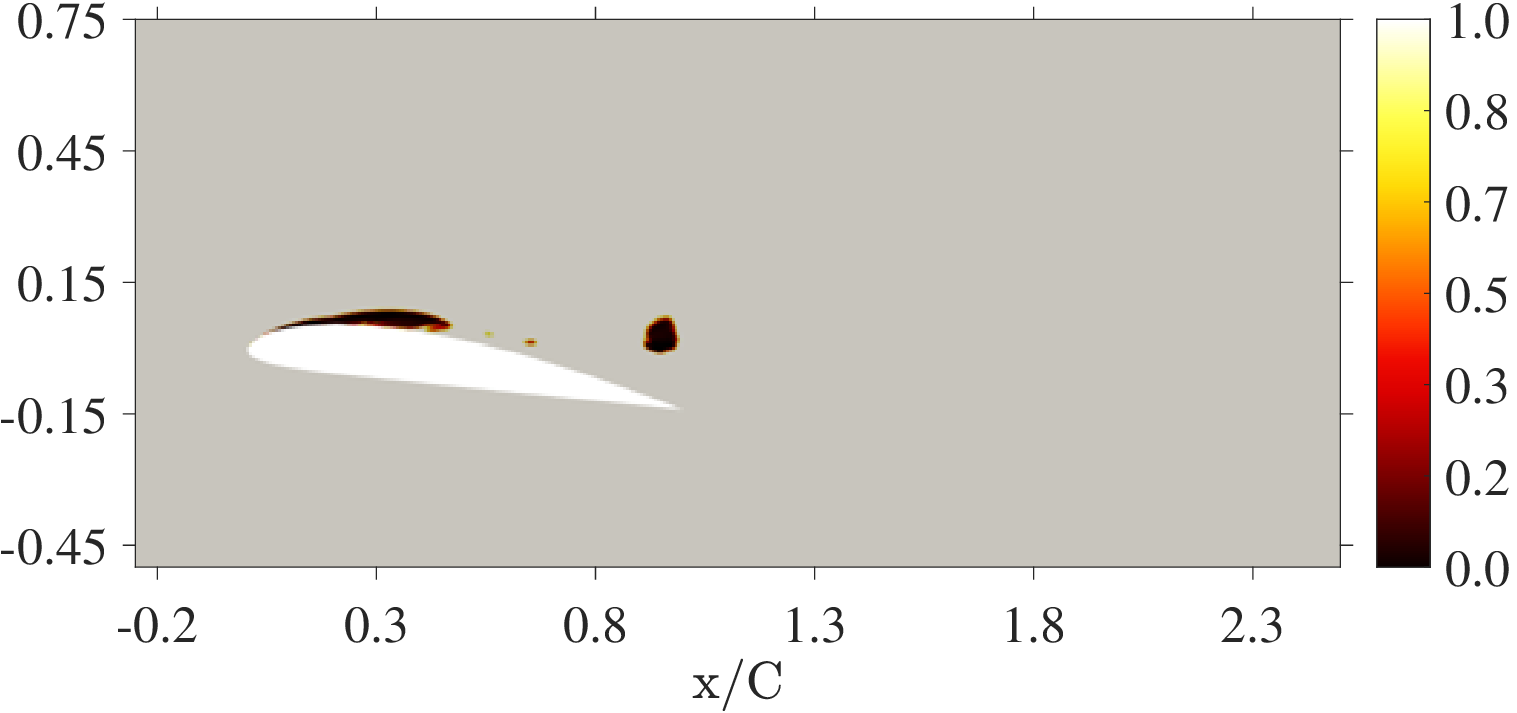} & \includegraphics[width=0.35\textwidth]{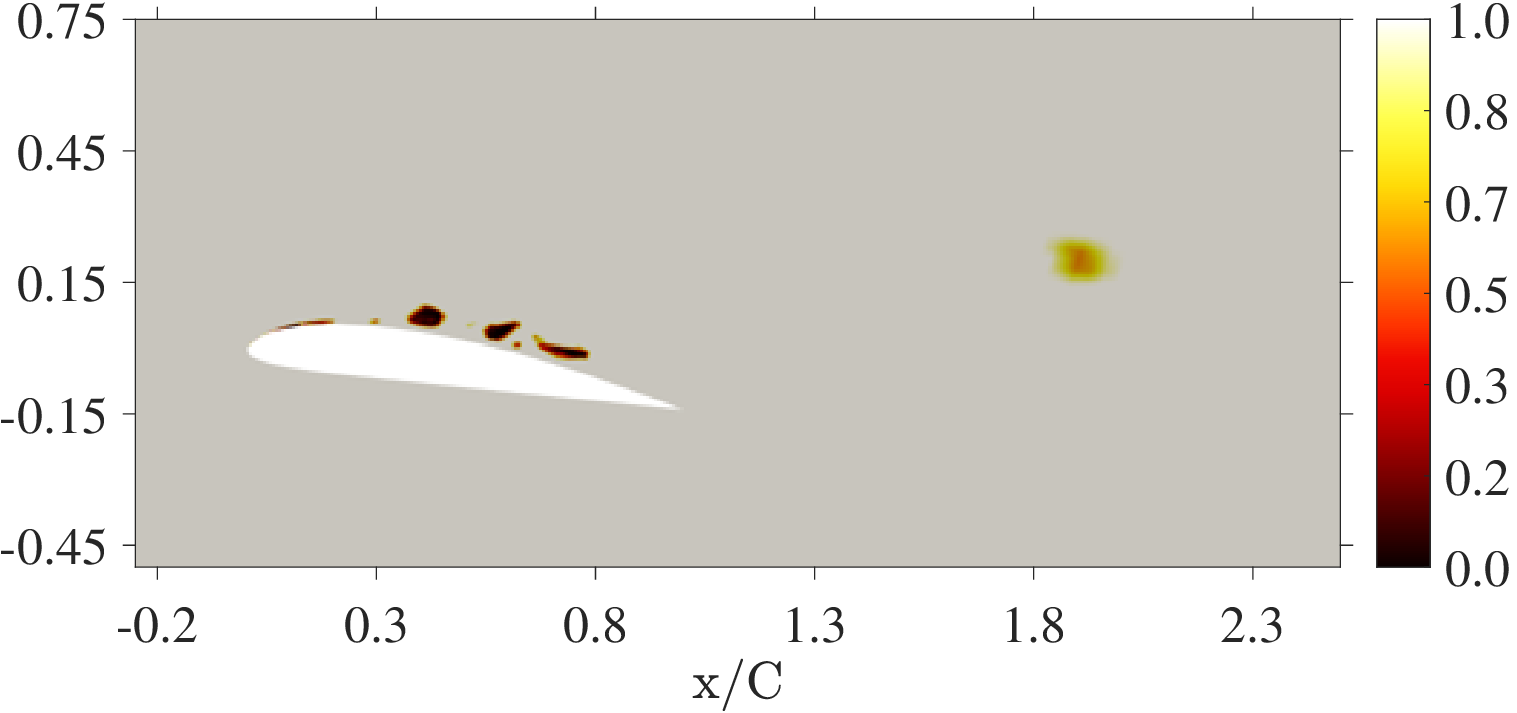} \\
\includegraphics[width=0.35\textwidth]{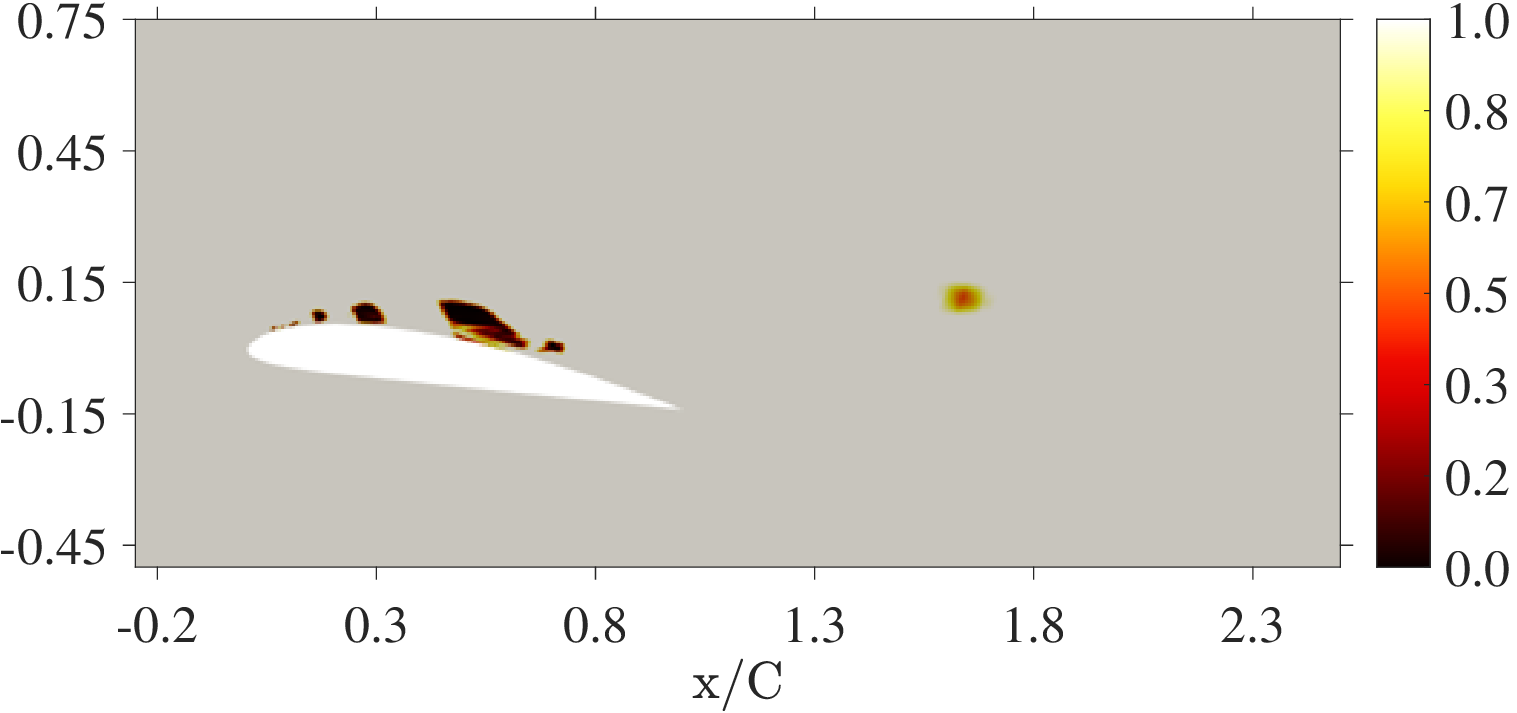} & \includegraphics[width=0.35\textwidth]{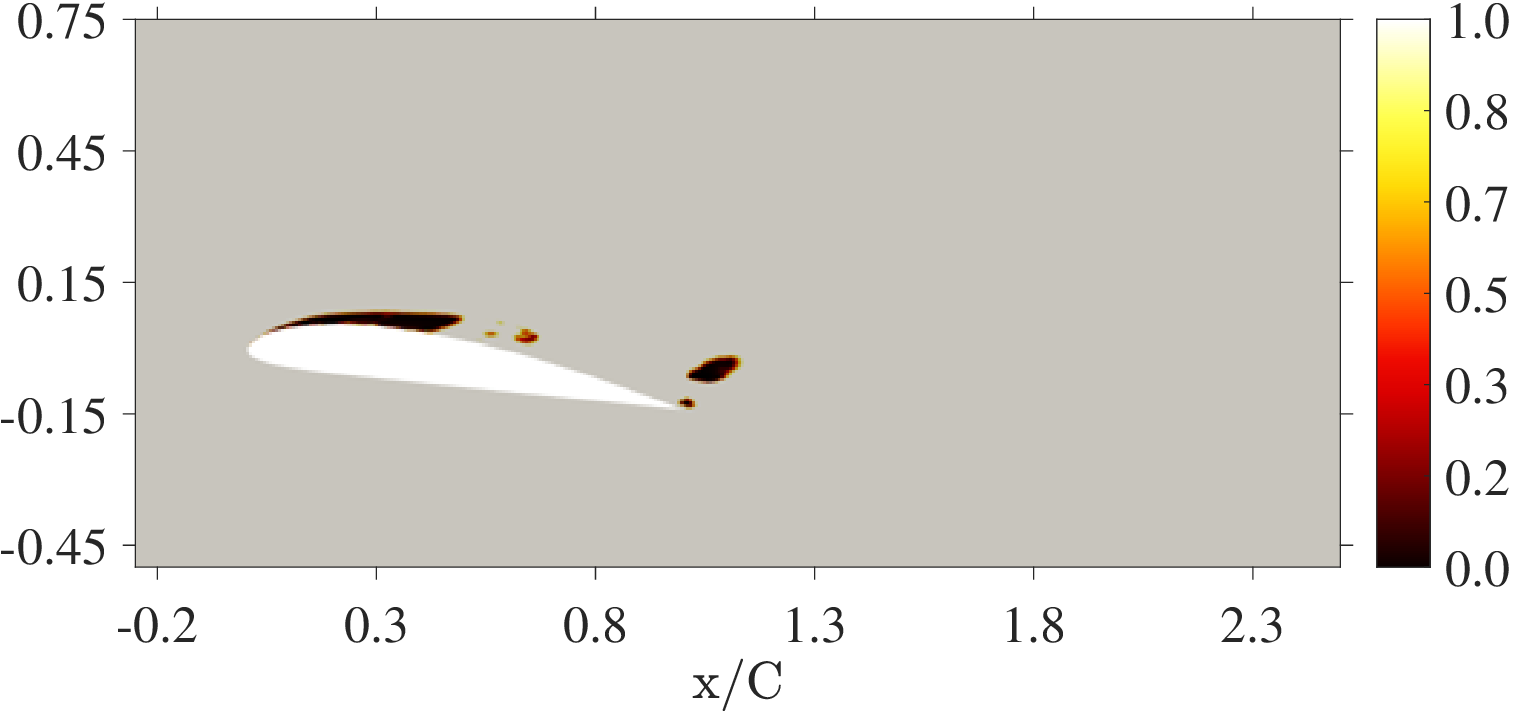} \\
\includegraphics[width=0.35\textwidth]{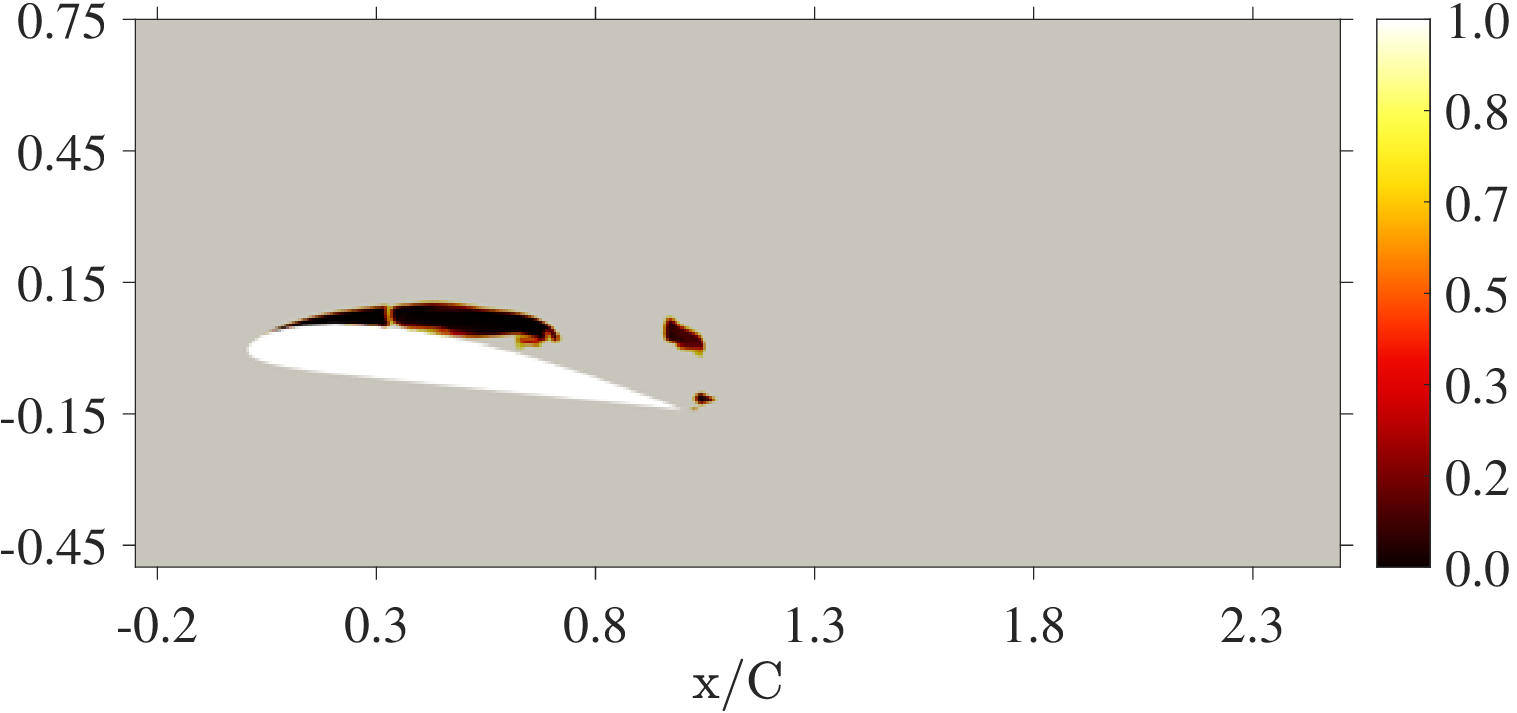} & \includegraphics[width=0.35\textwidth]{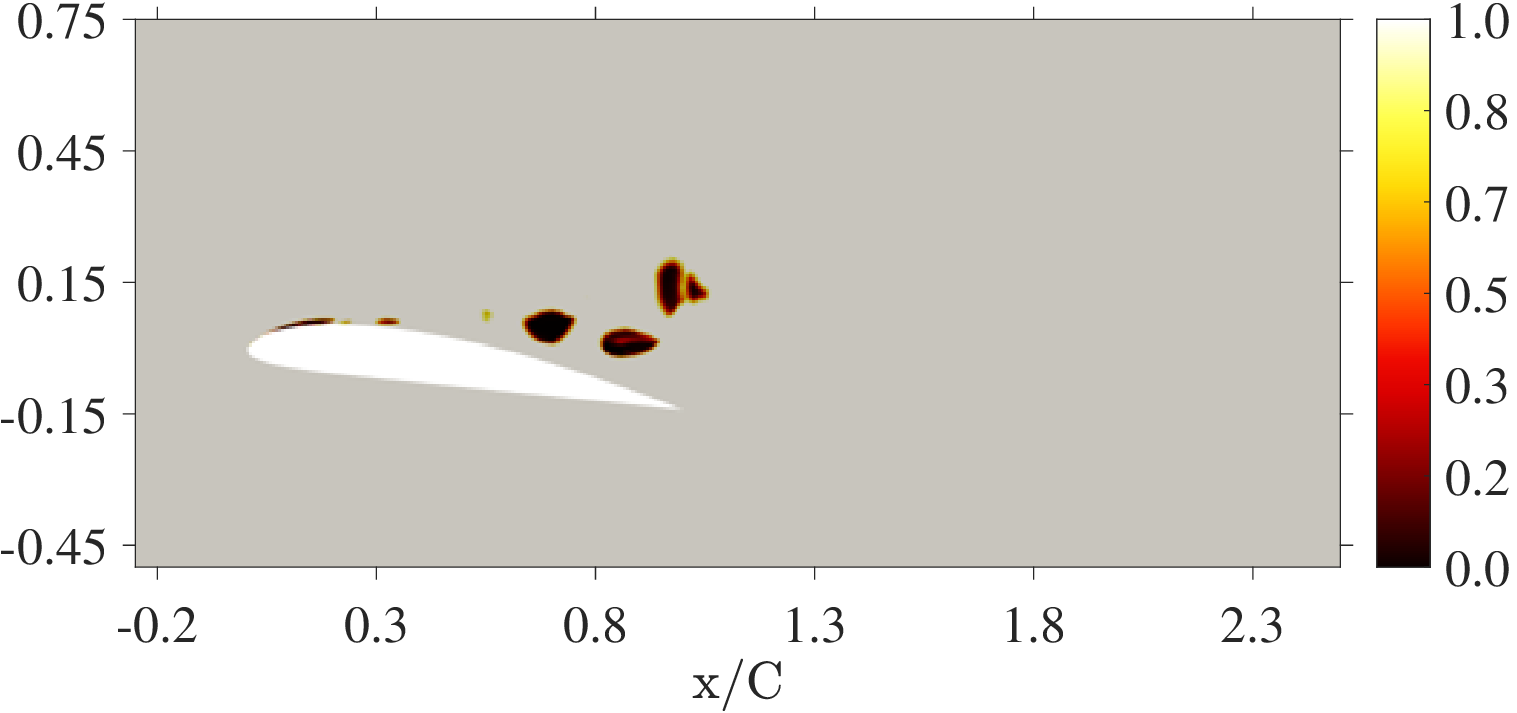} \\
\includegraphics[width=0.35\textwidth]{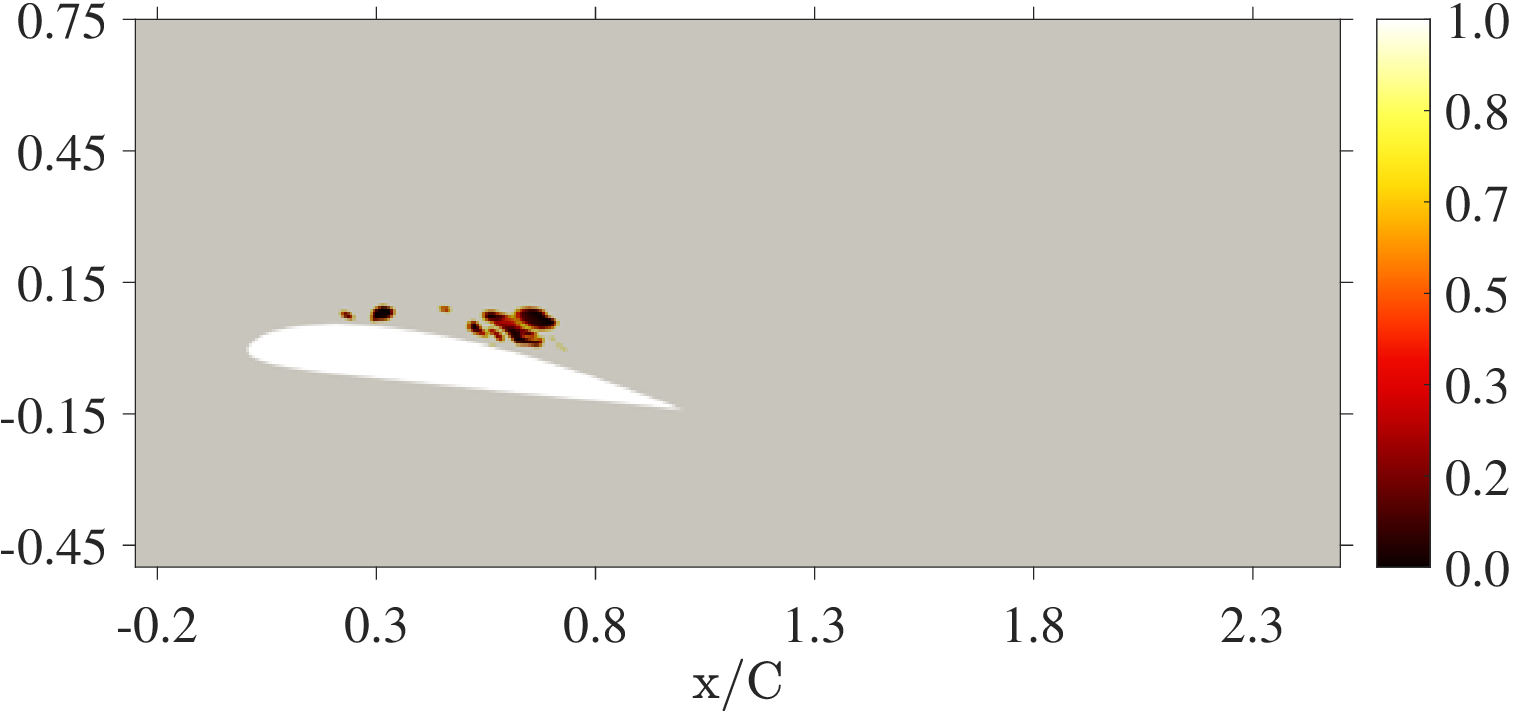} & \includegraphics[width=0.35\textwidth]{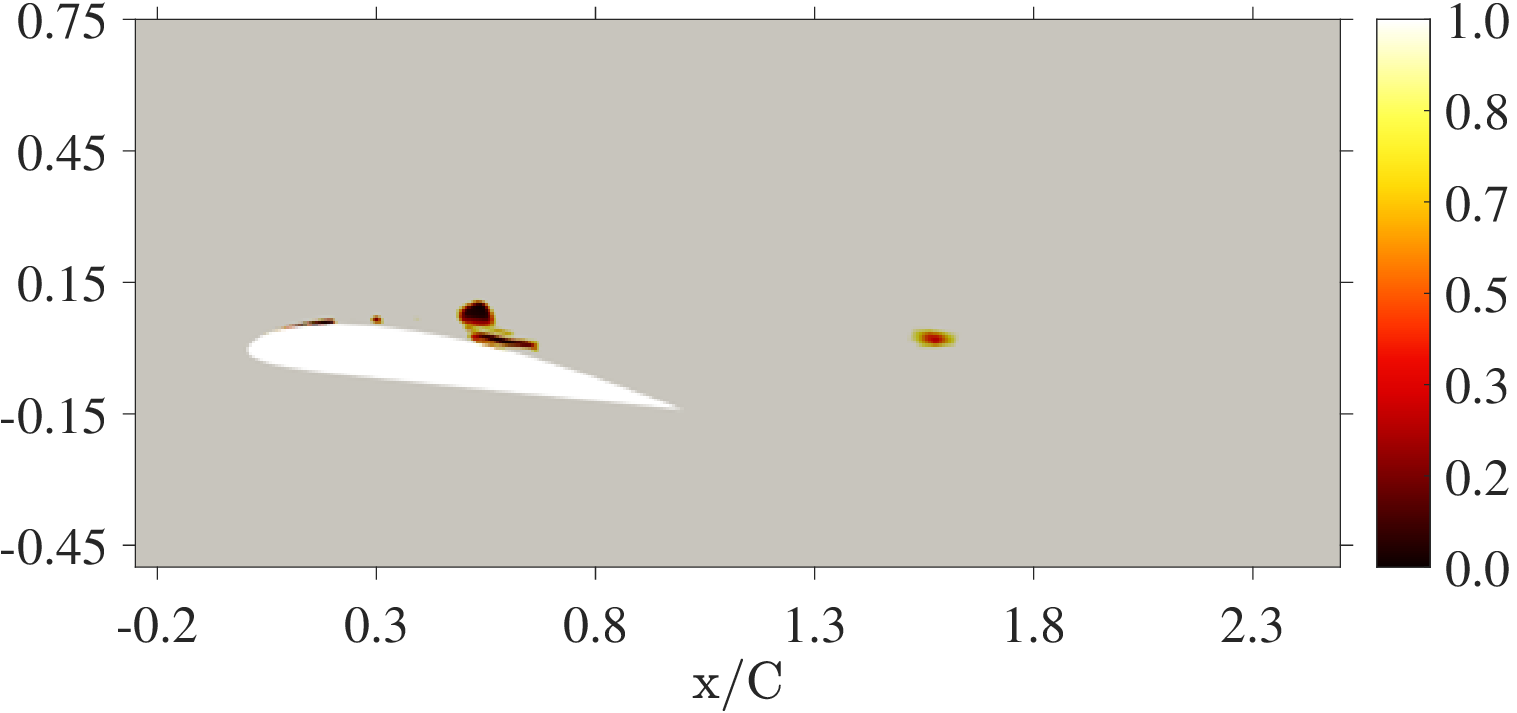} \\
\includegraphics[width=0.35\textwidth]{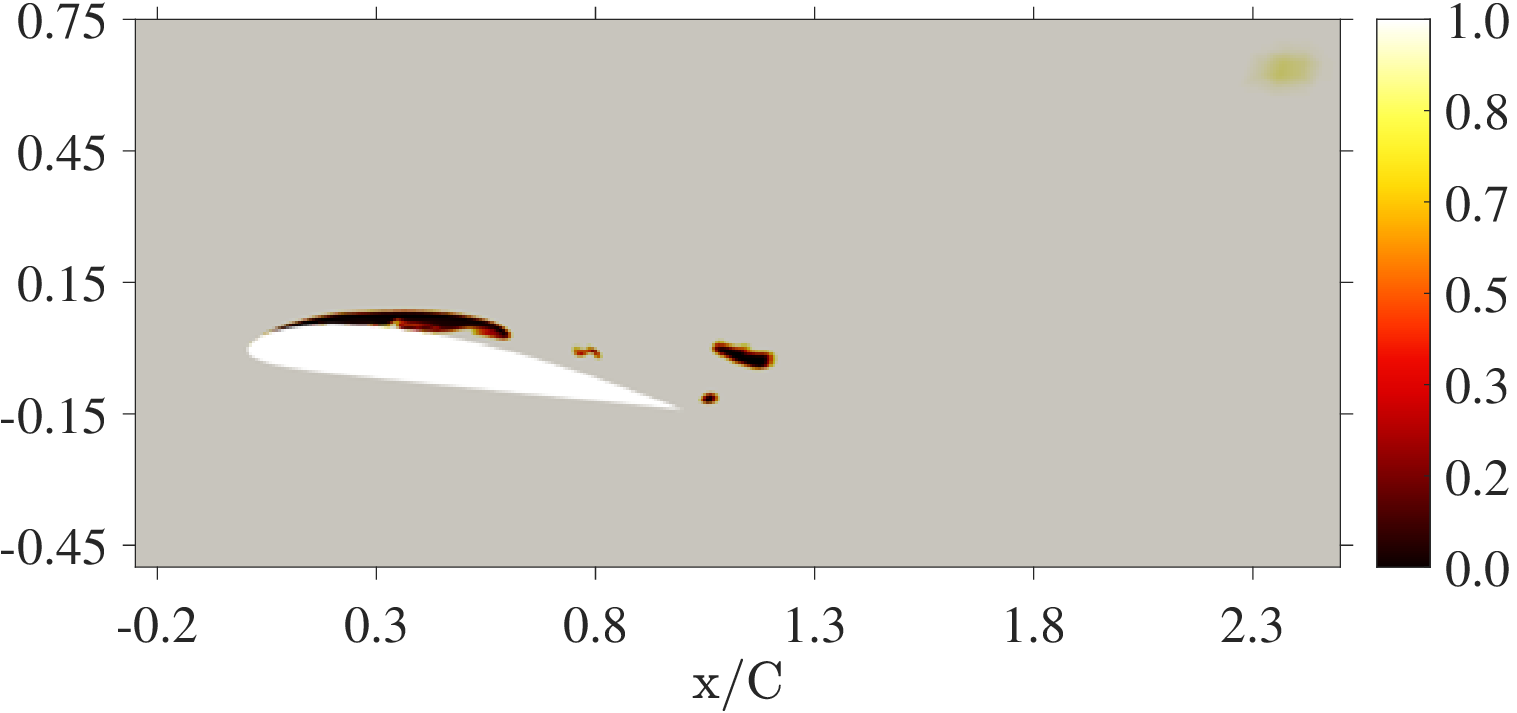} & \includegraphics[width=0.35\textwidth]{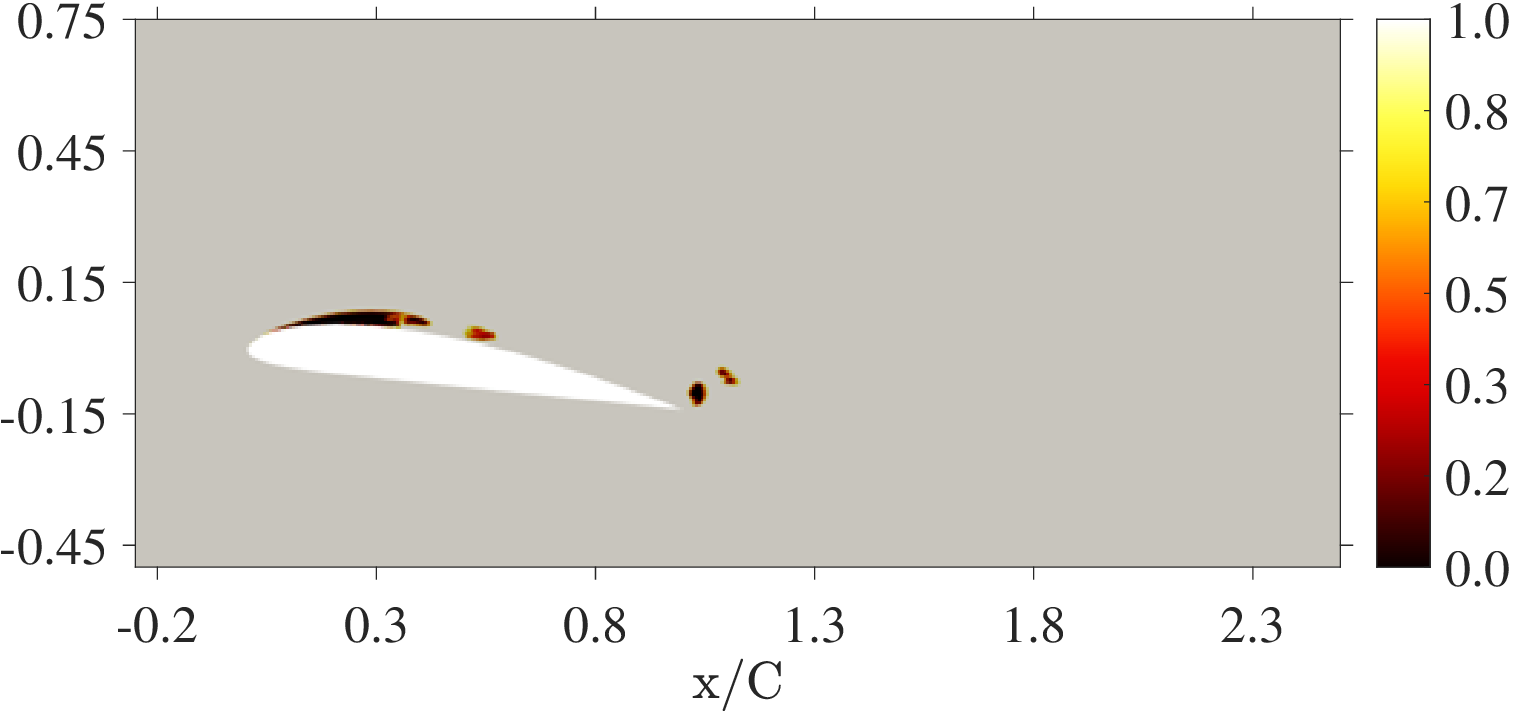}\\
\text{WCA=$0^\circ$} & \text{WCA=$160^\circ$} \\

\end{tabular}
\caption{Time-resolved snapshots of liquid volume fraction ($\alpha_{\text{water}}$) for $\sigma = 1.6$ over the interval $t = 1.00$ to $1.25$ s (10 time steps) for flow around a Clark Y hydrofoil at 8 degrees of angle of attack. Left column: WCA = $0^\circ$; Right column: WCA = $160^\circ$. The superhydrophobic surface leads to smaller, detached vapor bubbles that remain stable near the leading edge, whereas the hydrophilic surface produces elongated, partially attached vapor structures with larger volume and stronger temporal variation.}
\label{fig:lvf_sigma16}
\end{figure}

Figure~\ref{fig:lvf_sigma16} presents time-resolved snapshots of liquid volume fraction ($\alpha_{\text{water}}$) from $t = 1.00$ to 1.25 s, highlighting the distinct cavitation responses on superhydrophilic ($\text{WCA} = 0^{\circ}$) and superhydrophobic ($\text{WCA} = 160^{\circ}$) surfaces.

On the hydrophilic surface (left column, Fig.~\ref{fig:lvf_sigma16}), vapor cavities form intermittently and demonstrate strong adherence to the surface over extended durations. These structures are elongated and maintain relatively stable attachment to the wall, reflecting the high surface energy and strong liquid-solid adhesion that promotes vapor-wall contact. This adherent behavior creates smooth pressure distributions along the surface, as the vapor structures remain consistently attached and do not generate significant pressure oscillations through frequent detachment-reattachment cycles.

The superhydrophobic case (right column, Fig.~\ref{fig:lvf_sigma16}), in stark contrast, displays consistently smaller and predominantly detached bubbles. These vapor structures exhibit minimal surface contact and frequently detach and reattach near the leading edge, creating dynamic bubble behavior. This unstable attachment-detachment process is a direct consequence of the extremely low surface energy of the superhydrophobic coating, which promotes reduced liquid-solid contact while simultaneously preventing stable vapor adherence. The absence of a strong liquid-solid attraction leads to continuous bubble mobility and frequent detachment events, generating pressure fluctuations at the wall interface.

\begin{figure}[H]
    \centering
    \includegraphics[width=0.6\textwidth]{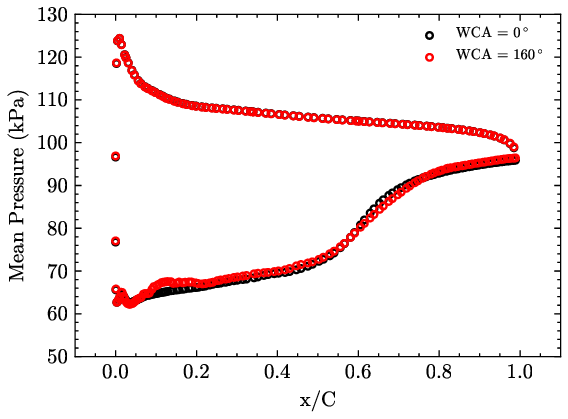}
\caption{Mean pressure distribution along the chord for WCA = $0^\circ$ and $160^\circ$ at $\sigma = 1.6$ for flow around a Clark Y hydrofoil at 8 degrees of angle of attack. The hydrophilic surface exhibits larger pressure fluctuation ranges due to unsteady cavity shedding and re-entrant jet activity.}
    \label{fig:mean_pressure_sigma16}
\end{figure}

In addition, Fig.~\ref{fig:mean_pressure_sigma16} shows the mean pressure distribution around the hydrofoil for both WCA of $0^{\circ}$ and $160^{\circ}$ at cavitation number of 1.6. Evidently, for this high cavitation number, there are pressure fluctuations over the wall for WCA of $160^{\circ}$ at leading edge part while the distribution remains smooth for WCA of $0^{\circ}$, consistent with the different bubble attachment behaviors observed.

These fundamental differences suggest that while both configurations operate under conditions generally unfavorable to sustained cloud cavitation (unlike $\sigma=0.4$ as seen in Fig.~\ref{fig:alpha_snapshots_sigma04}), surface wettability profoundly governs the local interface response and the stability of individual vapor structures. The hydrophilic surface favors vapor elongation and sustained wall adherence due to strong surface forces, creating stable vapor-wall contact and smooth pressure distributions. In contrast, the hydrophobic surface, characterized by weak adhesion and the absence of strong liquid-solid attraction, promotes frequent bubble detachment and reattachment cycles, leading to smaller, more mobile vapor structures and corresponding pressure fluctuations at the interface. Consequently, the spatial extent, detachment frequency, and overall vapor coverage pattern are all significantly influenced by WCA, emphasizing its critical role even under weak cavitation regimes where individual bubble dynamics are paramount.

\section{Conclusion}
This study systematically examined the influence of wall wettability, quantified through the wall contact angle (WCA), on cavitation patterns and flow stability around a Clark Y hydrofoil using high-fidelity LES. A dynamic contact angle model was employed to represent wall wettability effects, and simulations were conducted across three cavitation numbers ($\sigma = 1.6$, $0.8$, and $0.4$) to capture distinct cavitation regimes. The findings highlight the critical role of WCA as a passive control parameter in modulating cavitation inception, vapor structure evolution, and hydrodynamic performance.

At the high cavitation number ($\sigma = 1.6$), corresponding to incipient or sheet cavitation conditions, the influence of WCA was evident in shaping vapor bubble behavior and surface interactions. Hydrophilic surfaces ($\text{WCA} = 0^\circ$) promoted the formation of elongated vapor cavities with strong adhesion to the wall, resulting in smoother pressure distributions and minimal pressure fluctuations. In contrast, superhydrophobic surfaces ($\text{WCA} = 160^\circ$) led to the formation of smaller, detached, and highly mobile vapor bubbles that generated localized pressure variations. These findings suggest that at high cavitation numbers, surface wettability primarily governs individual bubble dynamics and localized pressure field characteristics.

At the moderate cavitation number ($\sigma = 0.8$), characteristic of cloud cavitation regimes, WCA demonstrated a pronounced effect on cavity development and shedding dynamics. As WCA increased, cavitation inception occurred earlier along the chord, and thicker, more extensive vapor cavities developed. Superhydrophobic surfaces facilitated larger vapor volume fractions, intensified vapor shedding, delayed velocity recovery in the wake, and amplified pressure fluctuations. The wake became increasingly asymmetric and turbulent as WCA rose, underscoring the destabilizing influence of hydrophobic surfaces on flow stability in this cavitation regime.

At the low cavitation number ($\sigma = 0.4$), indicative of supercavitation conditions, the impact of WCA on cavity morphology and stability was most significant. Hydrophilic surfaces exhibited intermittent cavity detachment and reattachment, driven by vigorous re-entrant jet activity, leading to strong temporal variations in vapor structure and associated pressure fluctuations. Conversely, superhydrophobic surfaces maintained stable, wall-adhered vapor layers along the hydrofoil chord, effectively suppressing re-entrant jets and stabilizing the cavitation interface. This resulted in sustained low-pressure zones and reduced unsteady loading on the hydrofoil surface.

This study demonstrates that increasing WCA consistently enhances cavitation onset, cavity thickness, and unsteadiness across all cavitation regimes, while hydrophilic surfaces offer greater flow stability by suppressing cavitation dynamics. These insights provide a foundation for using surface wettability as a passive and effective strategy for cavitation control in marine, hydraulic, and turbomachinery applications.

 \bibliographystyle{elsarticle-num} 

 \bibliography{references}
\end{document}